\RequirePackage{lineno}
\documentclass[
twocolumn,
eqsecnum,
prd,
floatfix,
superscriptaddress,
showpacs]{revtex4} 

\usepackage{graphicx}
\usepackage{dcolumn}
\usepackage{amsmath}
\usepackage{amssymb}
\usepackage{epstopdf}
\makeatletter
\def\btt#1{\texttt{\@backslashchar#1}}%
\DeclareRobustCommand\bblash{\btt{\@backslashchar}}%
\makeatother

\begin{document}
\title{Solar Neutrino Measurements in Super--Kamiokande--IV}
\newcommand{\AFFicrr}{\affiliation{Kamioka Observatory, Institute for Cosmic Ray Research, University of Tokyo, Kamioka, Gifu 506-1205, Japan}}
\newcommand{\AFFkashiwa}{\affiliation{Research Center for Cosmic Neutrinos, Institute for Cosmic Ray Research, University of Tokyo, Kashiwa, Chiba 277-8582, Japan}}
\newcommand{\AFFipmu}{\affiliation{Kavli Institute for the Physics and
Mathematics of the Universe (WPI), The University of Tokyo Institutes for Advanced Study,
University of Tokyo, Kashiwa, Chiba 277-8583, Japan }}
\newcommand{\AFFmad}{\affiliation{Department of Theoretical Physics, University Autonoma Madrid, 28049 Madrid, Spain}}
\newcommand{\AFFubc}{\affiliation{Department of Physics and Astronomy, University of British Columbia, Vancouver, BC, V6T1Z4, Canada}}
\newcommand{\AFFbu}{\affiliation{Department of Physics, Boston University, Boston, MA 02215, USA}}
\newcommand{\AFFbnl}{\affiliation{Physics Department, Brookhaven National Laboratory, Upton, NY 11973, USA}}
\newcommand{\AFFuci}{\affiliation{Department of Physics and Astronomy, University of California, Irvine, Irvine, CA 92697-4575, USA }}
\newcommand{\AFFcsu}{\affiliation{Department of Physics, California State University, Dominguez Hills, Carson, CA 90747, USA}}
\newcommand{\AFFcnm}{\affiliation{Department of Physics, Chonnam National University, Kwangju 500-757, Korea}}
\newcommand{\AFFduke}{\affiliation{Department of Physics, Duke University, Durham NC 27708, USA}}
\newcommand{\AFFfukuoka}{\affiliation{Junior College, Fukuoka Institute of Technology, Fukuoka, Fukuoka 811-0295, Japan}}
\newcommand{\AFFgifu}{\affiliation{Department of Physics, Gifu University, Gifu, Gifu 501-1193, Japan}}
\newcommand{\AFFgist}{\affiliation{GIST College, Gwangju Institute of Science and Technology, Gwangju 500-712, Korea}}
\newcommand{\AFFuh}{\affiliation{Department of Physics and Astronomy, University of Hawaii, Honolulu, HI 96822, USA}}
\newcommand{\AFFkek}{\affiliation{High Energy Accelerator Research Organization (KEK), Tsukuba, Ibaraki 305-0801, Japan }}
\newcommand{\AFFkobe}{\affiliation{Department of Physics, Kobe University, Kobe, Hyogo 657-8501, Japan}}
\newcommand{\AFFkyoto}{\affiliation{Department of Physics, Kyoto University, Kyoto, Kyoto 606-8502, Japan}}
\newcommand{\AFFmiyagi}{\affiliation{Department of Physics, Miyagi University of Education, Sendai, Miyagi 980-0845, Japan}}
\newcommand{\AFFnagoya}{\affiliation{Institute for Space-Earth Enviromental Research, Nagoya University, Nagoya, Aichi 464-8602, Japan}}
\newcommand{\AFFkmi}{\affiliation{Kobayashi-Maskawa Institute for the Origin of Particles and the Universe, Nagoya University, Nagoya, Aichi 464-8602, Japan}}
\newcommand{\AFFpol}{\affiliation{National Centre For Nuclear Research, 00-681 Warsaw, Poland}}
\newcommand{\AFFsuny}{\affiliation{Department of Physics and Astronomy, State University of New York at Stony Brook, NY 11794-3800, USA}}
\newcommand{\AFFokayama}{\affiliation{Department of Physics, Okayama University, Okayama, Okayama 700-8530, Japan }}
\newcommand{\AFFosaka}{\affiliation{Department of Physics, Osaka University, Toyonaka, Osaka 560-0043, Japan}}
\newcommand{\AFFregina}{\affiliation{Department of Physics, University of Regina, 3737 Wascana Parkway, Regina, SK, S4SOA2, Canada}}
\newcommand{\AFFseoul}{\affiliation{Department of Physics, Seoul National University, Seoul 151-742, Korea}}
\newcommand{\AFFshizuokasc}{\affiliation{Department of Informatics in
Social Welfare, Shizuoka University of Welfare, Yaizu, Shizuoka, 425-8611, Japan}}
\newcommand{\AFFskk}{\affiliation{Department of Physics, Sungkyunkwan University, Suwon 440-746, Korea}}
\newcommand{\AFFtokyo}{\affiliation{The University of Tokyo, Bunkyo, Tokyo 113-0033, Japan }}
\newcommand{\AFFtodai}{\affiliation{Department of Physics, University of Tokyo, Bunkyo, Tokyo 113-0033, Japan }}
\newcommand{\AFFtoronto}{\affiliation{Department of Physics, University of Toronto, 60 St., Toronto, Ontario, M5S1A7, Canada }}
\newcommand{\AFFtriumf}{\affiliation{TRIUMF, 4004 Wesbrook Mall, Vancouver, BC, V6T2A3, Canada }}
\newcommand{\AFFtokai}{\affiliation{Department of Physics, Tokai University, Hiratsuka, Kanagawa 259-1292, Japan}}
\newcommand{\AFFtsinghua}{\affiliation{Department of Engineering Physics, Tsinghua University, Beijing, 100084, China}}
\newcommand{\AFFuw}{\affiliation{Department of Physics, University of Washington, Seattle, WA 98195-1560, USA}}

\AFFicrr
\AFFkashiwa
\AFFmad
\AFFbu
\AFFubc
\AFFbnl
\AFFuci
\AFFcsu
\AFFcnm
\AFFduke
\AFFfukuoka
\AFFgifu
\AFFgist
\AFFuh
\AFFkek
\AFFkobe
\AFFkyoto
\AFFmiyagi
\AFFnagoya
\AFFkmi
\AFFpol
\AFFsuny
\AFFokayama
\AFFosaka
\AFFregina
\AFFseoul
\AFFshizuokasc
\AFFskk
\AFFtokai
\AFFtokyo
\AFFtodai
\AFFipmu
\AFFtoronto
\AFFtriumf
\AFFtsinghua
\AFFuw

\author{K.~Abe}
\AFFicrr
\AFFipmu
\author{Y.~Haga}
\AFFicrr
\author{Y.~Hayato}
\AFFicrr
\AFFipmu
\author{M.~Ikeda}
\AFFicrr
\author{K.~Iyogi}
\AFFicrr 
\author{J.~Kameda}
\author{Y.~Kishimoto}
\AFFicrr
\AFFipmu 
\author{Ll.~Marti}
\AFFicrr
\author{M.~Miura} 
\author{S.~Moriyama} 
\author{M.~Nakahata}
\AFFicrr
\AFFipmu 
% ...

\collaboration{The Super-Kamiokande Collaboration}
\noaffiliation

\date{\today}

\begin{abstract}
Upgraded electronics, improved water system dynamics, better calibration and
analysis techniques allowed Super-Kamiokande-IV to clearly observe very
low-energy $^8$B solar neutrino interactions, with recoil electron kinetic energies
as low as 3.49 MeV. Super-Kamiokande-IV data-taking
began in September of 2008; this paper includes data until February 2014,
a total livetime of 1664 days. The measured solar neutrino flux is
$(2.308\pm0.020(\text{stat.})^{+0.039}_{-0.040}(\text{syst.}))\times10^6/(\text{cm}^{2}\text{sec})$
assuming no oscillations. The observed recoil electron energy spectrum is
consistent with no distortions due to neutrino oscillations.
An extended maximum likelihood fit to the amplitude of the expected solar zenith
angle variation of the neutrino-electron elastic scattering rate in SK-IV
results in a day/night asymmetry of
$(-3.6\pm1.6(\text{stat.})\pm0.6(\text{syst.}))\%$.
The SK-IV solar neutrino data determine the solar mixing angle as
$\sin^2\theta_{12}=0.327^{+0.026}_{-0.031}$, all SK solar data (SK-I, SK-II, SK III
and SK-IV) measures this angle to be $\sin^2\theta_{12}=0.334^{+0.027}_{-0.023}$,
the determined mass-squared splitting is $\Delta m^2_{21}=4.8^{+1.5}_{-0.8}\times10^{-5}$ eV$^2$.
\end{abstract}


\maketitle

\section{Introduction}
Solar neutrino flux measurements from Super-Kamiokande (SK)~\cite{sk1} and
the Sudbury Neutrino Observatory (SNO)~\cite{firstsno}
have provided clear evidence for solar neutrino flavor conversion in
which electron flavor neutrinos convert to either muon or tau flavor
neutrinos.
This flavor conversion is well described by flavor oscillations of three neutrinos.
In particular, the extracted oscillation parameters agree with nuclear
reactor anti-neutrino measurements~\cite{kamland}. However, while
oscillations of reactor antineutrinos at the solar frequency were observed,
there is still no clear evidence that the solar neutrino flavor conversion
is indeed due to neutrino oscillations and not caused by another mechanism.
Currently there are two types of testable signatures unique to neutrino
oscillations, the first being the observation and precision test of the
Mikheyev--Smirnov--Wolfenstein (MSW) resonance curve~\cite{msw}, the 
characteristic energy dependence of the flavor conversion (assuming
oscillation parameters extracted from solar neutrino and reactor anti-neutrino
measurements):
higher energy solar neutrinos (higher energy $^8$B and $hep$
neutrinos) undergo adiabatic resonant conversion within the Sun (present
data imply a survival probability of about $30\%$), while the
flavor changes of the lower energy solar neutrinos ($pp$, $^7$Be, $pep$, CNO
and lower energy $^8$B neutrinos) arise only from vacuum oscillations. These
averaged vacuum oscillations lead to an average survival probability which
-- for sufficiently small $1-3$ mixing -- must exceed $50\%$ (present data imply about $60\%$).
The transition from the matter-dominated oscillations within the Sun to the
vacuum-dominated oscillations should occur near three MeV. This makes $^8$B
neutrinos the best choice when looking for a transition point within the energy
spectrum. A second signature unique to oscillations arises from the effect of
the terrestrial matter density on solar neutrino oscillations. 
This effect is tested directly by comparing solar neutrinos that pass
long distances through the Earth at nighttime to those which do not pass
through the Earth during the daytime.
Those neutrinos which pass through the Earth
will generally have an enhanced electron neutrino content, leading
to an increase in the nighttime electron elastic
scattering rate (or any charged-current interaction rate), and hence a negative
``day/night asymmetry'' $(r_D-r_N)/r_{\mbox{\tiny ave}}$, where
$r_D$ ($r_N$) is the daytime (nighttime) rate and
$r_{\mbox{\tiny ave}}=\frac{1}{2}(r_D+r_N)$ is the average rate. SK is sensitive
to $^8$B and hep solar neutrinos in the energy range around 4 to $~18.7$ MeV
and precisely measures the neutrino interaction time.
It is therefore a good detector to search for both solar neutrino oscillation
signatures.

SK~\cite{sk} is a large, cylindrical, water Cherenkov detector containing
of 50,000 tons of ultra-pure water.  It is located 1,000 m beneath the peak of
Mount Ikenoyama, in Kamioka Town, Japan. The SK detector is optically
separated into a 32.5 kton cylindrical inner detector (ID) surrounded by a $\sim 2.5$ meter
water shield, $\sim 2$ m of which is the active veto outer detector (OD). The structure dividing the detector
regions contains an array of photo-multiplier tubes (PMTs).
SK started data-taking in April of 1996, with 11,146 ID and 1,885
OD PMTs, and was then shut down for maintenance in June of 2001. This
period is called SK-I \cite{sk1}. While refilling the tank with water in
November of 2001, a PMT implosion caused a chain reaction which destroyed
$60\%$ of the PMTs. The surviving and new PMTs were redistributed and
covered with
fiber-reinforced plastic (FRP) and acrylic cases, in order to avoid another
accidental chain reaction.  Data-taking re-started with 5,182 ID
and 1,885 OD PMTs in December of 2002, and the period until
October of 2005 is called SK-II \cite{sk2}. In October of 2006, newly
manufactured PMTs replaced those which had been destroyed, and with 11,129
ID and 1,885 OD PMTs data-taking resumed as the SK-III phase \cite{sk3}.
The fourth phase of SK (SK-IV) began in September of 2008, with new front-end
electronics (QTC Based Electronics with Ethernet, QBEE~\cite{qbee}) for both
the ID and OD, new data acquisition system, and continues to this day.
This paper will include data taken up until the beginning of February 2014.

Improvements in the front-end electronics, the water circulation system,
calibration techniques and the analysis methods have allowed the SK-IV solar
neutrino measurements to be made with a lower energy threshold and
smaller systematic uncertainties, compared to SK-I, II and III. The hardware and
software improvements are summarized in section II, while the SK-IV data set,
data reduction, and its systematic uncertainty estimations on the total flux
are detailed in section III.
The simulation of solar neutrino events in SK is described also in section III.
Unfortunately, the simulation code for the SK-III period used in~\cite{sk3}
was inaccurate, which affected the input recoil electron spectrum.
The details (and the correction applied) as well as a reanalysis of the
SK-III data are briefly described in section III and Appendix A.

In section IV, the energy spectrum results of SK-IV as well as all SK phases
combined are discussed. Section V presents the SK-IV day/night asymmetry analysis.
Finally, section VI contains an oscillation analysis
of SK-IV data by themselves and in combination with other SK phases,
and also a global analysis which combines the SK results
with other relevant experiments.

In previous SK solar neutrino publications~\cite{sk1,sk2,sk3} ``energy''
meant total recoil electron energy, while in this paper we subtract the
electron mass $m_e=511$ keV to obtain kinetic energy. The kinetic energy
threshold of the SK-IV data analysis is thus $3.49$ MeV, corresponding to
the total energy of $4.00$ MeV.

\section{Detector Performance}
\subsection{Electronics, Data acquisition system}
To ensure stable observation and to improve the
sensitivity of the detector, new front-end electronics called QBEEs were
installed, allowing for the development of a new online data acquisition
system. The essential components on the QBEEs used for the analog signal
processing and digitization are the QTC (high-speed Charge-to-Time Converter)
 ASICs \cite{qbee}, which achieve very high speed signal processing and allow
the integration of the charge and recording of the time of every PMT signal.
These PMT signal times and charge integrals are sent to online computers, where
a software trigger searches for timing coincidences within 200 ns to pick out
events in a similar fashion as the hardware ``hitsum trigger'' did in
SK-I through III~\cite{sk1,sk2,sk3}. The energy threshold of this coincidence trigger is determined by the number of coincident PMT signals that are required:
a smaller
coincidence level will be more sensitive to lower energy events, but will
result in larger event rates. The definitions of the  different trigger types
and the corresponding typical event rates are summarized in Table~\ref{tab:trigrates}. 
Since all PMT signals are digitized and recorded, there is no deadtime of the
detector from a large trigger rate, so the efficiency of triggering on HE
events does not limit the maximum possible rate of SLE triggers; only the processing
capability of the online computers limits this maximum rate. The software trigger
system uses flexible event time periods (1.3 $\mu$sec for SLE, 40 $\mu$sec for LE and HE).
The trigger efficiencies for the thresholds are $\sim84\%$ ($\sim99\%$) between 3.49 and 3.99 MeV
(3.99 and 4.49 MeV) and $100\%$ above 4.49 MeV.


\begin{table}[h]
\begin{center}
\caption{Normal data-taking trigger types along with the threshold of hits and
average trigger rates.}
\begin{tabular}{c c c} \hline\hline
Trigger Type  & Hits in 200 ns & Trigger Rate    \\ \hline
Super Low Energy (SLE)  & 34 & 3.0-3.4 kHz \\ 
Low Energy (LE)  & 47 & $\sim40$ Hz \\ 
High Energy (HE)  & 50 & $\sim10$ Hz \\ \hline\hline
\end{tabular}
\label{tab:trigrates}
\end{center}
\end{table}

\subsection{Water system}
To keep the long light attenuation length of the SK water stable, the water
is continuously purified with a flow rate of 60 ton/hour. Purified water
supplied to the bottom of the detector replaces water drained from its top.
A higher temperature of the supply water than the detector temperature
results in convection throughout the detector volume. This convection transports
radioactive radon gas, which is produced by radioactive
decays from the U/Th chain near the edge of the detector into the central region
of the detector. Radioactivity coming from the decay products of radon gas (most
commonly $^{214}$Bi beta decays) mimics the lowest energy solar neutrino events.  
In January of 2010, a new automated temperature control
system was installed, allowing for control of the supply water temperature
at the $\pm0.01$ degree level. By controlling the water flow rate
and the supply water temperature with such high precision, convection within
the tank is kept to a minimum and the background level in the central
region has since become significantly lower.

\begin{figure}
 \begin{center}
 \includegraphics[width=3.5cm,clip]{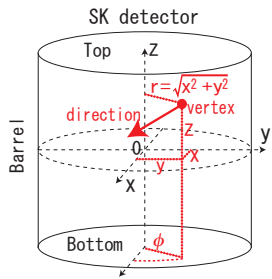}
 \caption{Definition of the SK detector coordinate system.}
 \label{fig:coord}
 \end{center}
\end{figure}

\begin{figure}
 \begin{center}
 \includegraphics[width=8cm,clip]{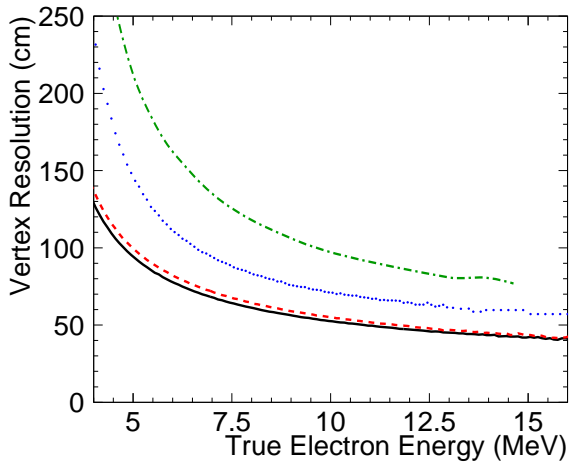}
 \caption{Vertex resolution for SK-I, II, III and IV shown by the dotted (blue),
dashed-dotted (green), dashed (red) and solid (black) lines, respectively. The SK-III vertex
resolution improvement over SK-I comes from using an improved vertex
reconstruction while the slightly improved timing resolution and better agreement between
data and simulated events are responsible for the further improvement in
SK-IV.}
 \label{fig:vres}
 \end{center}
\end{figure}

\subsection{Event reconstruction}
\label{sec:recon}
The methods used for the vertex, direction, and energy reconstructions are
the same as those used for SK-III \cite{sk3}. The Cartesian coordinate
system for the SK detector is shown in Fig.~\ref{fig:coord}.

\subsubsection{Vertex}
\begin{figure}
 \begin{center}
 \includegraphics[width=8cm,clip]{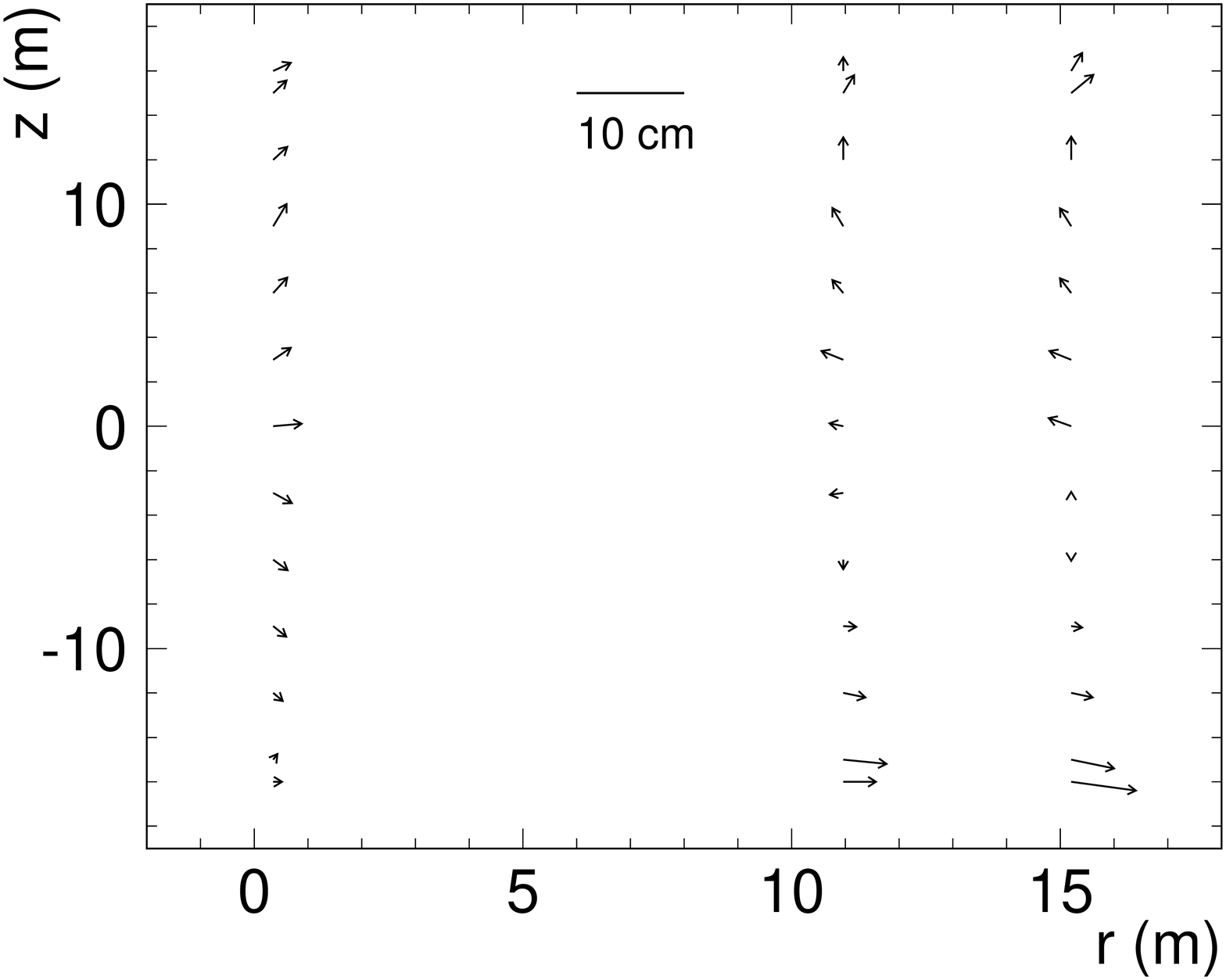}
 \caption{Vertex shift of the Ni calibration events in SK-IV.  The start of
the arrow is at the true Ni-Cf source position and the direction indicates
the averaged vertex shift at that position.  The length of the arrow
indicates the magnitude of the vertex shift. To make the vertex shifts easier
to see this length is scaled up by a factor of 20.}
 \label{fig:vshift_sk4}
 \end{center}
\end{figure}

The vertex reconstruction
is a maximum likelihood fit to the arrival times of the Cherenkov light at the
PMTs~\cite{sk2}. Fig.~\ref{fig:vres} shows the vertex resolution for each SK phase.
The large improvement in SK-III compared to SK-I is the result of using an
advanced vertex reconstruction program, while the improved timing resolution
and slightly better agreement of the timing residuals between data and
Monte Carlo (MC) simulated events are responsible for the additional
improvement of SK-IV. We observed a bias in the reconstructed vertex called
the vertex shift. This vertex shift is measured with a gamma-ray source
at several positions within the SK detector: neutrons from spontaneous fission of
$^{252}$Cf are thermalized in water and then captured on nickel in a
spherical vessel~\cite{sk,skcalib}. The nickel then emits ~9 MeV gammas
(Ni calibration source). Fig.~\ref{fig:vshift_sk4} shows the shift of
the reconstructed vertex of these Ni gammas in SK-IV from their true position (assumed to be
the source position). The SK-IV vertex shift is improved compared with SK-I, II
and III~\cite{sk,sk2,sk3}.

\subsubsection{Direction}
A maximum likelihood fit comparing the Cherenkov ring pattern of data to MC
simulations is used to reconstruct event directions. During the SK-III phase an energy
dependence was included in the likelihood and the angular resolution was
improved by about $10\%$ (10 MeV electrons) compared to SK-I.  The angular
resolution in SK-IV is similar to that in SK-III.

\subsubsection{Energy}
\label{sec:energyrecon}
The energy reconstruction is based on the number of PMT hits within a 50 ns
time window, after the photon travel time from the vertex is subtracted.
This number is then corrected for water transparency, dark noise, late
arrival light (due to scattering and reflection), multi-photon hits,
etc., producing an effective number of hits $N_{\text{eff}}$ (see~\cite{sk3}).
Simulations of mono-energetic electrons are used to produce a function
relating $N_{\text{eff}}$ to the recoil electron energy (MeV).

\begin{figure}
 \begin{center}
 \includegraphics[trim=0cm 2cm 0cm 0.5cm,width=9cm,clip]{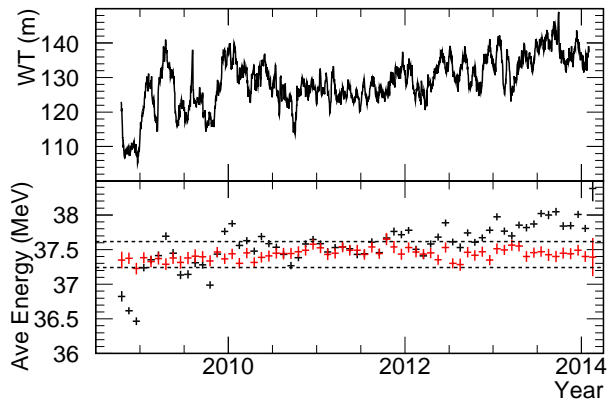}
 \caption{(Top) Time variation of the water transparency as
measured by decay electrons.  (Bottom) Time variation
of the mean reconstructed energy of $\mu$ decay electrons before (after) 
water-transparency correction in black (red).  Before the correction,
a water transparency of 90 m is assumed, then the mean value of 
the distribution is adjusted to that of the after correction.
After the correction the mean energy is stable within $\pm0.5\%$ (dashed lines).}
\label{fig:decaye}
 \end{center}
\end{figure}

The water transparency parameter used in the energy reconstruction is measured
using decay electrons from cosmic-ray muons. This method of obtaining
the water transparency is the same as for SK-I, II and III \cite{sk1,sk2,sk3}:
exploiting the azimuthal symmetry of the Cherenkov cone, we determine
the light intensity as a function of light travel distance and fit
it with an exponential light attenuation function. The top panel of
Fig.~\ref{fig:decaye} shows the time variation of the measured water transparency, while the
bottom panel shows the reconstructed mean energy of $\mu$ decay electrons
in black (red) before (after) water transparency corrections have been applied.
The stability of the water transparency
corrected energy reconstruction is within $\pm0.5\%$ (dashed lines).

\begin{figure}[t]
 \begin{center}
 \includegraphics[width=8cm,clip]{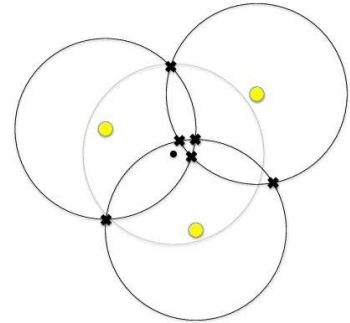}
 \caption{Schematic view of the event direction candidates used to calculate
the multiple scattering goodness. The yellow points represent PMT hits and the
black circles surrounding them are the projections of the $42^{\circ}$ cones
centered around each hit.  The black crosses give the intersection points
of the cones. The vectors from the event vertex position to these
intersection points are taken as
event direction candidates.  The black dot shows the event best fit
direction and the gray circle is the projection of its Cherenkov
cone onto the inner detector wall.  The intersections will cluster around the
event direction.}
 \label{fig:mcsg}
 \end{center}
\end{figure}

\begin{figure}[t]
 \begin{center}
 \includegraphics[width=8cm,clip]{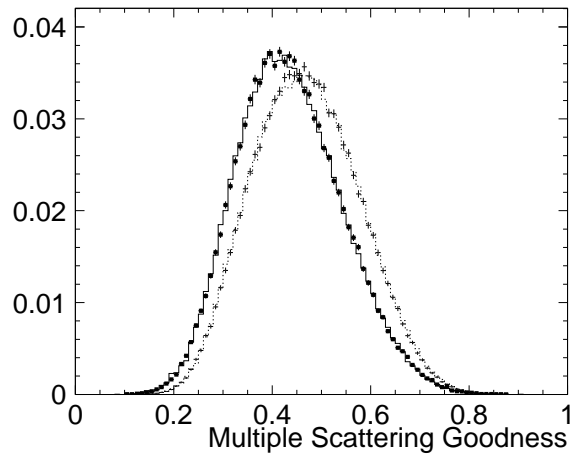}
 \caption{MSG for LINAC data (points) and MC (histogram), normalized by the number
of events. The solid (dotted) lines and points on that correspond to 4.38 MeV (8.16 MeV) electrons.}
 \label{fig:linac_mcsg} 
 \end{center}
\end{figure}

\subsubsection{Multiple scattering goodness (MSG)}
Even at the low energies of the recoil electrons from $^8$B solar
neutrino-electron scattering, the PMT hit pattern from the Cherenkov cone
reflects the amount of multiple Coulomb scattering recoil electrons experience.
Very low-energy electrons will
incur such scattering more than higher energy electrons and thus have a more
isotropic PMT hit pattern. Radioactive background events, such as $^{214}$Bi
beta decays, generally have less energy than $^8$B recoil electrons.
Radioactive background events with $\gamma$ emission will be more isotropic
still. The ``goodness'' of a directional fit characterizes this hit pattern
anisotropy: it is constructed by first projecting $42^\circ$ cones from
the vertex position, centered around each PMT that was hit within a 20 ns
time window (after time of flight subtraction).  Pairs of such cones are then
used to define ``event direction candidates'', which are vectors along the
intersection lines of the two cones. Only cone pairs which intersect twice are
used to define event direction candidates. Fig.~\ref{fig:mcsg} shows a
schematic view of how the event direction candidates are found.  The yellow
points represent hit PMTs, which will roughly be found around the Cherenkov
``ring'', the projection of the cone onto the inner detector wall shown by
the gray circle.  As seen in the figure, for pairs of PMTs with positions
located near the Cherenkov ring, one of the intersection lines shown by the
black crosses will fall close to the best fit direction vector shown as the
black point on the inner detector wall which this vector passes through.
Clusters of these event direction candidates are then found by associating
other event direction candidates which are within $50^\circ$ of a
``central event direction'' seeded by the candidates themselves.  Once an
event direction candidate has been associated to a cluster, it then
will not seed another cluster. The event direction candidate vectors of a
cluster are added together to adjust the central event direction. Several
iterations of this adjustment with subsequent cluster reassignment will
center the clusters and maximize the magnitude of the vector sum. The vector
sum with the largest magnitude is kept as the ``goodness direction''. 
The multiple scattering goodness (MSG) is then defined by the ratio of
this magnitude and the number of event direction candidates within the 20 ns
time window. The filled squares (error bars) and solid (dotted) lines of
Fig.~\ref{fig:linac_mcsg} compare the LINAC data and MC MSG distributions
for 4.38 MeV (8.16 MeV) electrons. As expected, higher energy electrons have
a larger mean MSG.

\begin{figure}
 \begin{center}
 \includegraphics[width=8cm,clip]{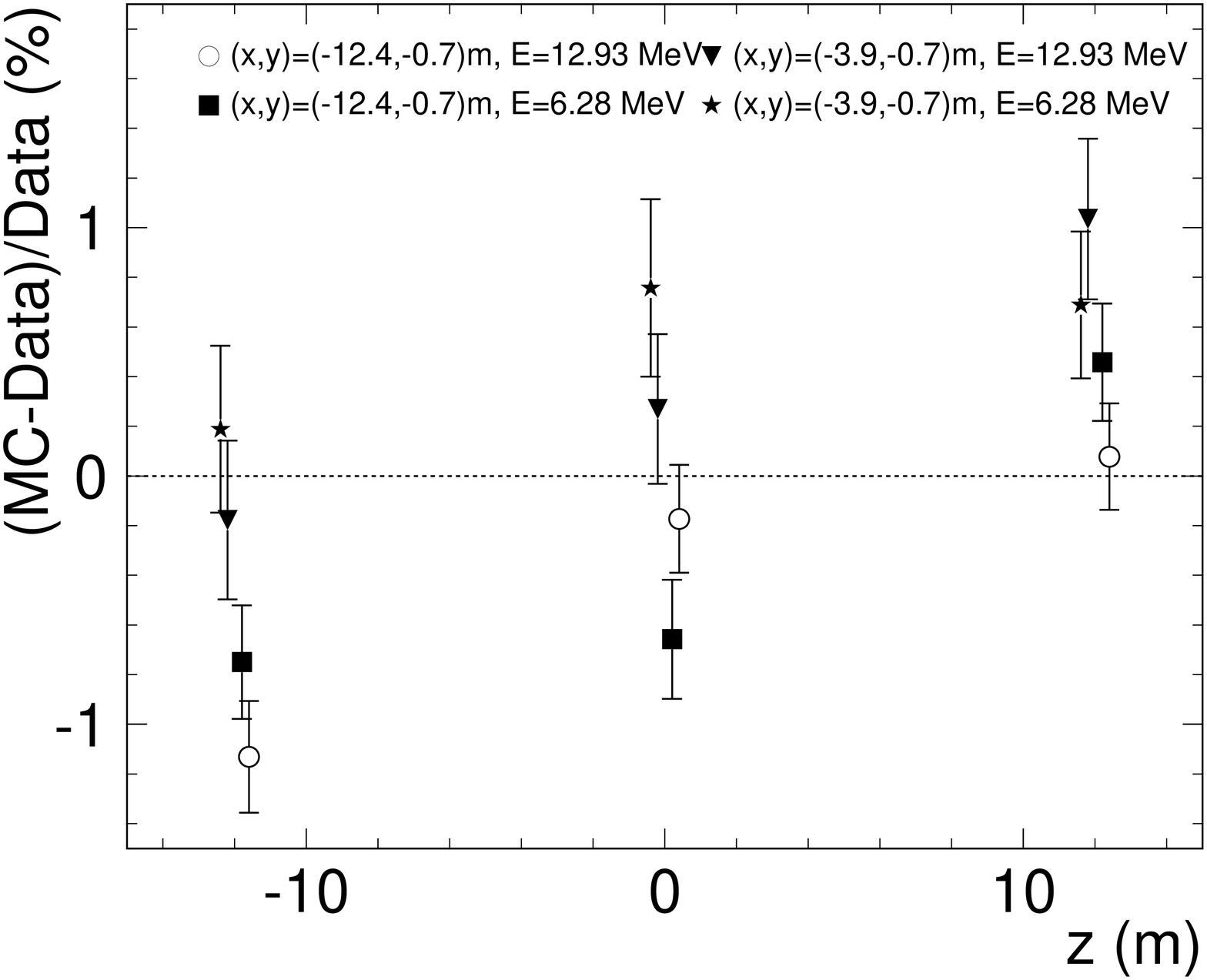}
 \caption{LINAC calibration $z$ position dependence of the absolute energy
scale of SK-IV.}
 \label{fig:esca_zdep}
 \end{center}
\end{figure} 

\begin{figure}[t]
 \begin{center}
 \includegraphics[width=8cm,clip]{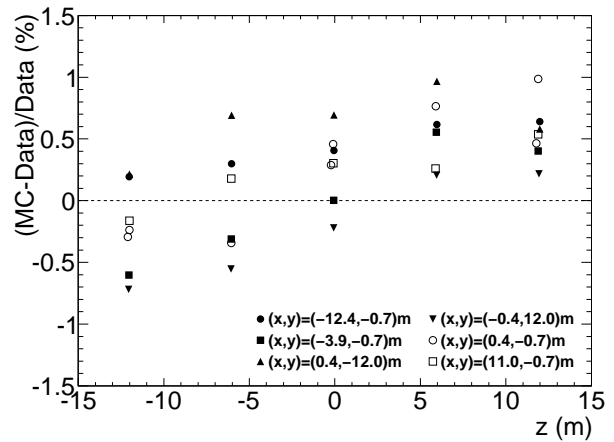}
 \caption{Difference of the mean reconstructed energy between data and
simulated events, at each position, coming from the SK-IV DT calibration.}
 \label{fig:dt_zdep}
 \end{center}
\end{figure}

\subsection{Energy calibration}\label{sec:energy}
\begin{figure}
 \begin{center}
 \includegraphics[width=8cm,clip]{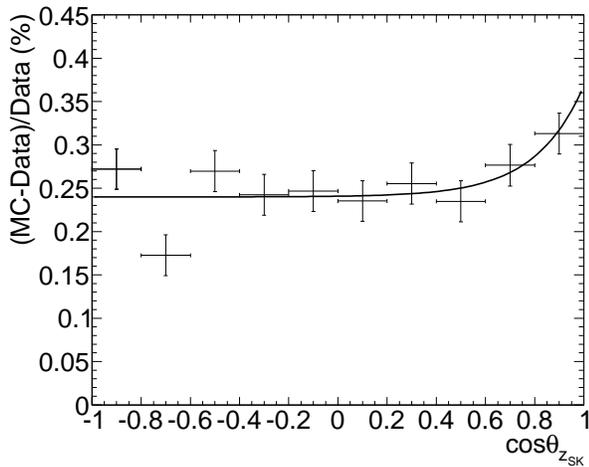}
 \caption{Difference of the mean energy between data and simulated events
as a function of the zenith angle in the SK-IV detector for DT calibration.
After subtracting the absolute offset, the uncertainty is estimated to be
$\pm0.1\%$.}
 \label{fig:dt_dirdep}
 \end{center}
\end{figure}

The absolute energy scale is determined by an electron linear accelerator
(LINAC) \cite{LINAC}. The LINAC calibration system injects single monoenergetic
electrons into SK
in the downward direction. 
The energy of the momentum-selected electrons is precisely measured by a
germanium (Ge) detector using a thin titanium window similar to that
used under the water.
To determine the energy scale, 6.28
and 12.93 MeV electron data are compared to simulated events.
Fig.~\ref{fig:esca_zdep} shows the $z$ dependence of this comparison.
We cross-check the energy scale obtained from the LINAC energy with $^{16}$N $\beta$/$\gamma$
decays, which originate from the (n,p) reaction of
$^{16}$O with neutrons produced by a deuterium-tritium (DT) fusion neutron
generator~\cite{DT}. The 10.5 MeV endpoint $^{16}$N decays of the DT
calibration are isotropic, with $66\%$ of the decays emitting a ~6 MeV
$\gamma$ in conjunction with an electron. DT-produced $^{16}$N data are taken
at a much larger number of positions in SK than LINAC data.
Fig.~\ref{fig:dt_zdep} compares the reconstructed energy of $^{16}$N
simulated events with data, as a function of the $z$ position of the production.
Fig.~\ref{fig:dt_dirdep} shows the directional dependence of the energy
scale, with respect to the detector zenith angle. The two bins between
$\cos\theta_{z_{SK}}=0.6$ and 1 are affected by increased shadowing
from the DT generator. Conservatively, we fit the entire data with a linear
combination of a constant and an exponential function to estimate the
systematic uncertainty on the day/night asymmetry due to the
directional dependence of the bias of the reconstructed energy.

The systematic
uncertainty of the energy scale due to position (direction) dependence
is estimated to be $0.44\%$ ($0.1\%$). The effect of the water transparency
variation during LINAC calibration is estimated to be $0.2\%$, while the
uncertainty of the LINAC electron beam energy (as measured by the Ge
detector), is estimated to be $0.21\%$. The total systematic uncertainty of
the absolute energy scale thus becomes $0.54\%$, calculated by adding all the
contributions in quadrature, and is summarized in Table~\ref{tab:escale_sys}.
These uncertainties are similar to those in SK-III ($0.53\%$).

\begin{table}[h]
\begin{center}
\caption{Systematic uncertainty of the energy scale\label{tab:escale_sys}.}
\begin{tabular}{l c} \hline\hline               
Position Dependence  & $0.44\%$\\ 
Direction Dependence & $0.10\%$\\ 
Water Transparency   & $0.20\%$\\
LINAC Energy         & $0.21\%$\\ 
\hline
Total                & $0.54\%$\\ \hline \hline
\end{tabular}
\end{center}
\end{table}

The detector's energy resolution is determined using the same method as
described in \cite{sk3}.  Monoenergetic electrons are simulated and
used to determine the relationship between the effective number of hits in
the detector and the electron energy in MeV.  Using the width of Gaussian
fits to the energy distributions resulting from these simulated electrons,
the energy dependence of the energy resolution is well described by the function

\begin{equation}
\sigma(E)=-0.0839+0.349\sqrt{E}+0.0397E,
\end{equation}

\noindent in units of MeV, where $E$ is electron total energy.
This is comparable to the SK-III energy
resolution, given as $\sigma(E)=-0.123+0.376\sqrt{E}+0.0349E$ in \cite{sk3}.

\subsection{Light propagation in water}
\subsubsection{Water parameters}
The water transparency in the MC simulation is determined using absorption and scattering
coefficients as a function of wavelength (full details of this and other more
general detector calibrations can be found in \cite{skcalib}). These
coefficients are independently
measured by a nitrogen laser and laser diodes at five different wavelengths: 337 nm, 375 nm,
405 nm, 445 nm and 473 nm. Based on these measurements, the dominant
contribution to the variation of the water transparency is a
variation in the absorption length.
The absorption coefficient 
is time and position dependent, as explained below.
This SK-IV solar neutrino analysis only varies the absorption, and uses a single 
set of time independent scattering coefficients, as measured by the
laser diodes~\cite{skcalib}.

\subsubsection{Time dependence}
To track the absorption time dependence, we measure the light attenuation of Cherenkov
light from decay electrons (from cosmic-ray muons stopping throughout the SK
inner detector volume). This measurement uses the azimuthal symmetry of the
emitted Cherenkov cone to compare different light propagation path lengths
within the same event and assumes a simple exponential attenuation. This
effective attenuation length is one of the energy reconstruction parameters.
The top panel of Fig~\ref{fig:decaye} shows the decay electron water transparency parameter as 
a function of time.

\begin{figure}[t]
 \begin{center}
 \includegraphics[width=8cm,clip]{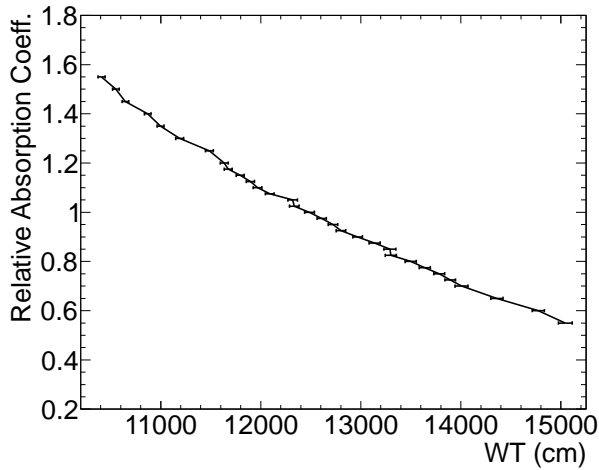}
 \caption{Change in the absorption coefficient, relative to the
coefficient when the absolute energy scale calibration was done, as a function
of the $\mu$ decay electron measured water transparency.\label{fig:abs_vs_wt}}
 \end{center}
\end{figure}

In orer to connect the absorption time dependence in the MC to the water
transparency parameter measured by decay electrons we generate mono-energetic
electron samples throughout the detector for a wide range of absorption
coefficients with nine different energies between 4 and 50 MeV. Each MC sample
is assigned a particular decay electron water transparency parameter that
minimizes the difference between input energy and average reconstructed energy.
As expected, the relationship between water transparency and MC absorption
coefficent does not significantly depend on the generated energy. The same
procedure establishes the relationship between the (corrected) number of PMT 
hits and energy.  Fig.~\ref{fig:abs_vs_wt} shows the obtained relationship
between absorption coefficient and water transparency parameter. For
convenience we measure the absorption coefficient
relative to the coefficient at the time of the LINAC calibration data-taking,
which defines the energy scale (see \cite{skcalib}). We employ a linear
interpolation between the data points. The mean energy of these decay
electrons is used to evaluate the systematic uncertainty of the time dependence
of the energy scale (see bottom panel of Fig.~\ref{fig:decaye}). After
correction for the time variation of the absorption coefficient, the apparent
time dependence of the $\mu$ decay electron mean energy becomes smaller than
$\pm0.5\%$.

\begin{figure}
 \begin{center}
 \includegraphics[width=8cm,clip]{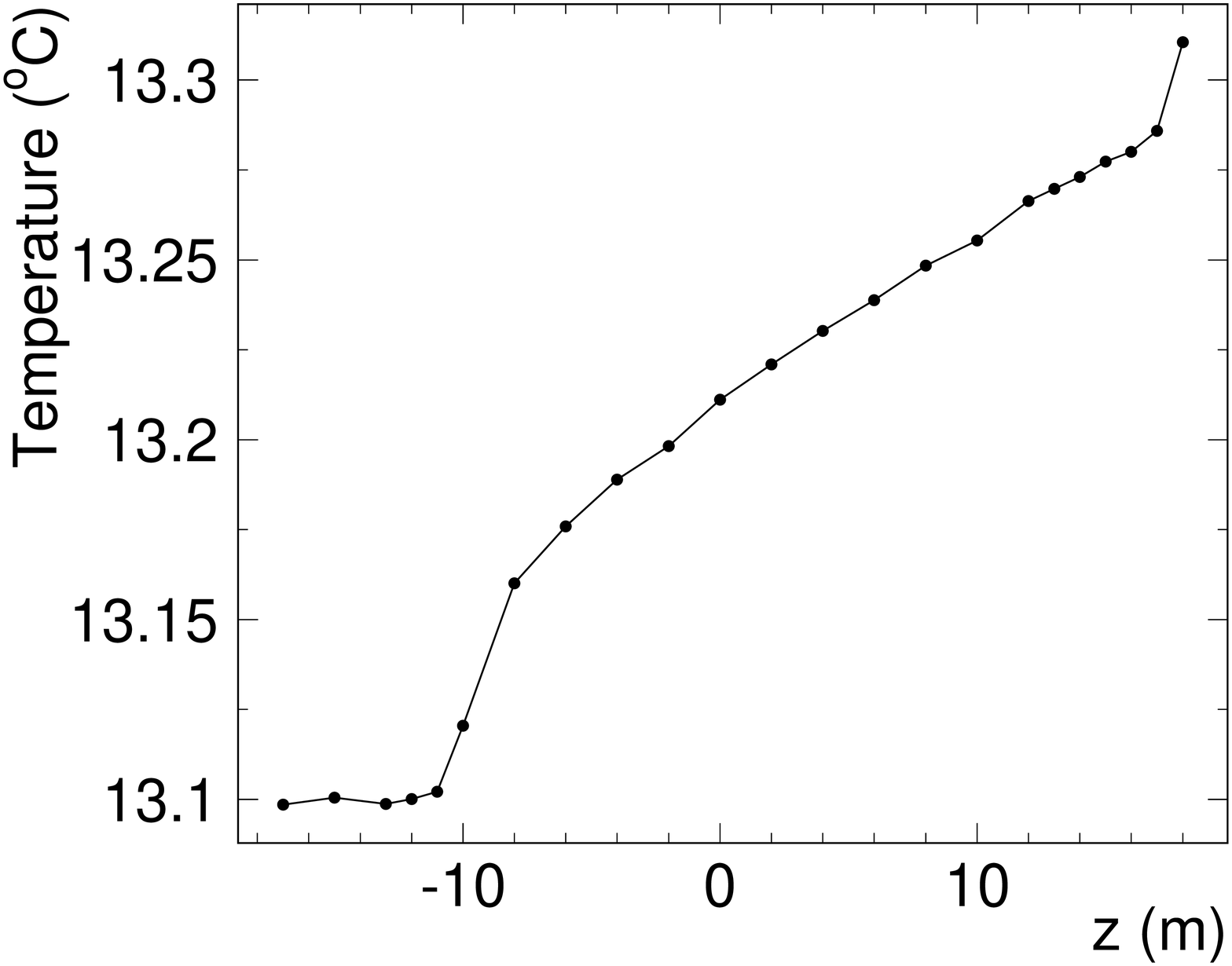}
 \caption{Typical $z$ dependence of the water temperature in SK detector.
Below $-11$ m the temperature is constant due to convection,
and so the absorption coefficient is assumed to be constant below this point.
 \label{fig:ztemp}}
 \end{center}
\end{figure}

\subsubsection{Position dependence}
As already explained, the water in the SK detector is continuously
recirculated through the SK water purification system. 
Water is drained from
the detector top, purified, and re-injected at the bottom. Due to careful
temperature control of the injected water, the
convection inside the SK tank is suppressed everywhere but at the bottom part of
the tank below $z=-11$ m. Fig.~\ref{fig:ztemp} shows the typical water
temperature as a function of $z$ in the SK detector.
The temperature is uniform
below $z=-11$ m, where convection is occurring and increases steadily above that.
We assume that absorption is strongly correlated with the amount of convection and model the
position dependence of the absorption length as constant below $-11$ m
and linearly changing above $-11$ m:

\begin{align}
\alpha_{abs}(\lambda,z,t)=\qquad\qquad\qquad\qquad\qquad\qquad\qquad \nonumber \\
\left\{
 \begin{array}{ll}
  \alpha(\lambda,t)(1+\beta(t) \cdot z) , &\quad  \text{for}\;z\geq -11\text{ m} \\
  \alpha(\lambda,t)(1-\beta(t) \cdot 11) , &\quad \text{for}\;z\leq -11\text{ m,}
 \end{array}
\right.
\end{align}

where $\beta$ parametrizes the $z$-dependence of the absorption.
The $\beta$ parameter is determined by studying the distribution
of hit PMTs of Ni calibration data (see section~\ref{sec:recon}) ~\cite{skcalib}
in the ``top'', ``bottom'' and ``barrel'' regions of the detector
(see Fig.~\ref{fig:coord}). After other detector asymmetries like
quantum efficiency variations of the PMTs are taken into account, the hit
rate of the top region in the detector is $3\sim5\%$ lower than that of the
bottom region. $\beta$ is then fit using the hit asymmetry of Ni calibration
events.  Since the Ni calibration hit pattern varies with
time, both $\alpha$ and $\beta$ depend on time. The Xe flash lamp scintillator
ball calibration system~\cite{skcalib} tracks the $\beta$ time dependence:
a Xe flash lamp powers a scintillator ball located near the middle of
the detector. The time dependence of $\beta$ is also monitored by
Ni calibration data. The introduction of $\beta$ into the MC
simulation has helped to reduce the systematic uncertainty on the energy
scale, as it addresses a significant contribution to its directional dependence.
This is important for the solar neutrino day/night asymmetry analysis.

\section{Data Analysis}
After installation of the new front-end electronics,
SK-IV physics data-taking started on October 6, 2008. This paper includes data
taken from October 6, 2008 until February 1, 2014.
%
The total livetime is 1664 days. The entire data period was taken with a
new very low energy threshold of 34 hits within $200$ ns (cf. Table
\ref{tab:trigrates}).
To reduce the required data storage capacity,
obvious backgrounds are removed using faster and less-stringent
implementations of the analysis cuts on fiducial volume, energy,
ambient events and external events, before the data is permanently stored.
By applying these pre-cuts, the data load was reduced to $\sim1\%$ of its
original size.

\subsection{Event selection}
Most of the cuts used are the same as those used in SK-III \cite{sk3}, but 
some of the cut values and the energy regions in which they are applied
are changed to optimize the significance: if $S$ ($BG$) is the number of
signal (background) events, we define the significance as $S/\sqrt{BG}$.
Also, as was the case in SK-III, below 4.99 MeV the fiducial
volume is reduced since backgrounds appear localized at the
bottom of the detector and at large radii.

\subsubsection{Ambient background reduction}
As in~\cite{sk1,sk2,sk3}, several cuts remove low-energy radioactive
backgrounds. These backgrounds originate mostly from the PMT enclosures, the
PMT glass, and the detector wall structure. While the true vertices
lie outside the fiducial volume, some radioactive background events are
mis-reconstructed inside the fiducial volume.
The quality of the event reconstruction is tested by variables describing
its goodness.  The first variable is a timing
goodness $g_t$ testing the ``narrowness'' of the PMT hit timing residuals,
which is defined in~\cite{sk2} (section III.B, equation 3.1).  The second is
a hit pattern goodness $g_p$ testing the azimuthal symmetry of the
Cherenkov cone ($g_p=0$ is perfectly symmetric, $g_p=1$ is completely
asymmetric).  Good single electron events must have $g_t^2-g_p^2$ greater than 0.2.
Events below 6.99 MeV (4.99 MeV) must have $g_t^2-g_p^2$ greater than
0.25 (0.29). The same cut was applied for SK-III.


\begin{figure}
 \begin{center}
 \includegraphics[width=8.5cm,clip]{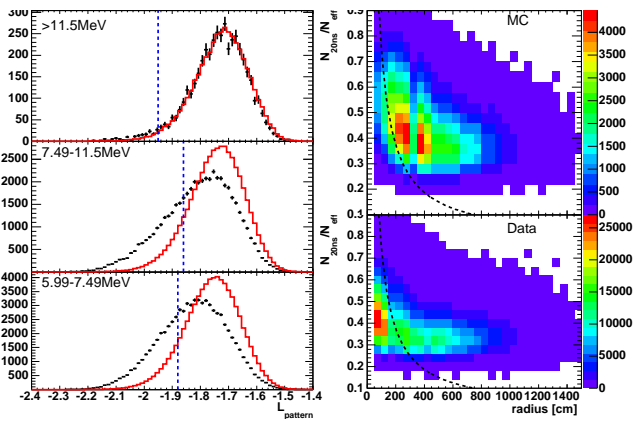}
 \caption{Left: Hit pattern likelihood distributions in
three different energy ranges for data (black error bars)
and MC (red histogram). The cut point is shown by the blue
dashed line. Right: Removal of small hit clusters. The top
panel shows the MC cluster size versus the cluster radius,
the bottom panel is the data. Events below the dashed black line
are removed.}
 \label{fig:patlikclus}
 \end{center}
\end{figure}

We also check the consistency of the observed light pattern with a single $42^\circ$
Cherenkov cone as in~\cite{sk1} (section VII.C, equation 7.4). This cut will remove
events with multiple Cherenkov cones, e.g., from beta decays to an excited nuclear state
with subsequent gamma emission. The hit pattern is assigned a likelihood based on
the direction fit likelihood function. Fig~\ref{fig:patlikclus} shows the likelihood
and cut criteria in three different energy ranges. Further details are found in~\cite{D-Nakano}.

A small hit cluster cut targets radioactive background events in the PMT
enclosures or glass, which coincide with an upward fluctuation of the PMT dark
noise.
Only events with a reconstructed $r^2$ bigger than 155 m$^2$ (120 m$^2$),
a reconstructed $z$ smaller than $-7.5$ m ($-3$ m), or a reconstructed $z$
bigger than 13 m, for reconstructed energies in $4.49 \sim 4.99$ MeV
($3.49 \sim 4.49$ MeV), are subject to this cut. To characterize small hit clusters, we
select PMT hits with times coincident within 20 ns (after time-of-flight
subtraction, see section~\ref{sec:energyrecon}), and then find the smallest
sphere around any of the selected PMTs that encloses
at least $20\%$ of all selected PMTs. This radius is multiplied by the ratio of
PMT hits coincident within 20 ns (without time-of-flight correction) divided by
$N_{\text{eff}}$ (see section~\ref{sec:energyrecon}). Solar neutrinos near the edge
of the fiducial
volume have a bigger radius$\times$hit ratio (see also section III~C
in~\cite{sk3}, Fig.~17 and 18) than the radioactive background. As in SK-III,
we remove events with radius$\times$hit ratio less than 75 cm as shown
in Fig.~\ref{fig:patlikclus}.

\begin{table}[h]
\caption{Locations used by the calibration source cut. The sources are described
in detail in ~\cite{skcalib}.}
\centerline{\begin{tabular}{l c c c c}
\hline\hline
Source               & $x$ (cm)  & $y$ (cm)   & $z$ (cm) \cr
\hline
Xenon flasher        &   353.5   &  $-70.7$   &     0.0  \cr
LED                  &   35.5    &  $-350.0$  &   150.0  \cr
TQ Diffuser Ball     &  $-176.8$ &  $-70.7$   &   100.0  \cr
DAQ Rate Test Source &  $-35.3$  &   353.5    &   100.0  \cr
Water Temp. Sensors 1&  $-35.3$  &   1200     &  $-2000$ \cr
Water Temp. Sensors 2&   70.7    &  $-777.7$  &  $-2000$ \cr
\hline\hline
\end{tabular}}
\label{tab:calsrccut}
\end{table}


Finally, we remove spurious events due to various calibration sources
(mostly radioactiv decays), if they are below 4.99 MeV. A reconstructed position closer
than 2 m to the source, or closer than 1 m to the source or 
water temperature sensor cable (all cables
run along the $z$ axis from the top down to the source position) means the
event is removed. Table~\ref{tab:calsrccut} lists the various calibration
sources which are considered. The fiducial volume is reduced by about 0.48kton
due to this cut.

\subsubsection{External event cut}
To remove radioactive background coming from the PMTs or the detector wall
structure, we calculate the distance to the PMT-bearing surface
from the reconstructed vertex looking back along the reconstructed event
direction. Radioactive backgrounds tend to appear ``incoming'', so we remove
events where this distance is small. Solar neutrino candidates above 7.49 MeV
(above 4.99 MeV and below 7.49 MeV) must have a distance of at least 4 m
(6.5 m). In the energy region below 4.99 MeV we distinguish between the ``top''
(cylinder top lid),  ``barrel'' (cylinder side walls) and ``bottom'' (cylinder
bottom lid) surfaces, shown in Fig.~\ref{fig:coord}. Candidates 
which come from
 the ``top'' (``bottom'') must have a distance of at least 10 m
(13 m), while ``barrel'' event candidate distances must exceed 12 m. SK-III
applied the same cuts.


\subsubsection{Spallation cut}
Some cosmic-ray $\mu$'s produce radioactive elements by breaking up an oxygen
nucleus~\cite{libeacomspall}. A spallation event occurs when these radioactive
nuclei eventually decay and emit $\beta$'s and/or $\gamma$'s. A spallation
likelihood function is made from the distance of closest approach between the
preceding $\mu$ track(s) and a solar neutrino candidate, their time difference,
and the charge deposited by the preceding $\mu$(s).
By using the likelihood function spallation-like events are rejected,
see~\cite{sk1,sk-Li9} for details.

When lower energy cosmic-ray $\mu^-$'s are captured by $^{16}$O nuclei in the
detector, $^{16}$N can be produced which decays with gamma-rays and/or
electrons with a half-life of 7.13 seconds. In order to reject these
events, the correlation between stopping $\mu$'s in the detector and the
remaining candidate events are checked. The cut criteria for $^{16}$N events
is as follows; (1) reconstructed vertex is within 250 cm to the stopping
point of the $\mu$, (2) the time difference is between 100 $\mu$sec and 30
sec.

To measure their impact on the signal efficiency, the spallation and $^{16}$N cuts are applied
to events that cannot be correlated with cosmic-ray muons (e.g. candidates
preceding muons instead of muons preceding candidates). This ``random sample''
then measures the accidental coincidences rate between the muons and subsequent
candidate events. 
The spallation ($^{16}$N) cut reduces signal efficiency by about 20$\%$ (0.53$\%$).

\subsubsection{Fiducial volume cut}
\begin{figure}
 \begin{center}
 \includegraphics[width=9cm,clip]{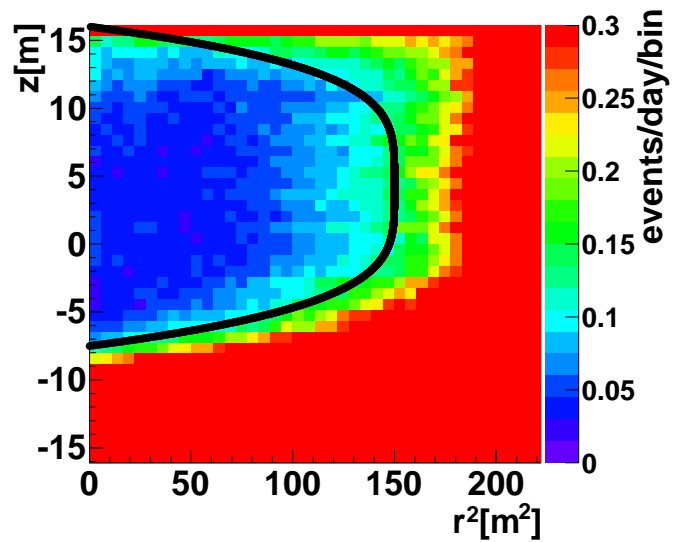}
 \caption{Vertex distribution for 3.49 to 3.99 MeV data.
Radioactive background leads to a large event rate at the bottom
and large radii. The black line indicates the reduced fiducial
volume in this energy region.}
 \label{fig:vertex45_50}
 \end{center}
\end{figure}

Events which occur near the wall of the detector (reconstructed
within 2 m from the ID edge) are rejected. 
The volume of this fiducial volume is 22.5 kton.
Below 4.99 MeV this cut is tightened.
Fig.~\ref{fig:vertex45_50} shows the $r^2$ $(=x^2+y^2)$ vs. $z$ data vertex
distribution for 3.49 to 3.99 MeV, after the above cuts. Each bin shows the
rate (events/day/bin), with blue showing a lower rate and red a higher rate.
We expect solar neutrino events to be uniformly distributed throughout the
detector volume, and the regions with high event rates are likely dominated by
background.  To increase the significance in the final data sample for this
energy region (3.49 to 4.49 MeV), we have reduced the fiducial volume to the
region shown by the black line in the figure and described by
\begin{equation}
 r^2 + \frac{150}{11.75^4}\times\left| z - 4.25\right|^4 \le 150,
\end{equation} 
\noindent where the coordinates are given in meters. This function was chosen
in order to approximately follow the contours of constant event rate.
For the energy range of
4.49 to 4.99 MeV, events which have $r^2>180\text{ m}^2$ or $z<-7.5$ m are cut.

\subsubsection{Other cuts}
Short runs ($<5$ minutes), runs with hardware and/or software problems, and
calibration runs are not used for this analysis. Cosmic-ray $\mu$ events are
removed by rejecting events with more than 400 hit PMTs, which corresponds
to about 60 MeV for electron type events.

\subsubsection{Summary}
\begin{figure}
 \begin{center}
 \includegraphics[width=8.5cm,clip]{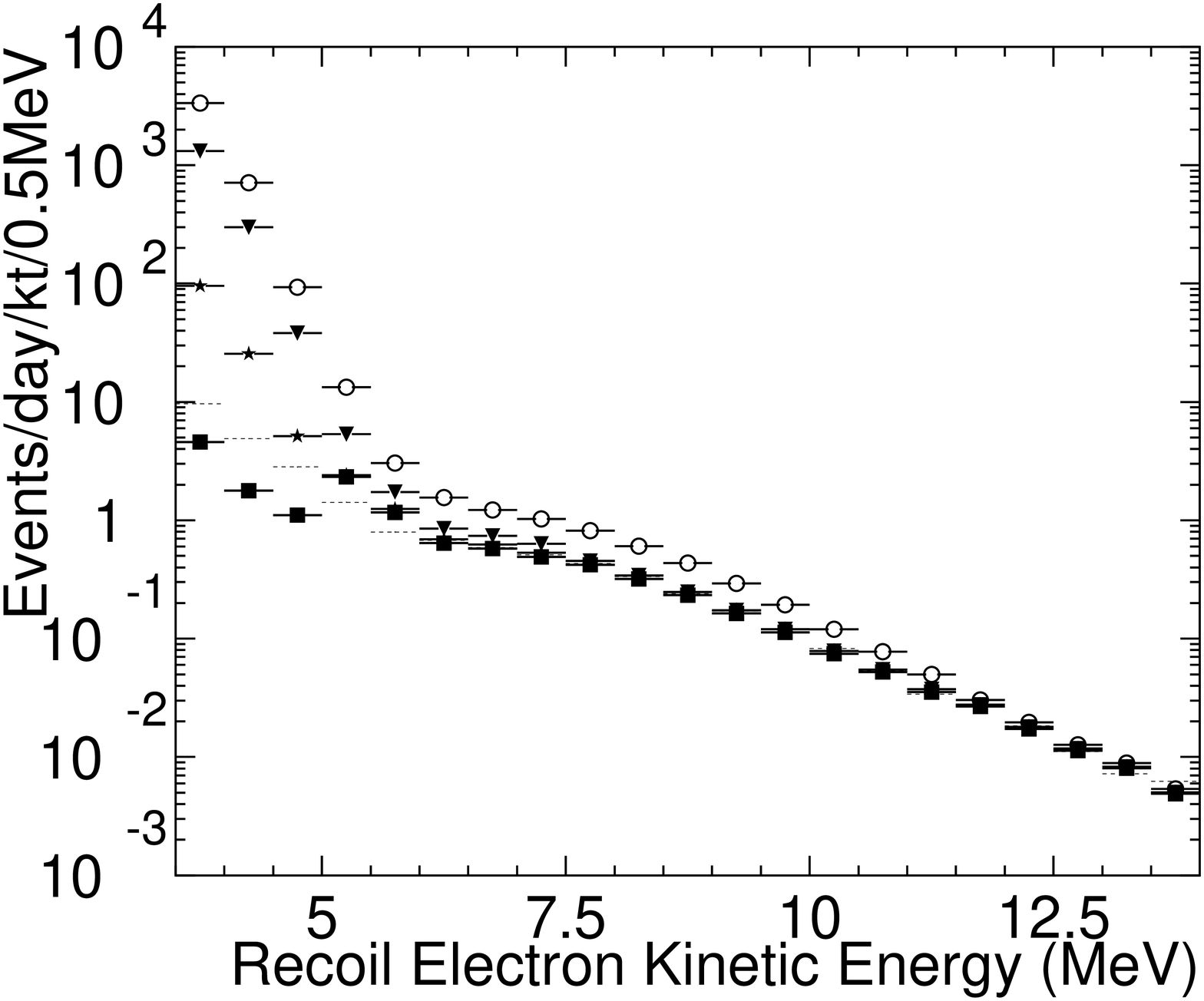}
 \caption{Energy spectrum after each reduction step in the 22.5 kton
  fiducial volume. The open circles
(filled inverted triangles) correspond to the reduction step after the spallation (ambient)
cut. The stars give the spectrum after the external event cut, and the final
SK-IV sample after the tight fiducial volume cut is given by the filled squares. The dashed line shows the final
sample of SK-III.}
 \label{fig:redstep}
 \end{center}
\end{figure}

\begin{figure}
 \begin{center}
 \includegraphics[width=8.5cm,clip]{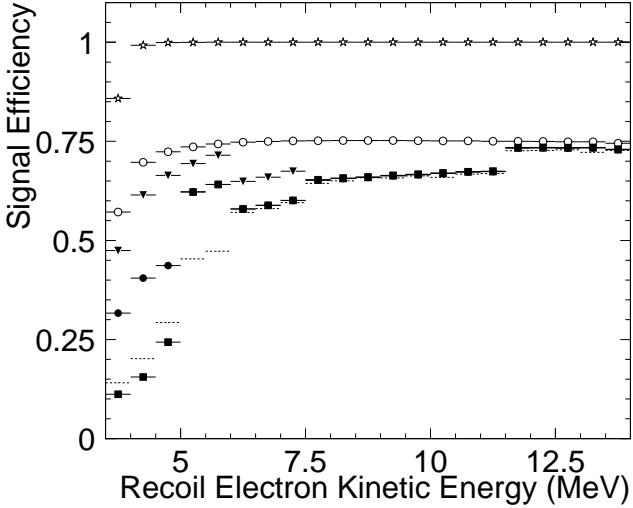}
 \caption{Signal efficiency after each reduction step. The open stars
are the trigger efficiency in the 22.5 kton fiducial volume, the open circles (filled inverted triangles)
correspond to the reduction step after the spallation (ambient) cut.
The filled squares give the final reduction efficiency, with the step from the filled
circles (after the external event cut) to the filled squares indicates the reduction in fiducial volume at low energy. The
dashed line shows the efficiency of SK-III.}
 \label{fig:redeff}
 \end{center}
\end{figure}

Fig.~\ref{fig:redstep}
shows the energy spectrum after each reduction step and Fig.~\ref{fig:redeff}
shows the reduction efficiency of the corresponding steps. The final sample
of SK-IV data is shown by the filled squares and for comparison the SK-III
final sample is superimposed (dashed lines). Above 5.99 MeV, the  efficiency
for solar neutrinos in the final sample is almost the same as in SK-III, while for
4.99 to 5.99 MeV, the SK-IV efficiency is better than SK-III.  The reason for the improvement
is the removal of a fiducial volume cut based on the ``second vertex
fit''~\cite{sk3,sk1} and making a looser ambient event cut.
The reduced fiducial volume and a tighter ambient event cut for
3.49 to 4.99 MeV results in a lower efficiency than SK-III,
but in exchange the background level has been reduced by $\sim40\%$.

\begin{figure}
 \begin{center}
 \includegraphics[width=8cm,clip]{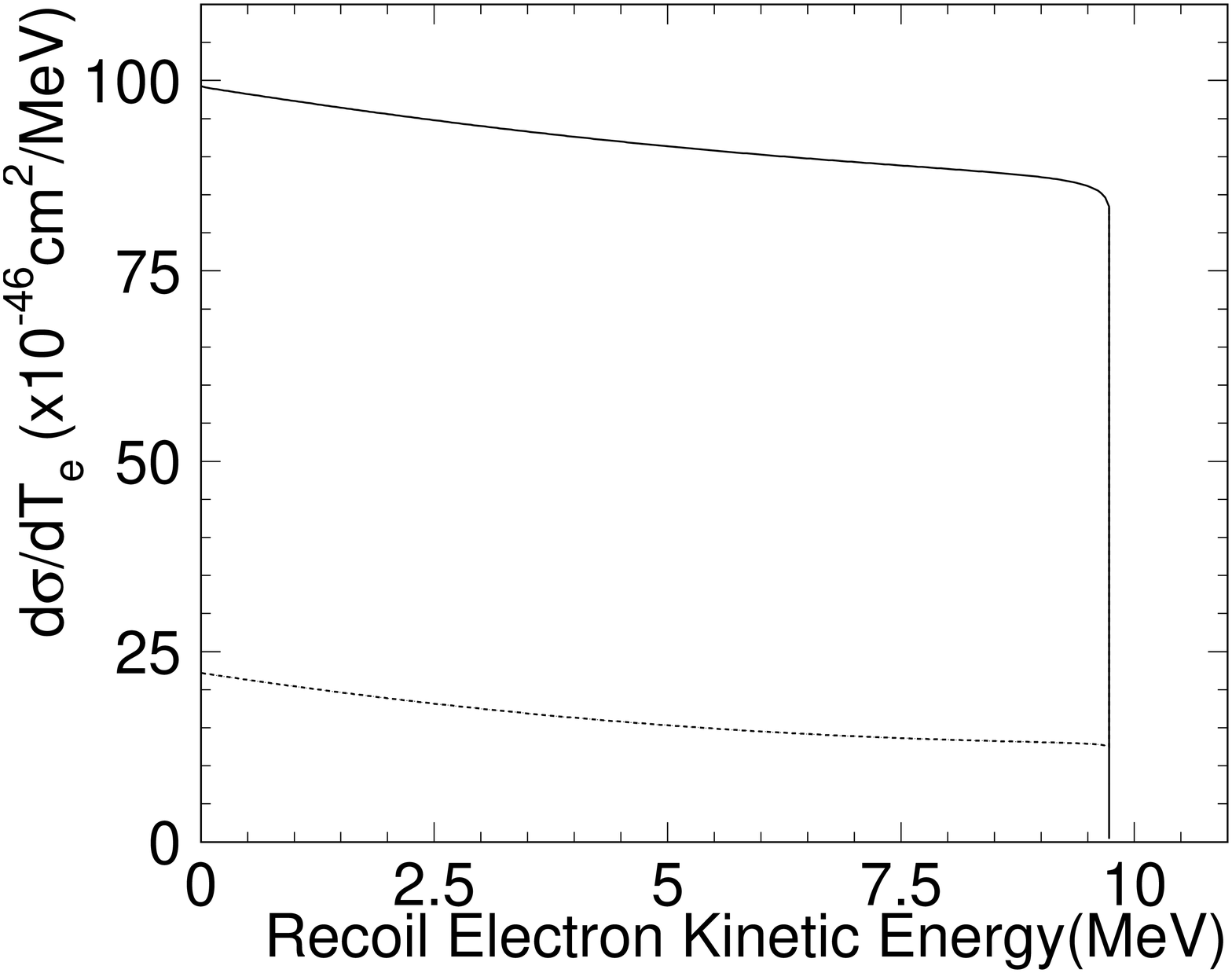}
 \caption{Differential cross section of $(\nu_e,e)$ (solid) and
$(\nu_{\mu,\tau},e)$ (dashed) elastic scattering for the case of 10 MeV
incident neutrino energy.}
 \label{fig:10mevneutrino}
 \end{center}
\end{figure}

\subsection{Simulation of solar neutrinos}
\label{sec:simulation}
There are several steps in simulating solar neutrino events at SK:
generate the solar neutrino fluxes and cross-sections, determine the recoil
electron kinematics, track the Cherenkov light in water and simulate the
response of the PMTs and electronics.  We used the $^8$B solar neutrino
spectrum calculated by Winter et al~\cite{winter} and the $hep$ solar
neutrino spectrum from Bahcall et al~\cite{bahcall}. The
systematic uncertainties from these flux calculations are incorporated in the
energy-correlated systematic uncertainty of the recoil electron spectrum.
The simulated event times are chosen according to the livetime
distribution of SK-IV so that the solar zenith angle distribution of the
solar neutrinos is reflected correctly across the simulated events.
The recoil electron energy spectrum is calculated by integrating the
differential
cross section between zero and $T_{\text{max}}$. $T_{\text{max}}$ is the
maximum kinetic energy of the recoiling electron, which is limited by the
incident neutrino energy.

Because $\nu_e$'s scatter via both $W^\pm$ and $Z^0$ exchange,
while $\nu_{\mu,\tau}$'s interact only in the neutral-current channel,
the $(\nu_e$,$e^-)$ cross section is approximately six times larger than
$(\nu_{\mu,\tau}$,$e^-)$. For the total and differential cross sections of
those interactions,
we adopted the calculation from \cite{bahcall2}, in which the radiative
corrections are taken into account and where the ratio
$d\sigma_{\nu_e}/dE_e$ and $d\sigma_{\nu_{\mu,\tau}}/dE_e$ depends on the
recoil electron energy $E_e$.
Fig.~\ref{fig:10mevneutrino}
shows the differential cross section of $(\nu_e$,$e^-)$ (solid) and
$(\nu_{\mu,\tau},e^-)$ (dashed) elastic scattering, for the case
of 10 MeV incident neutrino energy. This recoil electron energy dependence of the cross
section was accidentally omitted in the SK-III flux calculation
in~\cite{sk3}. Therefore, wrong recoil electron kinematics were generated
for the SK-III analysis, primarily affecting the lowest energy.
We re-analyzed SK-III with the correct energy dependence (leaving everything
else unchanged), the results of which can be found in Appendix A.

\subsection{Total flux}
\label{sec:flux}

\begin{figure}[t]
 \begin{center}

 \includegraphics[width=8.5cm,clip]{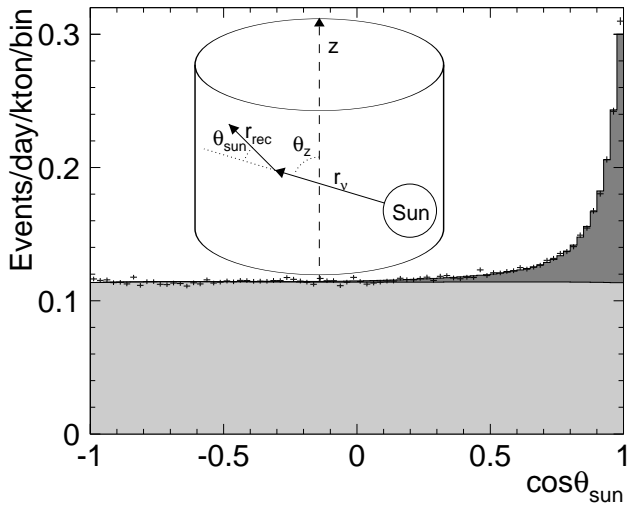}




 \caption{Solar angle distribution for 3.49 to 19.5 MeV.
$\theta_{\text{sun}}$
is the angle between the incoming neutrino direction $r_\nu$ and the
reconstructed recoil electron direction $r_\text{rec}$.  $\theta_z$ is the
solar zenith angle. Black points are data while the histogram is the best
fit to the data. The dark (light) shaded region is the solar neutrino
signal (background) component of this fit.
\label{fig:solang}}
 \end{center}
\end{figure}

In the case of $(\nu,e^-)$ interactions of solar neutrinos in SK, the incident
neutrino and recoil electron directions are highly correlated.
Fig.~\ref{fig:solang} shows the $\cos\theta_{\text{sun}}$ distribution for
events in the energy range 3.49 to 19.5 MeV, as well as the
definition of
$\cos\theta_{\text{sun}}$. In order to obtain the number of solar neutrino
interactions, an extended maximum likelihood fit is used. This method is also
used in the SK-I~\cite{sk1}, II~\cite{sk2}, and III~\cite{sk3} analyses.
The likelihood function is defined as
\begin{equation}
\mathcal{L} = e^{-(\sum_i B_i +S)} \prod _{i=1} ^{N_{\text{bin}}} \prod _{j=1} ^{n_i} (B_i \cdot b_{ij}+S \cdot Y_i \cdot s_{ij}),
\label{eq:emlf}
\end{equation}
where $N_{\text{bin}}$ is the number of energy bins. The flux analysis of
SK-IV has $N_{\text{bin}}=23$
energy bins; 20 bins of 0.5 MeV width between 3.49 and 13.5 MeV, two
energy bins of 1 MeV between 13.5 MeV and 15.5 MeV, and one bin between
15.5 MeV and 19.5 MeV. $n_{i}$ is the number of observed events in
the $i$-th energy bin. $S$ and $B_{i}$, the free parameters of this likelihood
function, are the number of solar neutrino interactions in all bins and the
number of background events in the $i$-th energy bin, respectively.
$Y_{i}$ is the fraction of signal events in the $i$-th energy bin,
calculated from solar neutrino simulated events. The background
weights $b_{ij}=\beta_i(\cos\theta_{ij}^{\text{sun}})$ and the signal weights
$s_{ij}=\sigma(\cos\theta_{ij}^{\text{sun}},E_{ij})$ are calculated
from the expected shapes of the background and solar neutrino signal,
respectively (probability density functions). The background shapes $\beta_i$
are based on the zenith and azimuthal angular distributions of real data, while
the signal shapes $\sigma$ are obtained from the solar neutrino simulated
events. The values of $S$ and $B_{i}$ are obtained by maximizing
the likelihood. The histogram of Fig.~\ref{fig:solang} is the best fit
to the data, the dark (light) shaded region is the solar neutrino signal
(background) component of that best fit.
The systematic uncertainty for this method of signal extraction is estimated
to be $0.7\%$.

\subsubsection{Vertex shift systematic uncertainty}
\begin{figure}[t]
 \begin{center}
 \includegraphics[width=8cm,clip]{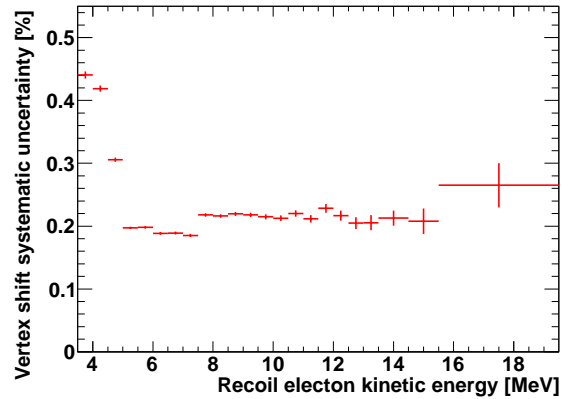}
 \caption{Vertex shift systematic uncertainty on the flux.  The increase
below 4.99 MeV comes from the tight fiducial volume cut. (see text)
\label{fig:vshift_sys}}
 \end{center}
\end{figure}

The systematic uncertainty resulting from the fiducial volume cut comes from
event vertex
shifts. To calculate the effect on the elastic scattering rate, the
reconstructed vertex positions of solar neutrino MC events are artificially
shifted following the arrows in Fig.~\ref{fig:vshift_sk4}, and the number
of events passing the fiducial volume cut with and without the artificial
shift are compared. Fig.~\ref{fig:vshift_sys} shows the energy dependence of
the systematic uncertainty coming from the shifting of the vertices. The
increase below 4.99 MeV comes from the reduced fiducial volume (smaller
surface to volume ratio), not from an energy dependence of the
vertex shift. The systematic uncertainty on the total rate is $\pm0.2\%$.

\subsubsection{Trigger efficiency systematic uncertainty}
The trigger efficiency depends on the vertex position, water transparency,
number of hit PMTs, and response of the front-end electronics. The systematic
uncertainty from the trigger efficiency is estimated by comparing
Ni-calibration data (see section~\ref{sec:recon}) with MC simulation.
For 3.49-3.99 MeV and 3.99-4.49 MeV, the difference between
data and MC is $-3.43\pm0.37 \%$ and $-0.86\pm0.31\%$, respectively~\cite{D-Nakano}. 
Above 4.49 MeV the
trigger efficiency is $100\%$ and its uncertainty is negligible. The resulting
total flux systematic uncertainty due to the trigger efficiency is $\pm0.1\%$.

\subsubsection{Angular resolution systematic uncertainty}
The angular resolution of electrons is defined as the angle which includes
$68\%$ of events in the distribution of the angular difference between their
reconstructed direction and their true direction.  
The MC prediction of the angular resolution is checked and the systematic uncertainty 
is estimated by comparing the difference in the reconstructed and true directions of 
LINAC data and LINAC (see~\cite{LINAC}) simulated events.
This difference is shown in Table~\ref{tab:sysares} for various
energies.
To estimate the systematic uncertainty on the total flux, the signal shapes
$s_{ij}^{\text{ang+}}$ and $s_{ij}^{\text{ang-}}$ are varied by shifting the
reconstructed directions of the simulated solar neutrino events by the
uncertainty in the angular resolution.  These new signal shapes are used
when extracting the total flux, and the resulting $\pm0.1\%$
change in the extracted flux is taken as the systematic uncertainty
from angular resolution.

\begin{table}[h]
\begin{center}
\caption{Angular resolution difference between LINAC data and simulated
LINAC events for each SK phase.  The energy refers to the electron's in-tank
kinetic energy.}
\begin{tabular}{c c c c c}
\hline \hline
Energy (MeV)    & SK-I($\%$) & SK-II($\%$) & SK-III($\%$) & SK-IV($\%$) \\ \hline
 4.0            & --         & --          & --           & 0.64      \\         
 4.4            & $-1.64$    & --          & 0.74         & 0.68      \\         
 5.3            & $-1.38$    & --          & --           & --        \\ 
 6.3            & 2.32       & $5.93$      & --           & 0.02      \\ 
 8.2            & 2.33       & $7.10$      & 0.40         & 0.06      \\ 
10.3            & 1.52       & --          & --           & --        \\ 
12.9            & 1.07       & $6.50$      & $-0.27$      & 0.22      \\ 
15.6            & 0.88       & --          & 0.39         & --        \\ 
18.2            & --         & --          & --           & 0.31      \\
\hline \hline
\end{tabular}
\label{tab:sysares}
\end{center}
\end{table}

\begin{table}
\caption{Summary of the systematic uncertainty on the total rate for each
SK phase. The details are also explained in \protect\cite{sk3,D-Nakano}. }
\centerline{\begin{tabular}{l c c c c}
\hline\hline
                                     & SK-I & SK-II & SK-III & SK-IV \cr
Threshold (MeV)                      & 4.49  & 6.49   & 3.99   & 3.49 \cr
\hline
Trigger Efficiency                   & $0.4\%$ & $0.5\%$ & $0.5\%$ & $0.1\%$ \cr
Angular Resolution                   & $1.2\%$ & $3.0\%$ & $0.7\%$ & $0.1\%$ \cr
Reconstruction Goodness     & $^{+1.9}_{-1.3}\%$ & $3.0\%$ & $0.4\%$ & $0.1\%$ \cr
Hit Pattern                          & $0.8\%$ & $-$     & $0.3\%$ & $0.5\%$ \cr
Small Hit Cluster                    & $-$     & $-$     & $0.5\%$ & $^{+0.5}_{-0.4}\%$ \cr
External Event Cut                   & $0.5\%$ & $1.0\%$ & $0.3\%$ & $0.1\%$ \cr
Vertex Shift                         & $1.3\%$ & $1.1\%$ & $0.5\%$ & $0.2\%$ \cr
Second Vertex Fit                    & $0.5\%$ & $1.0\%$ & $0.5\%$ & $-$ \cr
Background Shape                     & $0.1\%$ & $0.4\%$ & $0.1\%$ & $0.1\%$ \cr
Multiple Scattering Goodness         & $-$     & $0.4\%$ & $0.4\%$ & $0.4\%$ \cr
Livetime                             & $0.1\%$ & $0.1\%$ & $0.1\%$ & $0.1\%$ \cr
Spallation Cut                       & $0.2\%$ & $0.4\%$ & $0.2\%$ & $0.2\%$ \cr
Signal Extraction                    & $0.7\%$ & $0.7\%$ & $0.7\%$ & $0.7\%$ \cr
Cross Section                        & $0.5\%$ & $0.5\%$ & $0.5\%$ & $0.5\%$ \cr
\hline
Subtotal                             & $2.8\%$ & $4.8\%$ & $1.6\%$ & $1.2\%$ \cr
\hline
Energy Scale               & $1.6\%$ & $^{+4.2}_{-3.9}\%$ & $1.2\%$ & $^{+1.1}_{-1.2}\%$ \cr
Energy Resolution                    & $0.3\%$ & $0.3\%$ & $0.2\%$ & $^{+0.3}_{-0.2}\%$\cr
$^8$B Spectrum                       & $^{+1.1}_{-1.0}\%$ & $1.9\%$
                                                         & $^{+0.3}_{-0.4}\%$ & $^{+0.4}_{-0.3}\%$\cr
\hline
Total                                & $^{+3.5}_{-3.2}\%$ & $^{+6.7}_{-6.4}\%$
                                                         & $2.2\%$ & $1.7\%$ \cr
\hline\hline
\end{tabular}}
\label{tab:totsyst}
\end{table}

\subsubsection{Result}

The systematic uncertainty on the total flux (between 3.49 and
19.5 MeV) is summarized in Table \ref{tab:totsyst}. The combined systematic
uncertainty is calculated as the quadratic sum of all components, and found to
be $1.7\%$. This is the smallest systematic uncertainty of all phases of SK.
In particular, the systematic uncertainties that are energy-correlated (arising
from the energy scale and resolution uncertainty) are smallest: while SK-IV's
livetime is the same for all energy bins, previous phases have less livetime
below 5.99 MeV recoil electron kinetic energy. For example, SK-III data below
5.99 MeV has only about half the livetime as the full SK-III phase. The
improved livetime below 5.99 MeV, a higher efficiency in that energy region,
and the additional data below 4.49 MeV all lessen the impact of energy
scale and resolution uncertainties on the flux determination compared to
previous phases. Other contributions to the reduction come from the removal
of the fiducial volume cut based on an alternate vertex fit, and better
control of vertex shift, trigger efficiency and
angular resolution systematic effects. The number of solar
neutrino events (3.49-19.5 MeV) extracted from
Fig.~\ref{fig:solang} is $31,918^{+283}_{-281}$(stat.)$\pm543$(syst.). This
number corresponds to a $^8$B solar neutrino flux of

\begin{align*}
\Phi_{^8\text{B}}(\text{SK-IV})=\qquad\qquad\qquad\qquad\qquad\qquad\qquad\qquad\qquad\\
\!\!(2.308\pm0.020(\text{stat.})^{+0.039}_{-0.040}(\text{syst.}))\times 10^6 /(\text{cm}^2\text{sec}),
\end{align*}

\noindent assuming a pure $\nu_e$ flavor content.

\begin{table}[h]
  \caption{SK measured solar neutrino flux by phase.}
  \begin{tabular}{l c}
  \hline\hline
            & Flux ($\times10^6$/(cm$^2$sec)) \\ \hline
  SK-I      & $2.380\pm0.024^{+0.084}_{-0.076}$ \\
  SK-II     & $2.41\pm0.05^{+0.16}_{-0.15}$ \\
  SK-III    & $2.404\pm0.039\pm0.053$ \\
  SK-IV     & $2.308\pm0.020^{+0.039}_{-0.040}$ \\ \hline
  Combined  & $2.345\pm0.014\pm0.036$ \\
  \hline\hline
  \end{tabular}
  \label{tab:flux}
\end{table}

As seen in Table~\ref{tab:flux}, the SK-IV measured flux agrees with that
of previous phases within systematic uncertainty. It can then be combined
with the previous three SK flux measurements to give the SK measured flux as

\begin{align*}
\Phi_{^8\text{B}}(\text{SK})=\qquad\qquad\qquad\qquad\qquad\qquad\qquad\qquad\qquad\\
(2.345\pm0.014(\text{stat.})\pm0.036(\text{syst.}))\times 10^6 /(\text{cm}^2\text{sec}).
\end{align*}

\section{Energy Spectrum}
\label{sec:spectrum}
Present values of $\Delta m^2_{21}$ and $\sin^2\theta_{12}$ imply that
solar neutrino flavor oscillations above about three MeV
are dominated by the solar MSW~\cite{msw} resonance, while low-energy solar neutrino flavor
changes are mostly due to vacuum oscillations. Since the MSW effect rests
solely on standard weak interactions, it is rather interesting to compare
the expected resonance curve with data. Unfortunately multiple Coulomb
scattering prevents the kinematic reconstruction
of the neutrino energy in neutrino-electron elastic scattering interactions.
However, the energy of the recoiling electron still provides a lower
limit to the neutrino's energy. Thus, the neutrino spectrum is inferred
statistically from the recoil electron spectrum. Moreover, the differential
cross section of $\nu_{\mu,\tau}$'s is not just a factor of about six smaller
than the one for $\nu_e$'s, but also has a softer energy dependence.  In this
way, the observed recoil electron spectrum shape depends both on the flavor
composition and the energy dependence of the composition of the solar
neutrinos (see section~\ref{sec:simulation} in particular
Fig.~\ref{fig:10mevneutrino}). Thus, even a flat composition of $33\%$ $\nu_e$
and $67\%$ $\nu_{\mu,\tau}$ would still distort the recoil electron spectrum
compared to one with $100\%$ $\nu_e$. The energy dependence of the day/night
effect and rare $hep$ neutrino interactions (with a higher endpoint than $^8$B
$\nu$'s) also distort the spectrum.

\begin{figure}[t]
 \begin{center}
 \includegraphics[width=9cm,clip]{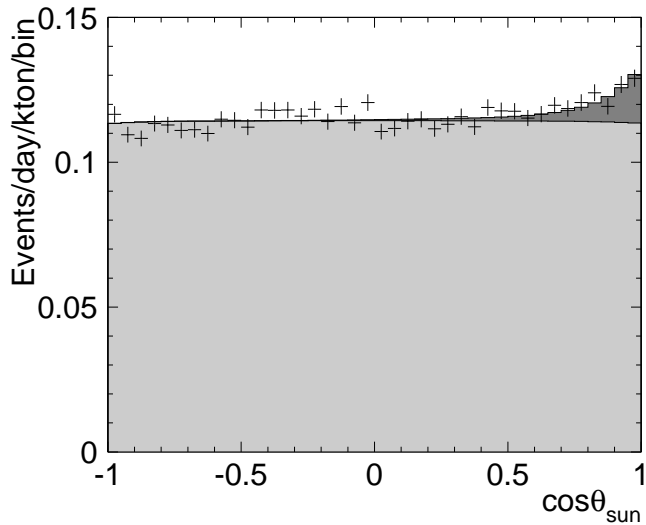}
 \caption{Solar angle distribution for events with electron energies between 3.49
and 3.99 MeV. 
The style definitions are same as FIG.~\protect{\ref{fig:solang}}.
\label{fig:solang_3.5}}
 \end{center}
\end{figure}

\begin{figure}[t]
 \begin{center}
 \includegraphics[width=8.5cm,clip]{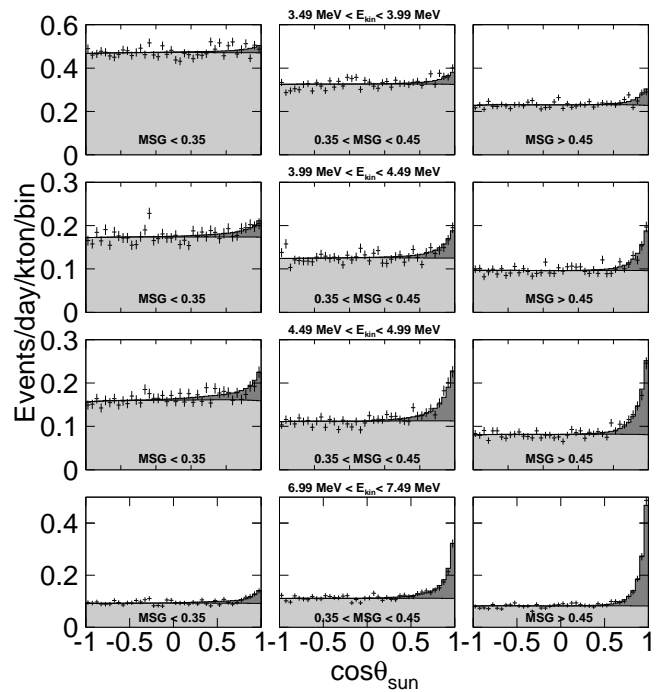}
 \caption{Solar angle distribution for the electron energy ranges
3.49-3.99 MeV, 3.99-4.49 MeV, 4.49-4.99 MeV and 6.99-7.49 MeV (from top to
bottom), for each MSG bin (left to right).
The style definitions are same as FIG.~\protect{\ref{fig:solang}}.
\label{fig:mcsg_cossun}}
 \end{center}
\end{figure}

Since the transition between MSW resonance and vacuum oscillations lies
around 3 MeV, the lowest energy solar neutrinos show the largest deviation
from the resonance electron survival probability. Here, we report
for the first time, a clear solar neutrino signal with high statistics in
the energy range 3.49-3.99 MeV observed over the entire data-taking
period of SK-IV.  Fig.~\ref{fig:solang_3.5} shows the
solar angle distribution for this energy bin, with a distinct peak (above
the background) coming from solar neutrinos. The number of solar
neutrino interactions (measured in this energy range from fits
to the distributions of Fig.~\ref{fig:mcsg_cossun} discussed below) is
%
\begin{equation*}
1063^{+124}_{-122}(\text{stat.})^{+55}_{-54}\text{(syst.) events.}
\end{equation*}

\subsection{SK-IV spectrum results}
As outlined in~\ref{sec:flux} (in particular Eq.~\ref{eq:emlf}), the
solar neutrino signal of SK-IV is extracted by an extended maximum likelihood
fit. While the $^8$B flux analysis uses all 23 energy bins at once (and
constrains the energy spectrum to the one expected from unoscillated
simulation via the $Y_i$ factors), we extract the solar neutrino energy
spectrum by
fitting one recoil electron energy bin $i$ at a time, with $Y_i=1$.
Below 7.49 MeV, each energy bin is split into three sub-samples according to
the MSG of the events,
with boundaries set at MSG=0.35 and 0.45. These three sub-samples are then fit
simultaneously to a single signal and three independent background components.
The signal fraction $Y_{ig}$ in each MSG bin $g$ is determined by solar
neutrino simulated events in the same manner as the $Y_i$ factors in the
$^8$B flux analysis. Similar to the $^8$B flux analysis, the signal and
background shapes depend on the MSG bin $g$: the signal shapes $\sigma_g$ are
calculated from solar neutrino simulated events and the background shapes
$\beta_{ig}$ are taken from data. Fig.~\ref{fig:mcsg_cossun} shows the measured
angular distributions (as well as the fits) for the energy ranges
3.49-3.99 MeV, 3.99-4.49 MeV, 4.49-4.99 MeV and 6.99-7.49 MeV (from top to
bottom), for each MSG bin (left to right).
As expected in the lowest energy bins, where the dominant part of the
background is due to very low-energy $\beta$/$\gamma$ decays, the background
component is largest in the lowest MSG sub-sample. Also as expected, the solar
neutrino elastic scattering peak sharpens as MSG is increased.

\begin{figure}[t]
 \begin{center}
 \includegraphics[width=8cm,clip]{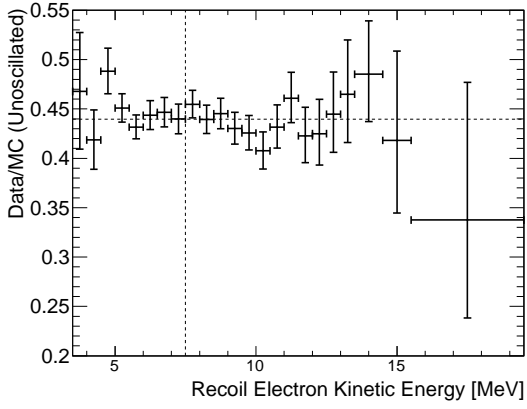}
 \caption{SK-IV energy spectrum using MSG sub-samples below 7.49 MeV, shown
as the ratio of the measured rate to the simulated unoscillated rate.
 The horizontal dashed line gives the SK-IV total average (0.440).
Error bars shown are statistical plus energy-uncorrelated systematic
uncertainties.
\label{fig:mcsg_spectrum_skiv}}
 \end{center}
\end{figure}

Using this method for recoil electron energy bins
below 7.49 MeV gives $\sim10\%$ improvement in the statistical uncertainty on
the number of extracted signal events (the additional
systematic uncertainty is small compared to the statistical gain).
Fig.~\ref{fig:mcsg_spectrum_skiv} shows the resulting SK-IV energy
spectrum, where below 7.49 MeV
MSG has been used and above 7.49 MeV the standard signal extraction
method without MSG is used.  Table~\ref{tab:spectrumsignal}
gives the measured and
expected rate in each energy bin, as well as that measured for the day and
night times separately, along with the 1 $\sigma$ statistical deviations.
We re-analyzed the SK-III spectrum below 7.49 MeV with the
same method, the same MSG bins and the same energy bins as SK-IV, down to 3.99
MeV. We also re-fit the entire SK-II (which has poorer resolution) spectrum
using the same three MSG sub-samples. The gains in precision are similar
to SK-IV.  The SK-II and III spectra are given in section~\ref{sk-1234-spec}.

To analyze the
spectrum, we simultaneously fit the SK-I, II, III and IV spectra to their
predictions, while varying the $^8$B and $hep$ neutrino fluxes within
uncertainties. The $^8$B flux is constrained to
$(5.25\pm0.20)\times10^6$ /(cm$^2$sec)
and the $hep$ flux to $(8\pm16)\times10^3$ /(cm$^2$sec)
(motivated by SNO's measurement~\cite{snopaper} and limit~\cite{snohep}). The
$\chi^2$ is described in detail in Section~\ref{sec:osc}.

\subsection{Systematic uncertainties on the energy spectrum}
\begin{table*}[]
\begin{center}
\caption{Energy-uncorrelated systematic uncertainties on the spectrum shape.
The systematic error of the (unlisted) small hit cluster cut (only applied below 4.99 MeV)
is negligible.
\label{tab:euncorr}}
\begin{tabular}{l c c c c c c c c c} \hline\hline
Energy (MeV)            & 3.49-3.99 & 3.99-4.49 & 4.49-4.99     & 4.99-5.49 & 5.49-5.99 & 5.99-6.49 & 6.49-6.99     & 6.99-7.49 & 7.49-19.5 \\
\hline
Trigger Efficiency  &$^{+3.6}_{-3.3}\%$&$\pm 0.8\%$& -             & -         & -         & -         & -             & -         & - \\
Reconstruction Goodness &$\pm 0.6\%$&$\pm 0.7\%$&$^{+0.6}_{-0.5}\%$&$\pm 0.4\%$&$\pm 0.2\%$&$\pm 0.1\%$&$\pm 0.1\%$    &$\pm 0.1\%$&$\pm 0.1\%$ \\
Hit Pattern             & -         & -         & -             & -         & -         &$\pm 0.6\%$&$\pm 0.6\%$    &$\pm 0.6\%$&$\pm 0.4\%$ \\
External Event Cut      &$\pm 0.1\%$&$\pm 0.1\%$&$\pm 0.1\%$    &$\pm 0.1\%$&$\pm 0.1\%$&$\pm 0.1\%$&$\pm 0.1\%$    &$\pm 0.1\%$&$\pm 0.1\%$ \\
Vertex Shift            &$\pm 0.4\%$&$\pm 0.4\%$&$\pm 0.2\%$    &$\pm 0.2\%$&$\pm 0.2\%$&$\pm 0.2\%$&$\pm 0.2\%$    &$\pm 0.2\%$&$\pm 0.2\%$ \\
Background Shape        &$\pm 2.9\%$&$\pm 1.0\%$&$\pm 0.8\%$    &$\pm 0.2\%$&$\pm 0.1\%$&$\pm 0.1\%$&$\pm 0.1\%$    &$\pm 0.1\%$&$\pm 0.1\%$ \\
Signal Extraction       &$\pm 2.1\%$&$\pm 2.1\%$&$\pm 2.1\%$    &$\pm 0.7\%$&$\pm 0.7\%$&$\pm 0.7\%$&$\pm 0.7\%$    &$\pm 0.7\%$&$\pm0.7\%$ \\
Cross Section           &$\pm 0.2\%$&$\pm 0.2\%$&$\pm 0.2\%$    &$\pm 0.2\%$&$\pm 0.2\%$&$\pm 0.2\%$&$\pm 0.2\%$    &$\pm 0.2\%$&$\pm0.2\%$ \\
MSG                     &$\pm 0.4\%$&$\pm 0.4\%$&$\pm 0.3\%$    &$\pm 0.3\%$&$\pm 0.3\%$&$\pm 1.7\%$&$\pm 1.7\%$    &$\pm 1.7\%$& - \\ \hline
Total               &$^{+5.1}_{-4.9}\%$&$\pm 2.6\%$&$^{+2.4}_{-2.3}\%$&$\pm 0.9\%$&$\pm 0.9\%$&$\pm 2.0\%$&$^{+2.0}_{-1.9}\%$&$\pm 1.9\%$& $^{+0.9}_{-0.8}\%$ \\
\hline\hline
\end{tabular}
\end{center}
\end{table*}

\begin{figure}[t]
 \begin{center}
 \includegraphics[width=8cm,clip]{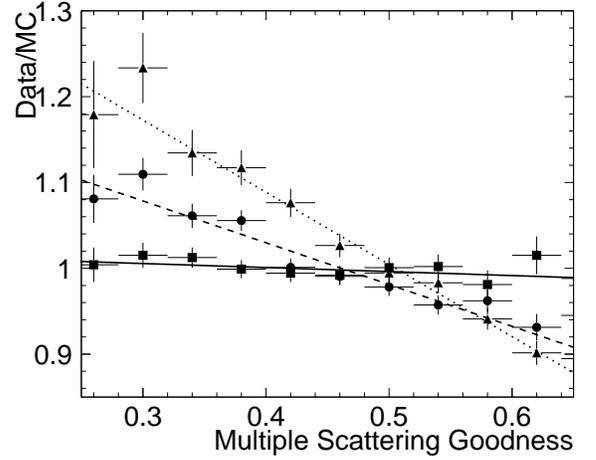}
 \caption{MSG scaling functions applied to simulated events to estimate the
systematic uncertainty on the energy spectrum.  The dotted, dashed and solid lines
correspond to 16.1, 8.67 and 4.89 MeV LINAC data over simulated events.}
 \label{fig:linac_scale}
 \end{center}
\end{figure}

Since we simultaneously fit multiple samples defined by the multiple Coulomb
scattering goodness in the lowest recoil electron energy region, a systematic
shift in this goodness of the data compared to solar $^8$B (or $hep$) neutrino
simulated events would affect the measured event rate in that energy region.
To estimate the systematic effect of using MSG sub-samples, MSG distributions
of LINAC data and simulated LINAC events were compared, as seen in
Fig.~\ref{fig:linac_mcsg}. The simulated solar neutrino MSG distributions
are adjusted using the observed ratio of the LINAC data and simulated events
at the nearest LINAC energy. This changes the solar signal shapes $\sigma_g$ and the
ratios of expected signal events $Y_{ig}$ for MSG bin $g$. 
The $\cos\theta_{\text{sun}}$
distributions are then re-fit, using the new angular distributions and signal
ratios and the change in the extracted number of signal events is taken as the
systematic uncertainty. The scaling functions for three
LINAC energies can be seen in Fig.~\ref{fig:linac_scale}.

The change for each energy bin and all other energy-uncorrelated
systematic uncertainties of the SK-IV recoil electron energy spectrum
are summarized in Table~\ref{tab:euncorr}. The total energy-uncorrelated
systematic uncertainty in this table is calculated as the sum in quadrature
of each of the components. Since we assume no correlations between the
energy bins in the SK-IV spectrum analysis, the combined uncertainty is added
in quadrature to the statistical error of that energy bin.

\begin{figure}[t]
 \begin{center}
 \includegraphics[width=8cm,clip]{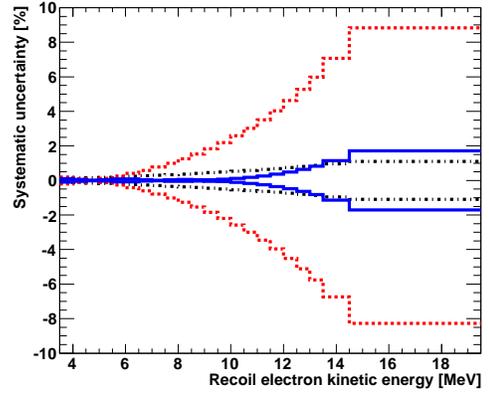}
 \caption{Energy-correlated systematic uncertainties. The dot-dashed, solid
and dashed distributions correspond to the systematic
uncertainties of the $^8$B spectrum shape, energy resolution and absolute
energy scale, respectively.\label{fig:ecor_sys}}
 \end{center}
\end{figure}

The $^8$B neutrino spectrum uncertainty (a shift of $\sim\pm100$ keV),
the SK-IV energy scale uncertainty ($\pm0.54\%$) and the SK-IV energy
resolution uncertainty 
($\pm1.0\% for < 4.89$ MeV, $0.6\%$ for $>6.81$ MeV)~\cite{D-Nakano},
 will shift all energy bins in a
correlated manner.  The size and correlation of these uncertainties are
calculated from the neutrino spectrum, the differential cross section, the
energy resolution function, and the size
of the systematic shifts. We vary each of these three parameters
($^8$B neutrino spectrum shift, energy scale, and energy resolution)
individually. Fig.~\ref{fig:ecor_sys}
shows the result of this calculation. When we analyze the spectrum, we apply
these shifts to the spectral predictions. When the SK-IV spectrum is combined
with the SK-I, II, and III spectra, the $^8$B neutrino spectrum shift is
common to all four phases, while each phase varies its energy scale and
resolution individually (without correlation between the phases).

\begin{figure}[h!]
 \begin{center}
 \includegraphics[width=8cm,clip]{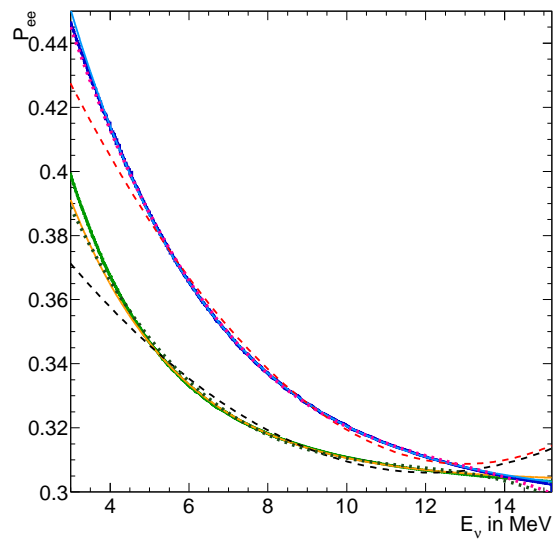}
 \caption{$\nu_e$ survival probability $P_{ee}$ based
on the oscillation parameters fit to SK (thick solid green)
and all solar neutrino and KamLAND data (thick solid blue).
The solid yellow (cyan) line is the best
exponential approximation to the thick solid green (blue) line.
The dashed black (dotted green) line is the best quadratic (cubic)
approximation to the thick solid green line and the dashed red
(dotted pink) line the best quadratic (cubic)
approximation to the thick solid blue.}
 \label{fig:pee_approx}
 \end{center}
\end{figure}

\begin{figure*}[]
\includegraphics[width=18cm,clip]{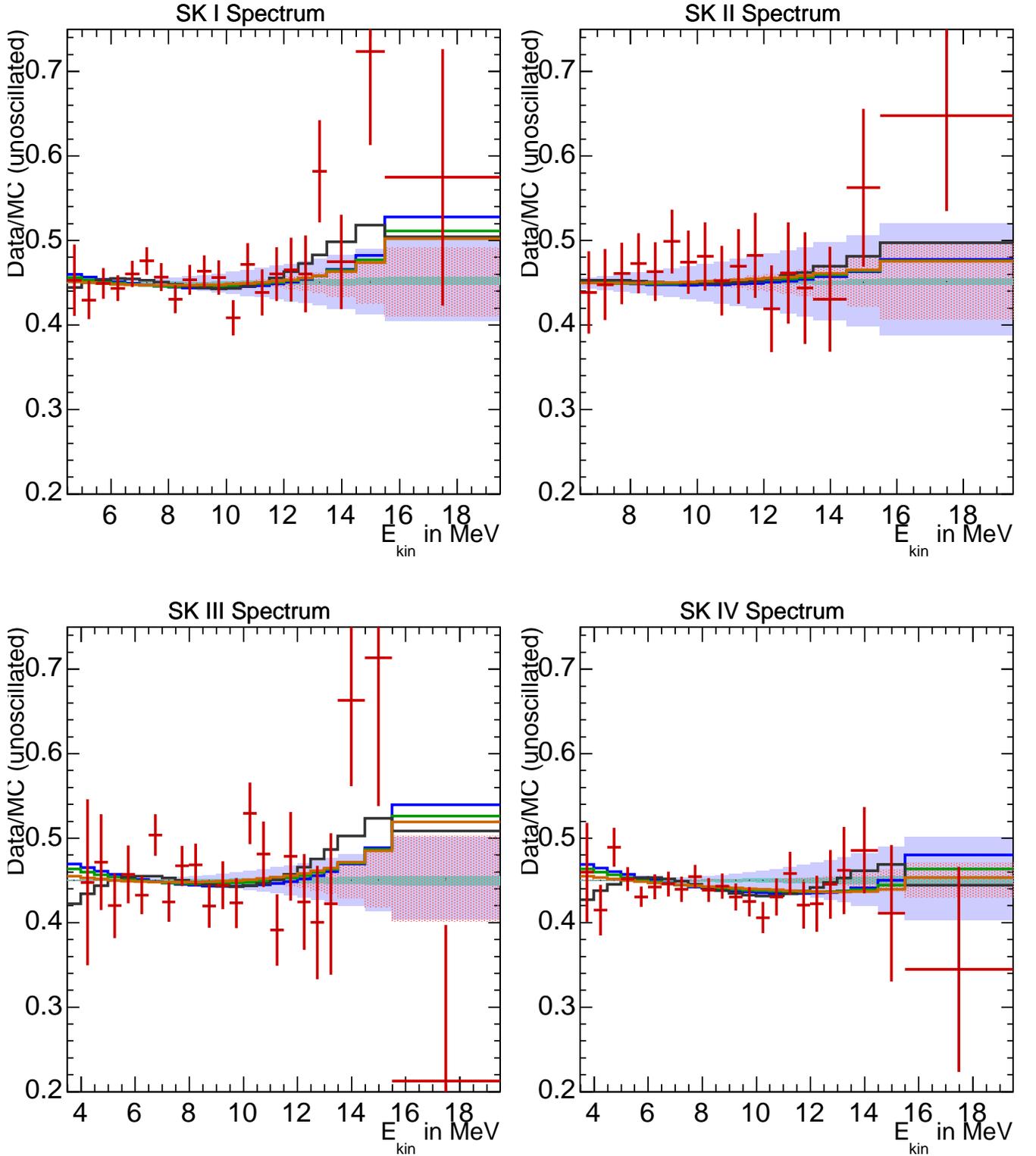}
\caption{SK-I, II, III and IV recoil
electron spectra divided by the non-oscillated expectation.
The green (blue) line represents the best fit to SK data using the
oscillation parameters from the fit to all solar (solar+KamLAND) data.
The orange (black) line is the best fit to SK data of a general exponential
or quadratic (cubic) $P_{ee}$ survival probability. 
Error bars on the data points give the
statistical plus systematic energy-uncorrelated uncertainties while the
shaded purple, red and green histograms give the energy-correlated
systematic uncertainties arising from energy scale, energy resolution, and
neutrino energy spectrum shift.}
\label{figspecfourspectrum}
\end{figure*}

\subsection{SK-I/II/III/IV combined spectrum analysis}
\label{sk-1234-spec}
In order to discuss the energy dependence of the solar neutrino flavor
composition in a general way, SNO~\cite{snopaper} has parametrized the
electron survival probability $P_{ee}$ using a quadratic function centered
at 10 MeV:

\begin{multline}
P_{ee}(E_\nu)=\\
c_0+c_1\left(\frac{E_\nu}{\mbox{\small MeV}}-10\right)+ 
c_2\left(\frac{E_\nu}{\mbox{\small MeV}}-10\right)^2,
\end{multline}
where $c_0$, $c_1$ and $c_2$ are polynomial parameters.

As seen in Fig.~\ref{fig:pee_approx}, this parametrization does not describe
well the MSW resonance based on the oscillation parameters of either best fit.
This is also true for alternative solutions such as non-standard
interactions~\cite{nsi} and mass-varying neutrinos~\cite{mavan}.
However, it is simple, and the SNO collaboration found that it introduces
no bias when determining oscillations parameters. In addition to this
quadratic function we have explored two different alternatives to parametrize
the survival probability in order to study any limitations the quadratic function
might have: an exponential fit and a cubic extension of the
quadratic fit. The exponential fit is parametrized as
\begin{equation}
P_{ee}(E_\nu)=e_0+\frac{e_1}{e_2}
\left(e^{e_2\left(\frac{E_\nu}{\mbox{\small MeV}}-10\right)}-1\right).
\end{equation}
This particular functional form allows direct comparison of $e_0$ and $e_1$
to the quadratic coefficients $c_0$ and $c_1$,
if $c_1$ and $e_1$ are small. The parameter $e_2$ controls the ``steepness''
of the exponential fall or rise. Both exponential and cubic parametrizations describe the MSW
resonance curve reasonably well as shown in Fig.~\ref{fig:pee_approx}.
This is true for both the SK-only and the solar+KamLAND best-fit oscillation
parameters discussed in the  oscillation section below. Table~\ref{tabpee_approx}
lists the exponential and cubic coefficients that best describe those two
MSW resonance curves. The definition of the spectrum $\chi^2$ and the
best-fit values are given in section \ref{sec:osc}.

To ease the comparison between SK spectral data and SNO's results, we
also performed a quadratic fit to SK data. Table~\ref{tabpee_approx} gives
the best quadratic coefficients for both the SK-only and the
solar+KamLAND results.
For each set of parameters, the expected rate in each energy bin
is adjusted according to the average day/night enhancement expected
from $\sin^2\theta_{12}=0.304$ and $\Delta m^2=4.90\times10^{-5}$ eV$^2$.
Fig.~\ref{figspecfourspectrum} shows the SK spectral data. They are
expressed as the ratio of the observed elastic
scattering rates of each SK phase over MC expectations, assuming no
oscillations (pure electron flavor composition), a $^8$B flux of
$5.25\times10^6$ /(cm$^{2}$sec) and a $hep$ flux of
$8\times10^3$ /(cm$^{2}$sec).  Table~\ref{tabspecdata} lists the data shown in
Fig.~\ref{figspecfourspectrum}, with the given errors including statistical
uncertainties as well as energy-uncorrelated systematic uncertainties.

\begin{table}[t!]
\footnotesize
\caption{Best approximations to the MSW resonances using exponential and
polynomial parametrizations of $P_{ee}$.}
\begin{tabular}{l c c c c}
\hline\hline
& \multicolumn{2}{c}{$\sin^2\theta_{12}=0.304$}
& \multicolumn{2}{c}{$\sin^2\theta_{12}=0.314$} \cr
& \multicolumn{2}{c}{$\Delta m^2_{21}=7.41\cdot10^{-5}$}
& \multicolumn{2}{c}{$\Delta m^2_{21}=4.90\cdot10^{-5}$} \cr
\hline
expon. $e_0$    &\multicolumn{2}{c}{$ 0.3205$}&\multicolumn{2}{c}{$ 0.3106$} \cr
expon. $e_1$    &\multicolumn{2}{c}{$-0.0062$}&\multicolumn{2}{c}{$-0.0026$} \cr
expon. $e_2$    &\multicolumn{2}{c}{$-0.2707$}&\multicolumn{2}{c}{$-0.3549$} \cr
expon. $\chi^2$, $\Delta\chi^2$ &
\multicolumn{2}{c}{$ 70.69$, $2.31$} & \multicolumn{2}{c}{$ 68.99$, $0.61$} \cr
\hline
polyn. $c_0$    & $ 0.3194$ & $ 0.3204$ & $0.3095$  & $ 0.3105$ \cr
polyn. $c_1$    & $-0.0071$ & $-0.0059$ & $-0.0033$ & $-0.0021$ \cr
polyn. $c_2$    & $+0.0012$ & $+0.0009$ & $+0.0008$ & $+0.0005$ \cr
polyn. $c_3$    & $0$       & $-0.0001$ & $0$       & $-0.0001$ \cr
polyn. $\chi^2$, $\Delta\chi^2$ &
$70.79$, $2.46$ & $70.71$, $7.07$ & $68.87$, $0.54$ & $69.06$, $5.43$\cr
\hline\hline
\end{tabular}
\label{tabpee_approx}
\end{table}



\begin{table*}[t!]
\caption{Spectrum fit $\chi^2$ comparison.}
\begin{tabular}{lccccccccccccccc}
\hline
\hline
Fit   & \multicolumn{3}{c}{MSW (sol+KamLAND)} & \multicolumn{3}{c}{MSW (solar)} & \multicolumn{3}{c}{exponential} &
\multicolumn{3}{c}{quadratic} & \multicolumn{3}{c}{cubic} \cr
\hline
Param. & 
\multicolumn{3}{c}{$\sin^2\theta_{12}$, $\sin^2\theta_{13}$, $\Delta m^2_{21}$} &
\multicolumn{3}{c}{$\sin^2\theta_{12}$, $\sin^2\theta_{13}$, $\Delta m^2_{21}$} &
\multicolumn{3}{c}{$e_0$ $e_1$, $e_2$} &
\multicolumn{3}{c}{$c_0$, $c_1$, $c_2$} &
\multicolumn{3}{c}{$c_0$, $c_1$, $c_2$, $c_3$} \cr
 &
\multicolumn{3}{c}{\parbox{2.8cm}{\vspace*{0.05in}$0.304$, $0.02$, $7.50\cdot10^{-5}$eV$^2$\vspace*{0.05in}}} &
\multicolumn{3}{c}{\parbox{2.8cm}{\vspace*{0.05in}$0.304$, $0.02$, $4.84\cdot10^{-5}$eV$^2$\vspace*{0.05in}}} &
\multicolumn{3}{c}{\parbox{2.8cm}{\vspace*{0.05in}$0.334$, -$0.001$, -$0.12$\vspace*{0.05in}}} &
\multicolumn{3}{c}{\parbox{2.8cm}{\vspace*{0.05in}$0.33$, $0$, $0.001$\vspace*{0.05in}}} &
\multicolumn{3}{c}{\parbox{2.8cm}{\vspace*{0.05in}$0.312$, $-0.031$, $0.0095$, $0.0044$ \vspace*{0.05in}}}
\cr
\hline
 & $\chi^2$ & $\Phi_{^8\text{B}}/$ & $\Phi_{\mbox\tiny hep}/$
 & $\chi^2$ & $\Phi_{^8\text{B}}/$ & $\Phi_{\mbox\tiny hep}/$
 & $\chi^2$ & $\Phi_{^8\text{B}}/$ & $\Phi_{\mbox\tiny hep}/$
 & $\chi^2$ & $\Phi_{^8\text{B}}/$ & $\Phi_{\mbox\tiny hep}/$
 & $\chi^2$ & $\Phi_{^8\text{B}}/$ & $\Phi_{\mbox\tiny hep}/$ \cr
 & & cm$^2$sec & cm$^2$sec & & cm$^2$sec & cm$^2$sec
 & & cm$^2$sec & cm$^2$sec & & cm$^2$sec & cm$^2$sec
 & & cm$^2$sec & cm$^2$sec \cr
\hline
SK-I   & 19.71 & 5.26$\cdot10^6$ & 39.4$\cdot10^3$ & 19.12 & 5.47$\cdot10^6$ & 41.0$\cdot10^3$ &
         18.82 & 5.22$\cdot10^6$ & 41.4$\cdot10^3$ & 18.94 & 5.24$\cdot10^6$ & 36.8$\cdot10^3$ &
         16.14 & 5.25$\cdot10^6$ &  5.1$\cdot10^3$ \cr
SK-II  &  5.39 & 5.33$\cdot10^6$ & 55.1$\cdot10^3$ &  5.35 & 5.53$\cdot10^6$ & 56.8$\cdot10^3$ &
          5.31 & 5.27$\cdot10^6$ & 56.9$\cdot10^3$ &  5.38 & 5.30$\cdot10^6$ & 51.5$\cdot10^3$ &
          5.15 & 5.34$\cdot10^6$ & 11.9$\cdot10^3$ \cr
SK-III & 29.06 & 5.34$\cdot10^6$ & 15.7$\cdot10^3$ & 28.41 & 5.55$\cdot10^6$ & 14.7$\cdot10^3$ &
         28.07 & 5.29$\cdot10^6$ & 13.8$\cdot10^3$ & 28.02 & 5.31$\cdot10^6$ & 10.9$\cdot10^3$ &
         26.59 & 5.30$\cdot10^6$ & -3.6$\cdot10^3$ \cr
SK-IV  & 14.43 & 5.22$\cdot10^6$ & 12.2$\cdot10^3$ & 14.00 & 5.44$\cdot10^6$ & 11.4$\cdot10^3$ &
         14.29 & 5.20$\cdot10^6$ & 10.8$\cdot10^3$ & 14.15 & 5.22$\cdot10^6$ &  8.2$\cdot10^3$ &
         14.07 & 5.22$\cdot10^6$ & -4.2$\cdot10^3$ \cr
\hline
comb.  & 71.04 & 5.28$\cdot10^6$ & 14.1$\cdot10^3$ & 69.03 & 5.49$\cdot10^6$ & 13.4$\cdot10^3$ &
         68.38 & 5.25$\cdot10^6$ & 13.1$\cdot10^3$ & 68.33 & 5.26$\cdot10^6$ & 11.9$\cdot10^3$ &
         63.63 & 5.25$\cdot10^6$ & -0.7$\cdot10^3$ \cr
\hline
\hline
\end{tabular}
\label{tabspecchi2comp}
\end{table*}

\begin{table}
\caption{Spectrum fit $\chi^2$ comparison for the ``flat suppresion''
of 0.4268 of the expected rate assuming no neutrino oscillation.}
\begin{tabular}{lcccccc}
\hline
\hline
Fit & \multicolumn{3}{c}{with D/N correction} & \multicolumn{3}{c}{without D/N correction}\cr
 & $\chi^2$ & $\Phi_{^8\text{B}}/$ & $\Phi_{\mbox\tiny hep}/$
 & $\chi^2$ & $\Phi_{^8\text{B}}/$ & $\Phi_{\mbox\tiny hep}/$ \cr
 & & cm$^2$sec & cm$^2$sec & & cm$^2$sec & cm$^2$sec \cr
\hline
SK-I    & 18.92 & $5.38\cdot10^6$ & $41.4\cdot10^3$ & 18.81 & $5.47\cdot10^6$ & $42.6\cdot10^3$ \cr
SK-II   &  5.30 & $5.43\cdot10^6$ & $56.3\cdot10^3$ &  5.27 & $5.52\cdot10^6$ & $58.4\cdot10^3$ \cr
SK-III  & 27.94 & $5.45\cdot10^6$ & $12.0\cdot10^3$ & 27.98 & $5.55\cdot10^6$ & $13.1\cdot10^3$ \cr
SK-IV   & 15.50 & $5.37\cdot10^6$ & $ 9.4\cdot10^3$ & 14.99 & $5.46\cdot10^6$ & $10.2\cdot10^3$ \cr
\hline
comb.   & $69.30$ & $5.41\cdot10^6$ & $12.3\cdot10^3$ & $68.75$ & $5.50\cdot10^6$ & $12.7\cdot10^3$ \cr
\hline
\hline
\end{tabular}
\label{tabspecchi2compflat}
\end{table}

\begin{figure}[t]
\includegraphics[width=8cm,clip]{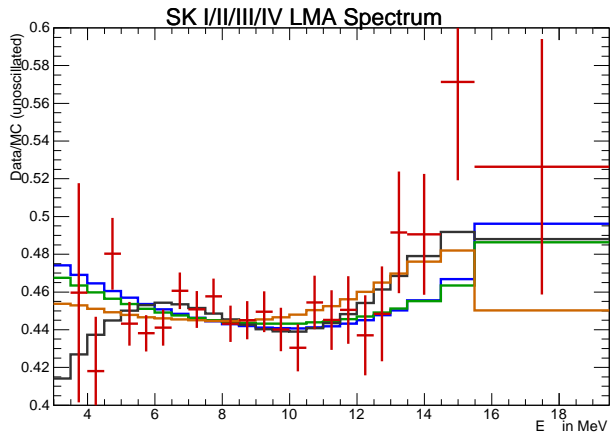}
\caption{SK-I+II+III+IV recoil
electron spectrum compared to the no-oscillation expectation.
The green (blue) shape is the MSW expectation using the
SK (solar+KamLAND) best-fit oscillation parameters.  The orange (black) line
is the best fit to SK data with a general exponential/quadratic (cubic)
$P_{ee}$ survival probability.}
\label{figspeccombspectrum}
\end{figure}


\begin{figure}[t]
\includegraphics[width=8.8cm,clip]{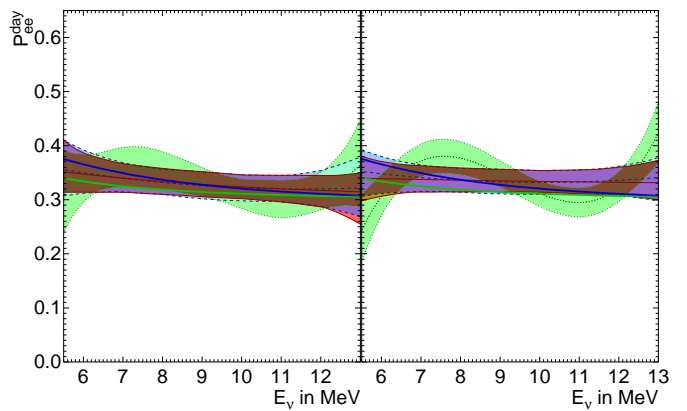}
\caption{Allowed survival probability 1 $\sigma$ band
from SK-IV data (left) and all SK data (right). The red (blue) area
is based on an
exponential (quadratic) fit and the green band
is based on a cubic fit. The $^8$B flux is constrained
to the measurement from SNO. The absolute value of the
$^8$B flux does not affect the shape constraint much,
just the average value.  Also shown are predictions
based on the oscillation parameters of a fit to all
solar data (green) and a fit to all solar+KamLAND data (blue).}
\label{figspecpeeexp}
\end{figure}

Table~\ref{tabspecpeecoef} gives the SK exponential and polynomial best-fit
coefficients and their correlations.  We compare the best $\chi^2$ of the
full MSW calculation to that of the best exponential, cubic and quadratic
function fits, as well as a simple energy-independent suppression of the
elastic scattering rate in SK. In the case of the flat (energy-independent)
suppression, 0.4268 was chosen as the ratio of observed elastic scattering
over expectation assuming no neutrino oscillations. The value 0.4268 corresponds
to a constant $P_{ee}=0.317$ if the cross section ratio was
$d\sigma_{\nu_\mu}/d\sigma_{\nu_e}=0.16$ independent of energy. In reality,
that ratio becomes larger at lower energy, leading to a small low-energy
``upturn'' even for a constant $P_{ee}=0.317$. The energy dependence of the
day/night effect (which is corrected for in the polynomial and exponential
fits) leads to a small ``downturn''. In case of this flat suppression we fit
with and without the day/night correction. Tables~\ref{tabspecchi2comp} and
\ref{tabspecchi2compflat} compare the various $\chi^2$,
while Table~\ref{tabpee_approx} gives the $\chi^2$ from the best exponential
(quadratic, cubic) approximations of the MSW resonance curve as well
as the difference in $\chi^2$ from the exponential (quadratic, cubic)
best fit. The exponential and quadratic fits are consistent with a flat
suppression as well as the MSW resonance ``upturn''. In either case an
``upturn'' fits slightly better (by about 1.0$\sigma$), but the coefficients
describing the MSW resonance are actually slightly disfavored
by 1.5$\sigma$ (exponential) and 1.6$\sigma$ (quadratic), for
the best-fit $\Delta m^2_{21}$ from KamLAND data, and by 0.8$\sigma$ (exponential)
and 0.7$\sigma$ (quadratic) for the best-fit $\Delta m^2_{21}$ from solar neutrino data.
The cubic fit disfavors the flat suppression by 2.3$\sigma$; as seen in
Fig.~\ref{figspecpeeexp} the fit prefers an inflection point in the spectrum
occurring near 8 MeV, a shape which cannot be accommodated
by the other two parametrizations. From Table~\ref{tabspecchi2comp} the SK-II
and SK-IV minimum $\chi^2$s of the cubic fit are similar to the quadratic and
exponential fit, however the SK-I (SK-III) data favor the cubic fit by about
$1.7 \sigma$ ($1.2 \sigma$). The reason for that preference is mostly
due to data above $\sim13$ MeV (see Figure~\ref{figspecfourspectrum}). We checked these 
data but found no reason to exclude them. However, conservatively, we disregard
the cubic best fit in our conclusions. Therefore, we find no significant
spectral ``upturn'' (or downturn) at low energy, but our data is consistent with the
``upturn'' predicted by the MSW resonance curve (disfavoring the one
based on solar+KamLAND best-fit parameters by about $1.5\sigma$).
Fig.~\ref{figspecfourspectrum} shows the predictions for the best MSW fits, the
best exponential/quadratic and the best cubic fit. Fig.~\ref{figspeccombspectrum}
statistically combines the different SK phases ignoring differences in energy
resolutions and systematic uncertainties. It is included only as an
illustration and should not be fit to predictions.


\begin{figure}[t]
\includegraphics[width=10cm,clip]{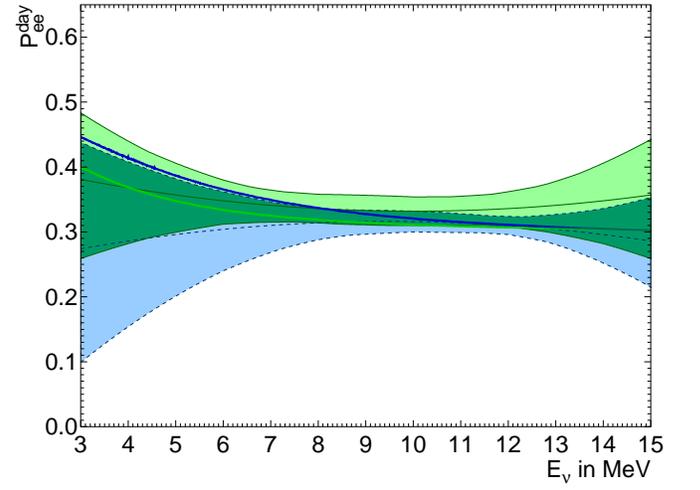}
\caption{Allowed survival probability 1 $\sigma$ band
from SK (solid green) and SNO (dotted blue) data. Also shown are predictions
based on the oscillation parameters of a fit to all
solar data (green) and a fit to all solar+KamLAND data (blue).}
\label{figspecpee}
\end{figure}


\begin{figure}[t]
\includegraphics[width=10cm,clip]{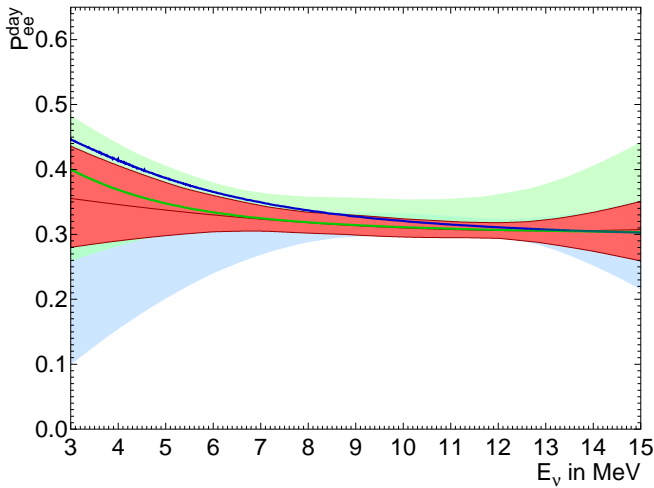}
\caption{Allowed survival probability 1 $\sigma$ band
from the combined data of SK and SNO (red). Also shown are predictions
based on the oscillation parameters of a fit to all
solar data (green) and a fit to all solar+KamLAND data (blue).
The pastel colored bands are the separate SK (green) and
SNO (blue) fits.}
\label{figspecpeecomb}
\end{figure}

\begin{figure}[t]
\includegraphics[width=8cm,clip]{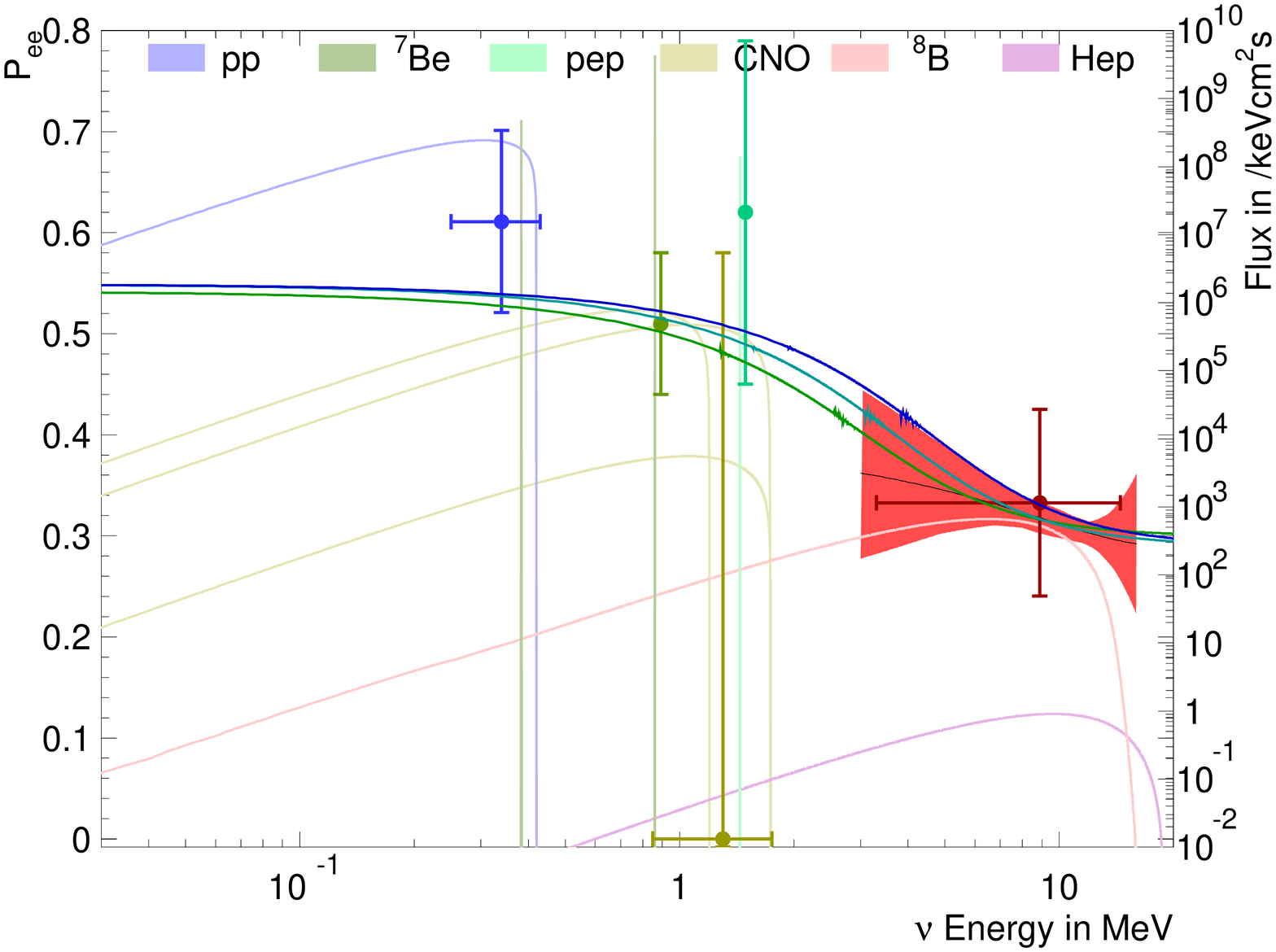}
\caption{Predicted solar neutrino spectra~\cite{ssm}. Overlaid are expected
MSW survival probabilities, green is that expected assuming oscillation
parameters from the SK best fit and blue from the solar+KamLAND best fit.  The
1 $\sigma$ band of $P_{ee}$ from the combined data of SK and SNO is shown
in red. Also shown are $P_{ee}$ measurements of the $^7$Be (green point),
the $pep$ (light green point) and the $^8$B flux (red point) by
Borexino~\cite{otherborexino}, as well as $pp$ (blue point) and
CNO values (gold point) extracted from other experiments~\cite{othersolar}.}
\label{figspecpeecomb_global}
\end{figure}

Section~\ref{appsec:peefit} of the appendix discusses the measured
coefficients, their
uncertainties, and their correlations of all three parametrizations of 
$P_{ee}$. It also compares the quadratic coefficients obtained from SK data
with those from SNO data, and the coefficients of the SK-SNO combined fit.
Fig.~\ref{figspecpeeexp} compares the allowed survival probability $P_{ee}$ based
on the exponential fit with that based on the cubic and quadratic fits.
Between about 5.5 and 12.5 MeV, the different parametrizations agree while
outside this energy region parametrization-dependent extrapolation effects
become significant.
While the strength of the SK data constraints on $P_{ee}$ is comparable to that of
SNO data, its low energy constraints are tighter and its high energy constraints
weaker. The reason for this is the absence of a nuclear threshold in
elastic electron-neutrino scattering, and the direct correlation of neutrino
energy and electron energy in neutrino-deuteron charged current interactions.
SK data prefers a slight ``upturn'', SNO data prefer a ``downturn''. The combined
fit favors an ``upturn'' more strongly than SK data by themselves since SK data prefer a higher average $P_{ee}$ than SNO data, and the tighter SK constraints
force the combined fit to this higher average probability at low energy, while
the tighter SNO constrains force the combined fit to the lower SNO value
at low energy.
Fig.~\ref{figspecpee} and
\ref{figspecpeecomb} (combined fit) display the 1 $\sigma$ allowed bands of
$P_{ee}(E_\nu)$. Fig.~\ref{figspecpeecomb_global} superimposes the same
combined
band (on a logarithmic scale) on the SSM~\cite{ssm} solar neutrino spectrum.
Also shown are the $pp$ and CNO neutrino flux constraints from all solar
data~\cite{othersolar} and the $^7$Be, the $pep$ and the $^8$B flux measurement
of
the Borexino experiment~\cite{otherborexino}. The SK and SNO combined allowed
band (and the other solar data) are in good agreement with the MSW curves
(based on different parameters: blue=solar+KamLAND best fit, 
data best fit, green=solar best fit).

\section{Day/Night Asymmetry}
The matter density of the Earth affects solar neutrino oscillations while the
Sun is below the horizon.  This so called ``day/night effect'' will lead to
an enhancement of the $\nu_e$ flavor content during the nighttime
for most oscillation parameters.  The most straightforward test of this effect
uses the solar zenith angle $\theta_z$ (defined in Fig.~\ref{fig:solang}) at
the time of each event to separately measure the solar neutrino
flux during the day $\Phi_D$ (defined as $\cos\theta_z \leq 0$) and
the night $\Phi_N$ (defined as $\cos\theta_z > 0$). The day/night
asymmetry $A_{\text{DN}}=(\Phi_D-\Phi_N)/\frac{1}{2}(\Phi_D+\Phi_N$)
defines a convenient measure of the size of the effect.

A more sophisticated method to test the day/night effect is given
in~\cite{dn,sk1}. For a given set of oscillation parameters, the interaction
rate as a function of the solar zenith angle is predicted. Only the shape of
the calculated solar zenith angle variation is used; the amplitude 
is scaled by an arbitrary parameter. The extended maximum likelihood fit
to extract the solar neutrino signal (see section~\ref{sec:flux}) is expanded
to allow time-varying signals. The likelihood is then evaluated as a function
of the average signal rates, the background rates and a scaling parameter,
termed the ``day/night amplitude''.  The equivalent day/night asymmetry
is calculated by multiplying the fit scaling parameter with the expected
day/night
asymmetry. In this manner the day/night asymmetry is measured more precisely
statistically and is less vulnerable to some key systematic effects.

Because the amplitude fit depends on the assumed shape of the day/night
variation (given for each energy bin in~\cite{dn} and \cite{sk1}), it
necessarily depends on the oscillation parameters, although with very little
dependence expected on the mixing angles (in or near the large mixing angle
solution and for $\theta_{13}$ values consistent with reactor neutrino
measurements~\cite{reactorexp}).  The fit is run for parameters covering the
MSW region of oscillation parameters
($10^{-9}$ eV$^2\le\Delta{m_{21}^2}\le10^{-3}$ eV$^2$ and $10^{-4}\le\sin^2\theta_{12} < 1$),
and values of $\sin^2\theta_{13}$  between 0.015 and 0.035.

\subsection{Systematic uncertainty on the solar neutrino amplitude fit day/night flux asymmetry}

\subsubsection{Energy scale}
True day (night) solar neutrino events will mostly be coming from the
downward (upward) direction, and so the directional dependence
of the SK light yield or energy scale will affect the observed interaction
rate as a function of solar zenith angle and energy. To quantify the
directional dependence of the energy scale, the energy of the DT-produced
$^{16}$N calibration data and its simulation are compared as a function of
the reconstructed detector zenith angle (Fig.~\ref{fig:dt_dirdep}). The
fit from Fig.~\ref{fig:dt_dirdep} is used to shift the energy of the $^8$B MC
events, while taking energy-bin correlations into account, and the unbinned
amplitude fit was re-run.  The resulting $0.05\%$ change in the
equivalent day/night asymmetry is taken as the systematic uncertainty coming
from the directional dependence of the energy scale.  The large reduction
compared to SK-I ($0.8\%$) comes from the use of a depth-dependent
water transparency parameter, introduced at the beginning of SK-III.  The
further reduction from SK-III ($0.2\%$) to SK-IV comes from an increase in
DT calibration statistics and the improved
timing agreement between data and MC, a result of the electronics upgrade.

\subsubsection{Energy resolution}
Throughout the different phases of SK, the energy resolution function
relating the true and reconstructed recoil electron energies was found by
two slightly different methods. During the SK-I and SK-IV phases, $^8$B
simulated events were used to set up a ``transfer matrix'' relating
reconstructed to true recoil electron energy (and reconstructed recoil
electron energy to neutrino energy.) This method, by construction, considers
the effect of all analysis cuts
on energy resolution. For the SK-II and III phases, dedicated mono-energetic
simulated events were produced to parametrize the energy resolution with a
Gaussian
function, modeling only some analysis cuts. The two methods produce slightly
different results; in particular, the predicted day/night asymmetries differ
by $0.05\%$.  To estimate the systematic uncertainty on the day/night asymmetry
coming from the energy resolution function, the amplitude fit was performed
using both methods, with the resulting $0.05\%$ difference taken as the
systematic uncertainty.

\subsubsection{Background shape}
Although there is only one background component fit in the day/night asymmetry
fit (any time dependence of the background should be much slower than the
day/night variation), different $\cos\theta_{\text{sun}}$ background shapes
must be used for different solar zenith angle bins.  We use one for the day
and six for
the night (in accordance with Table~\ref{tab:zenithspec}). The systematically
different shapes come from the detector's directional bias when reconstructing
background events (directions perpendicular to detector axes are preferred).
The background is first fit as functions of the detector zenith and azimuthal
angles. These fits also yield a covariance matrix $\mathbf{V}$ for the fit
parameters.
The parameters of each of the zenith and azimuthal fits are varied by the one
sigma statistical deviation, one at a time, giving a new background shape for
each solar zenith angle bin. Because the background distributions are
calculated
as projections of the detector zenith and azimuthal angles on the solar
direction, the shape deviations as a function of solar zenith angle are fully
correlated and must be varied simultaneously. The day/night amplitude fit is
then
re-run for each set of new background shapes. The difference in the central
value is taken as the error of the day/night asymmetry due to that particular
zenith or azimuthal fit parameter. These errors are then propagated to a total
systematic uncertainty using the covariance matrix $\mathbf{V}$ of the fit
to the detector zenith and azimuth angles.  The total uncertainty on the day/night
asymmetry coming from the background shapes is $0.6\%$,
and is the largest contribution to the total.

\subsubsection{Event selection}
Most of the analysis cuts affect the day and night solar neutrino interaction
rates equally, so their effect on the systematic uncertainty on the day/night
asymmetry can be neglected.  However, the vertex shift and angular resolution
difference between data and simulated events can cause a bias in the
external event cut efficiency when used in conjunction with the tight fiducial
volume cut.  To estimate the size of the effect, we artificially shift the
reconstructed vertex and direction and estimate the fraction of events which
are rejected by the cuts during daytime and during nighttime.
The associated estimated systematic uncertainty is $\pm0.1\%$.

\subsubsection{Earth model}
Different models of Earth's
density profile can change the signal rate zenith profiles, thus leading to
changes in the measured day/night asymmetry value.  For this reason it is
essential to model the earth as precisely as possible, most frequently done
using the PREM model~\cite{prem} and an Earth which is assumed to be spherical.
A spherical description of Earth using the equatorial radius leads to a
$\sim0.2\%$ change in the day/night effect from a spherical description using
an average radius. To better represent the Earth we have modeled
an ellipsoidal Earth, using the equatorial and polar radii as
the semi-major and semi-minor axes of an ellipse.  The ellipse is then rotated
around its minor axis to produce an ellipsoid and the spherical PREM model
density boundaries are mapped accordingly.

Due to SK's location on Earth and using the above procedure of modeling an
ellipsoidal Earth, the event rate is no longer rotationally symmetric about
the detector azimuthal angle and the
day/night zenith amplitude fit must take into account the change in the
expected signal rate as the azimuthal angle is varied.  This was done by
varying the azimuthal angle and the zenith angle when tracing
neutrinos through the Earth, and then using the detector livetime to average
over the azimuthal angle.  The resulting expected solar zenith angle dependent
signal rates were then used in the day/night amplitude fit and the results
compared to the results when assuming a spherical Earth with an average radius.
The $0.01\%$ change in the day/night
asymmetry is taken as the systematic uncertainty coming from the Earth shape.

As a final step in estimating the systematic uncertainty coming from the
model of the Earth, the PREM model was replaced with the more recent
PREM500 model \cite{prem500}, which gives an updated and more detailed
description of the density profile of Earth.  This resulted in a $0.01\%$
shift in the measured day/night asymmetry.  When added in quadrature to
the uncertainty coming from the Earth shape, $0.014\%$ gives the
total estimated uncertainty coming from the Earth model.

\subsubsection{Summary of the systematic uncertainty}
The total estimated systematic uncertainty on the measured day/night asymmetry
is calculated by adding the components in quadrature, the
result of which can be seen in Table~\ref{tab:DNsys}.  The large reduction
in systematics from SK-I \cite{sk1} to SK-IV comes from the introduction
of a $z$-dependent absorption into the simulation and
a better method of estimating the systematic uncertainty using DT data.  The
directional dependence of the energy scale is now better understood, bringing
the total systematic uncertainty to $\pm0.6\%$.

\begin{table}
  \caption{SK-IV amplitude fit day/night asymmetry systematic uncertainties.
The total is found by adding the contributions in quadrature.}
  \begin{tabular}{l c}
  \hline\hline
  Energy Scale       & $0.05\%$ \\
  Energy Resolution  & $0.05\%$  \\
  Background Shape   & $0.6\%$  \\
  Event Selection    & $0.1\%$  \\ 
  Earth Model        & $0.01\%$  \\ \hline
  Total              & $0.6\%$  \\
  \hline\hline
  \end{tabular}
  \label{tab:DNsys}
\end{table}

\subsection{SK day/night asymmetry results} \label{sec:dnamp}

\begin{figure}
\includegraphics[width=8.5cm,clip]{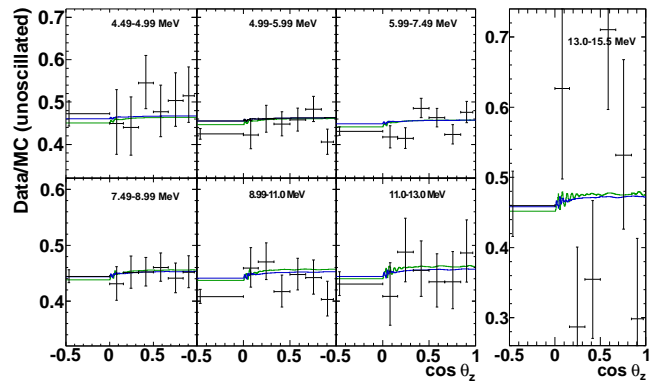}
  \caption{SK-IV data/MC (unoscillated) rate dependence on the solar zenith
angle, for various energy regions
(zenith angle and energy bins defined in Table~\ref{tab:zenithspec}, panels are ordered by energy with the upper, left panel being the lowest).
The unoscillated rate assumes a $^8$B (hep) flux of
$5.25\times 10^6$/(cm$^{2}$sec)
($8\times 10^3$/(cm$^{2}$sec)).  Overlaid
green (blue) lines are predictions when using the solar neutrino data (solar
neutrino data+KamLAND) best-fit oscillation parameters and the
assumed neutrino fluxes fit to best describe the data.   The error bars
shown are statistical uncertainties only.
  \label{fig:dn_enebin}}
\end{figure}

\begin{figure}[t]
 \begin{center}
 \includegraphics[width=8.5cm,clip]{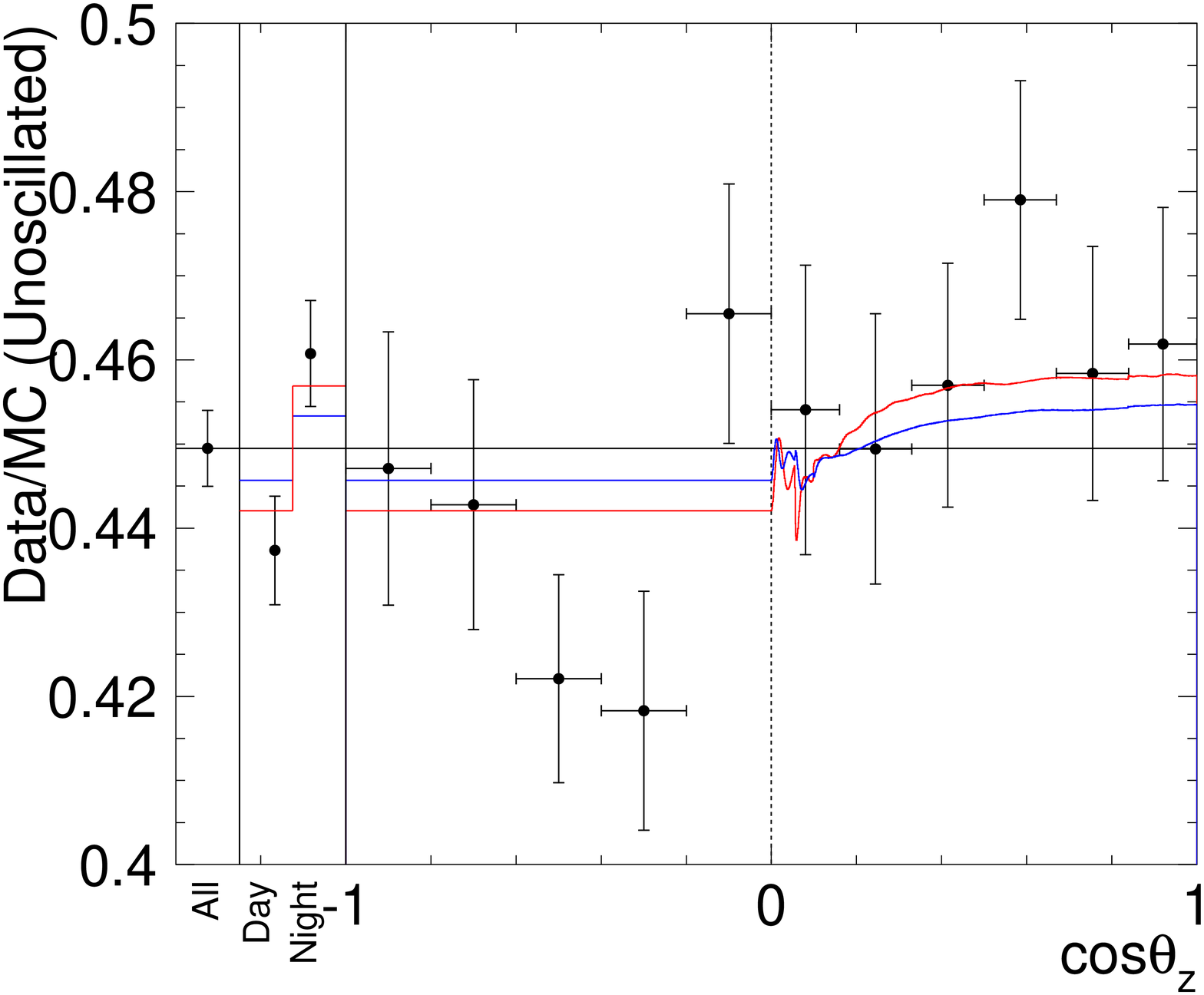}
\caption{SK-IV solar zenith angle dependence of the solar neutrino data/MC
(unoscillated) interaction rate ratio (4.49-19.5 MeV).  The day
data are subdivided into five bins, while the night data is divided into six
bins. Solar neutrinos in the last night bin pass through the Earth's outer
core. Overlaid red (blue) lines are predictions when using the solar neutrino
data (solar neutrino data+KamLAND) best-fit oscillation parameters and the
assumed neutrino fluxes fit to best describe the data. 
The error bars show are statistical uncertainties only.
\label{fig:dn_sk4}}
 \end{center}
\end{figure}

The SK-IV livetime during the day (night) is 797 days (866 days).
The solar neutrino flux between 4.49 and 19.5 MeV assuming no
oscillations is measured as
$\Phi_D=(2.250^{+0.030}_{-0.029}$(stat.)$\pm 0.038$(sys.)$)\times10^{6}$ /(cm$^2$sec)
during the day and
$\Phi_N=(2.364\pm0.029$(stat.)$\pm 0.040$(sys.)$)\times10^{6}$ /(cm$^2$sec)
during the night. Fig.~\ref{fig:dn_enebin} shows the solar zenith angle
variation of the ratio of the measured rate to the unoscillated
 simulated rate (assuming $5.25\times10^6$ /(cm$^2$sec) for the $^8$B flux)
in the seven energy regions shown in Table~\ref{tab:zenithspec}.
Overlaid is the expected zenith variation
for best-fit oscillation parameters coming from a fit to all solar neutrino
data (solar+KamLAND data) in red (blue).
Table~\ref{tab:zenithspec} lists the data used in
Fig.~\ref{fig:dn_enebin}, the errors are statistical uncertainties
only.  Fig.~\ref{fig:dn_sk4} shows the data over simulated rate ratio
between 4.49 and 19.5 MeV (assuming no oscillations) as a function of
$\cos\theta_z$, divided into five day and six night bins (corresponding to the
mantle 1-5 and core definitions of Table~\ref{tab:zenithspec}).
By comparing the separately measured day and night fluxes, the measured
day/night asymmetry for SK-IV is found to be
$A_{\text{DN}}=(-4.9\pm1.8(\text{stat.})\pm1.4(\text{syst.}))\%$.

\begin{table*}[t]
\begin{center}
\caption{The observed zenith angle dependence of event rates
(events/year/kton) in each energy
region, at 1 AU. The errors are statistical uncertainties only. The reduction
efficiencies are corrected and the expected event rates are for a flux of
$5.25\times10^6$ /(cm$^{2}$sec).}
\begin{tabular}{c c c c c c c c c c} \hline \hline
    & \multicolumn{7}{c}{Observed Rate} & \multicolumn{2}{c}{Unoscillated Rate}\\
 Energy & DAY & MANTLE1 & MANTLE2 & MANTLE3 & MANTLE4 & MANTLE5 & CORE                &  $^8$B & $hep$ \\
(MeV) & $\cos\theta_z=-1\sim0$ & $0\sim0.16$ & $0.16\sim0.33$ & $0.33\sim0.50$ & $0.50\sim0.67$ & $0.67\sim0.84$ & $0.84\sim1$ & & \\ \hline
 $ 4.49-4.99$ & $79.4^{+  5.1}_{-  5.0}$ & $75.5^{+ 13.4}_{- 12.2}$ & $74.5^{+ 12.1}_{- 11.1}$ & $91.6^{+ 10.9}_{- 10.2}$ & $80.3^{+ 10.6}_{- 9.9}$ & $85.1^{+ 11.1}_{- 10.3}$ & 
$86.9^{+ 11.4}_{- 10.6}$ & 167.8 & 0.323 \\
 $ 4.99-5.99$ & $124.2^{+  3.8}_{-  3.7}$ & $116.8^{+ 9.5}_{- 9.0}$ & $127.0^{+ 8.9}_{- 8.5}$ & $123.9^{+ 8.0}_{- 7.6}$ & $126.7^{+ 7.7}_{-  7.4}$ & $133.9^{+ 8.4}_{- 8.1}$ & 
$112.3^{+ 8.5}_{- 8.1}$ & 283.6 & 0.611 \\
 $ 5.99-7.49$ & $139.5^{+  3.3}_{-  3.2}$ & $134.2^{+ 8.6}_{- 8.2}$ & $133.3^{+ 8.1}_{-  7.7}$ & $155.7^{+  7.5}_{-  7.2}$ & $148.5^{+  7.1}_{-  6.9}$ & $136.1^{+  7.5}_{-  7.2}$ & 
$153.0^{+ 8.3}_{- 7.9}$ & 321.4 & 0.799 \\
 $ 7.49-8.99$ & $ 93.5^{+  2.7}_{-  2.7}$ & $ 89.3^{+  7.1}_{-  6.6}$ & $ 90.5^{+ 6.7}_{-  6.3}$ & $ 88.6^{+  5.9}_{-  5.6}$ & $ 94.0^{+  5.8}_{-  5.6}$ & $ 88.1^{+  6.2}_{-  5.9}$ & 
$102.2^{+  7.2}_{-  6.8}$ & 196.6 & 0.647 \\
 $ 8.99-11.0$ & $ 52.0^{+  1.8}_{-  1.8}$ & $ 55.7^{+  5.1}_{-  4.7}$ & $ 57.8^{+4.7}_{-  4.4}$ & $ 47.7^{+  4.0}_{-  3.7}$ & $ 54.4^{+  4.0}_{-  3.7}$ & $ 56.4^{+  4.4}_{-  4.1}$ & $ 
65.5^{+  5.1}_{-  4.8}$ & 122.2 & 0.619 \\
 $11.0-13.0$ & $ 15.5^{+  0.9}_{-  0.9}$ & $ 17.4^{+  2.6}_{-  2.2}$ & $ 17.3^{+2.5}_{-  2.1}$ & $ 15.3^{+  2.0}_{-  1.8}$ & $ 14.9^{+  2.0}_{-  1.7}$ & $ 15.2^{+  2.2}_{-  1.9}$ & $ 
17.7^{+  2.5}_{-  2.2}$ &  36.0 & 0.365 \\
 $13.0-15.5$ & $ 3.83^{+ 0.46}_{- 0.40}$ & $ 5.69^{+ 1.54}_{- 1.18}$ & $ 2.53^{+1.07}_{- 0.73}$ & $ 2.49^{+ 0.91}_{- 0.65}$ & $ 4.19^{+ 1.03}_{- 0.80}$ & $ 3.84^{+ 1.15}_{- 0.86}$ & $ 
4.48^{+ 1.33}_{- 1.01}$ &  7.45 & 0.204 \\
\hline \hline
\end{tabular}
\label{tab:zenithspec}
\end{center}
\end{table*}

The SK-IV day/night asymmetry resulting from the day/night amplitude fit
method, for an energy range of 4.49-19.5 MeV
and oscillations parameters preferred by SK
($\Delta{m_{21}^2}=4.84\times10^{-5}$ eV$^2$,
$\sin^2\theta_{12}=0.311$ and $\sin^2\theta_{13}=0.020$), is found to be

\begin{align*}
A_{\text{DN}}^{\text{fit, SK-IV}}=(-3.6\pm1.6(\text{stat.})\pm0.6(\text{syst.}))\%. 
\end{align*}

The expected day/night asymmetry for the above set of oscillation parameters
is $-3.3\%$.  For the case of a global fit to solar neutrino
data and KamLAND~\cite{kamland}, the mass squared splitting changes to
$\Delta{m_{21}^2}=7.50\times10^{-5}$ eV$^2$, and the expected day/night
asymmetry goes to $-1.7\%$.  However, the day/night amplitude fit measured
SK-IV day/night asymmetry is only slightly reduced to

\begin{align*}
A_{\text{DN}}^{\text{fit, SK-IV}}=(-3.3\pm1.5(\text{stat.})\pm0.6(\text{syst.}))\%.
\end{align*}

Within the LMA region, all measured values of the day/night asymmetry coming
from the day/night amplitude fit are within $\pm0.3\%$ of $-3.3\%$.  If the
above measurement is combined with the previous three phases
of SK, the SK combined measured day/night asymmetry is

\begin{align*}
A_{\text{DN}}^{\text{fit, SK}}=(-3.3\pm1.0(\text{stat.})\pm0.5(\text{syst.}))\%.
\end{align*}

Previously, we published
$A_{\text{DN}}^{\text{fit, SK}}=(-3.2\pm1.1(\text{stat.})\pm0.5(\text{syst.}))\%$
in ~\cite{skall_dn} which was the first significant indication that matter
effects influence neutrino oscillations. The slightly larger significance
here is due to a somewhat larger data set.

\section{Oscillation Analysis}
\label{sec:osc}
\begin{figure*}[t]
\includegraphics[width=8.8cm,clip]{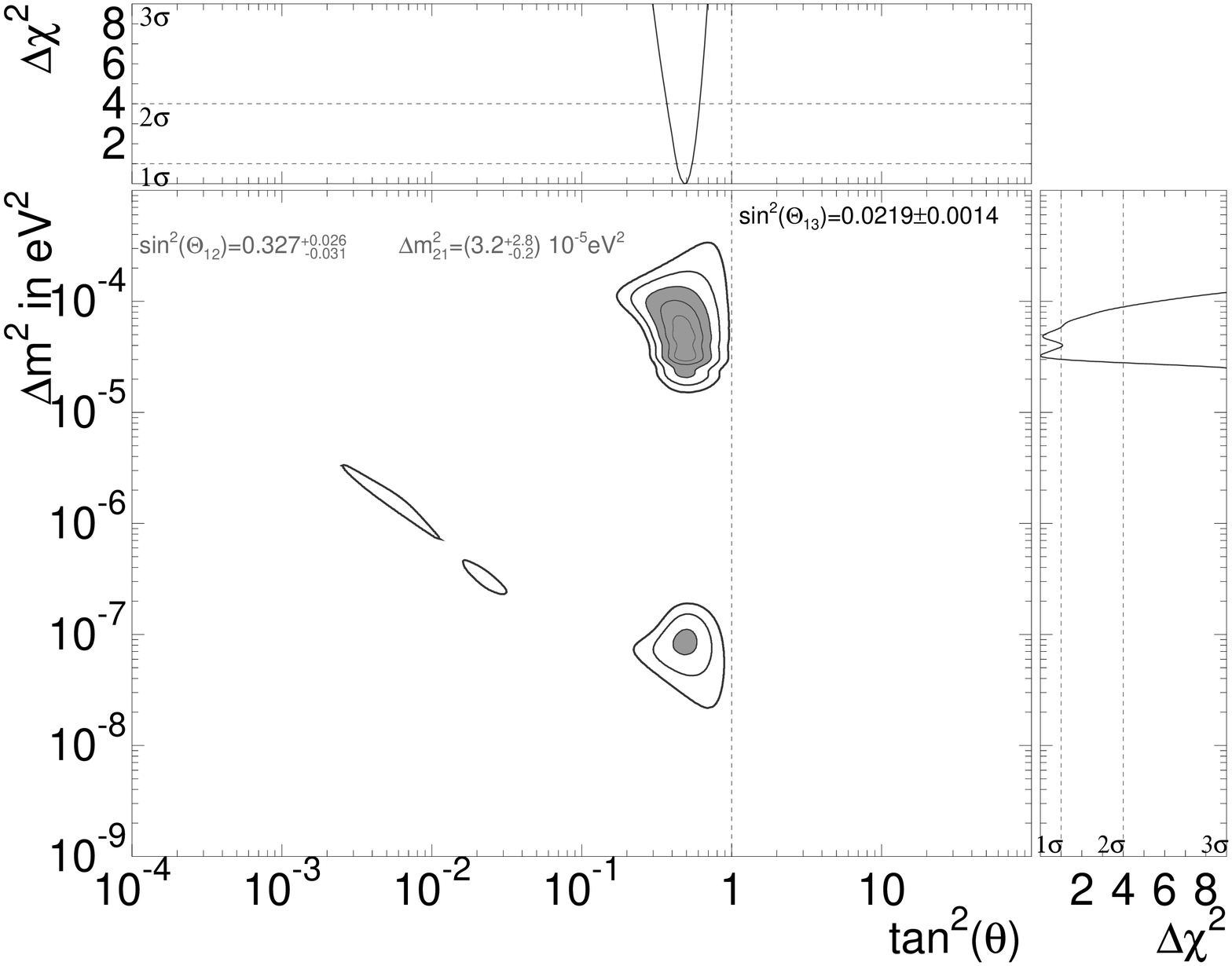}
\includegraphics[width=8.8cm,clip]{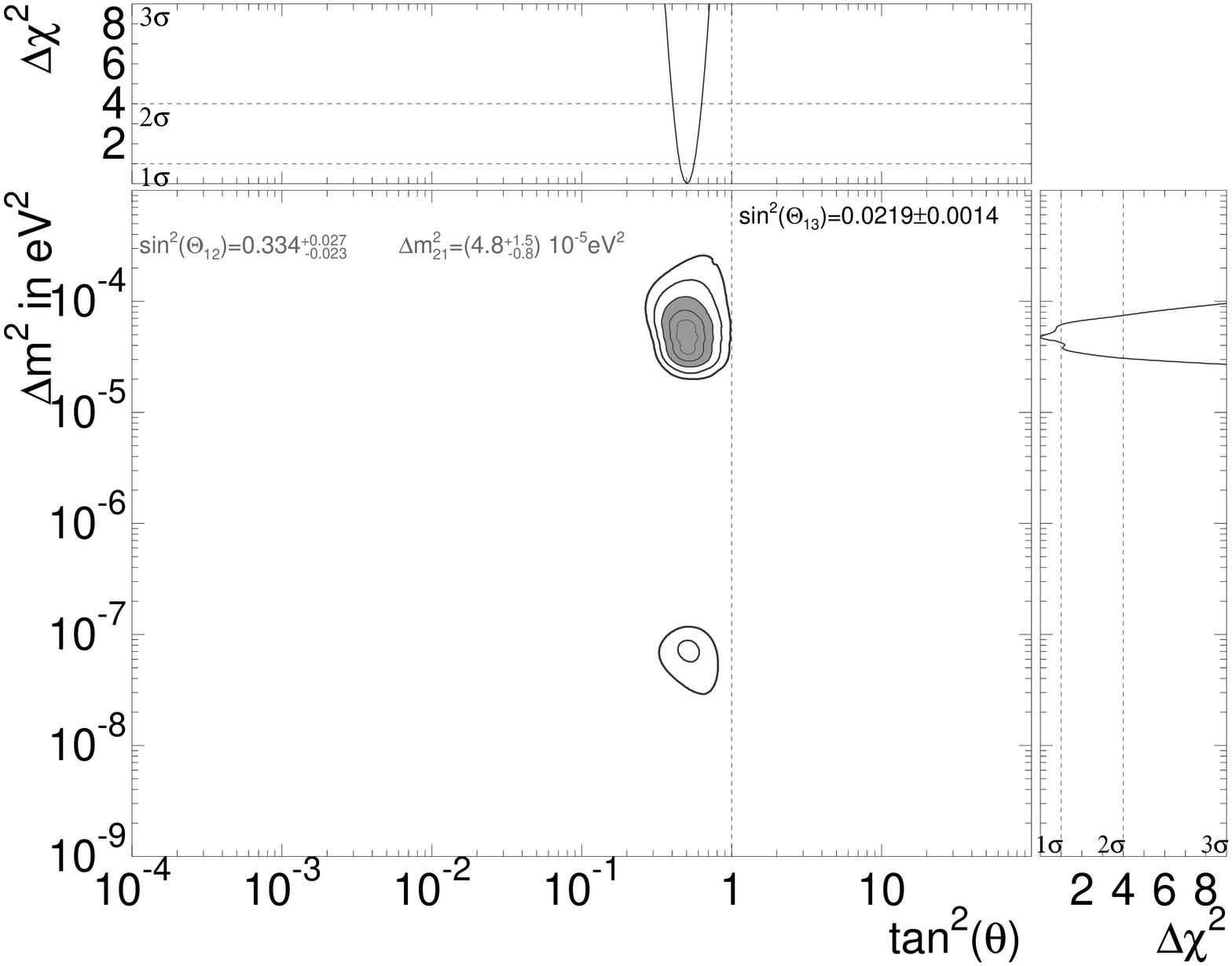}
\caption{Contours of $\Delta m^2_{21}$ vs. $\tan^2\theta_{12}$
from the SK-IV (left panel) and SK-I/II/III/IV (right panel)
spectral+day/night data with a $^8$B flux constraint of
$5.25\pm0.20\times10^6$ /(cm$^2$sec)
at the 1, 2, 3, 4 and 5 $\sigma$ confidence levels.
The filled regions give the 3 $\sigma$ confidence level results.
$\theta_{13}$ is constrained by
$\left(\frac{\sin^2\theta_{13}-0.0219}{0.0014}\right)^2$.}
\label{figoscdnspec3}
\end{figure*}

SK measures elastic scattering of solar neutrinos with electrons, the rate
of which depends on the flavor content of the solar neutrino flux, so it is
sensitive to neutrino flavor oscillations. To constrain the parameters
governing these oscillations, we analyze the integrated scattering rate, the
recoil electron spectrum (which statistically implies the energy-dependence
of the electron-flavor survival probability), and the time of the interactions
which defines the neutrino path through the earth during night time, and
therefore controls the earth matter effects on solar neutrino oscillations.
An expansion of the likelihood used in the extended maximum likelihood fit to
extract the solar neutrino signal (see section~\ref{sec:flux}) could make full
use of {\it all} information (timing, spectral information and rate), but is
CPU time intensive. Instead, we separate the log(likelihood) into a
time-variation (day/night variation) portion $\log\mathcal{L}_{\text{\tiny DN}}$
and a spectral portion:
$\log\mathcal{L}=\log\mathcal{L}_{\text{\tiny DN}}+\log\mathcal{L}_{\mbox{\tiny spec}}$ 
where $\mathcal{L}_{\mbox{\tiny spec}}$, the likelihood for assuming no
time variation, is replaced by $-\frac{1}{2}\chi^2_{\mbox{\tiny spec}}$.
This $\chi^2_{\mbox{\tiny spec}}$ fits the calculated elastic scattering rate
rate in energy bin $e$ of a particular SK phase $p$ to the measurement
$d_e^p\pm\sigma_e^p$. The calculated event rate $r_e^p$ is the sum of the
expeced elastic scattering rate $b_e^p$ from $^8$B neutrinos scaled by the
parameter $\beta$ and $h_e^p$ from hep neutrinos scaled by the parameter $\eta$:
$r_e^p=\beta b_e^p+\eta h_e^p$. The calculation includes neutrino flavor
oscillations of three flavors; they depend on the mixing angles
$\theta_{12}$, $\theta_{13}$ and the mass squared difference $\Delta m^2_{21}$.
$r_e^p$ is then multiplied by the spectral distortion factor $f_e^p(\tau,\epsilon_p,\rho_p)$ which
describes the effect of a systematic shift of the $^8$B neutrino spectrum
scaled by the constrained nuisance parameter $\tau$, a deviation
in the SK energy scale in phase $p$ described by the constrained
nuisance parameter $\epsilon_p$, and a systematic change in the SK energy
resolution based on a third constrained nuisance parameter $\rho_p$.
If $N_p$ is the number of energy bins of phase $p$, we minimize
\[
\chi^2_p(\beta,\eta)=
\sum_{e=1}^{N_p}\left(
\frac{d_e^p-f_e^pr_e^p(\sin^2\theta_{12},\sin^2\theta_{13},\Delta m^2_{21})}
{\sigma_e^p}
\right)^2
\]
over all systematic nuisance parameters and the flux scaling parameters:
\begin{equation}
\chi^2_{\mbox{\tiny spec},1}=\underset{\small\tau,\epsilon_p,\rho_p,\beta,\eta}{\mbox{Min}}
\left(\chi^2_{p,\mbox{\tiny dat}}+\tau^2+\epsilon_p^2+\rho_p^2+\Phi\right),
\label{eq:chi2}
\end{equation}
where
$\Phi=\left(\frac{\beta-\beta_0}{\sigma_\beta}\right)^2+\left(\frac{\eta-\eta_0}{\sigma_\eta}\right)^2$
constrains the flux parameters to prior uncertainties:
$\beta$ is constrained to result in a $^8B$ flux of  $(5.25\pm0.20)\times10^6$/(cm$^2$sec)
(motivated by the SNO NC measurement of the total $^8$B neutrino
flux~\cite{snopaper}), $\eta$ is only slightly constrained to correspond to a hep flux of
$(8\pm16)\times10^3$/(cm$^2$sec). The nuisance parameters $\tau$,
$\epsilon_p$, and $\rho_p$ are constrained to $0\pm1$ (i.e.
they are defined as standard Gaussian variables) by the ``penalty
terms'' $\left(\frac{\tau-0}{1}\right)^2$,
$\left(\frac{\epsilon_p-0}{1}\right)^2$, and
$\left(\frac{\rho_p-0}{1}\right)^2$. We rewrite
equation~\ref{eq:chi2} as a quadratic form with the $2\times2$
curvature matrix $\mathbf{C}_p$ and the best-fit flux parameters
$\beta^p_{\mbox{\tiny min}}$ and $\eta^p_{\mbox{\tiny min}}$ as
\begin{multline}
\chi^2_{\mbox{\tiny spec},\alpha_p}=\chi^2_{p,\mbox{\tiny min}}+ \\
\alpha_p
\left(\beta-\beta^p_{\mbox{\tiny min}},\eta-\eta^p_{\mbox{\tiny min}}\right)\cdot
\mathbf{C}_p\cdot\begin{pmatrix}\beta-\beta^p_{\mbox{\tiny min}}\cr \eta-\eta^p_{\mbox{\tiny min}}\end{pmatrix},
\end{multline}
\noindent for $\alpha_p=1$. The parameter $\alpha_p\neq1$ is introduced
to scale the a posteriori constraints on the flux parameters
by $1/\sqrt\alpha_p$ without affecting the $\chi^2$ minimum in order to
take into account additional systematic uncertainties of the total rate.
These uncertainties are not covered by $\sigma_e^p$ or $f_e^p$.
Table~\ref{tab:totsyst} (subtotal) lists these additional uncertainties
integrated over all energies. To incorporate them we choose
$\alpha_p=\frac{\sigma_{p,\mbox{\tiny stat}}^2}{\sigma_{p,\mbox{\tiny stat}}^2+\sigma_{p,\mbox{\tiny syst}}^2}$.
To best compare this three-flavor analysis to two-flavor analyses performed
for previous phases, we
also perform an analysis with an a priori constraint on $\theta_{13}$
coming from reactor neutrino experiments~\cite{reactorexp}. Unlike the
two-flavor analyses, $\theta_{13}$ is not fixed to zero, but constrained
to a non-zero value by $\left(\frac{\sin^2\theta_{13}-0.0219}{0.0014}\right)^2$.
The time-variation likelihood $\log\mathcal{L}_{\text{\tiny DN}}=\log\mathcal{L}_{\text{\tiny with}}-\log\mathcal{L}_{\text{\tiny without}}$ is simply
the difference between the likelihoods with and without the predicted
day/night variation {\it assuming} the best-fit flux and nuisance parameters
from the spectrum $\chi^2$ minimization. As the uncertainties in each
spectral bin are closely approximated by Gaussian uncertainties, the
total $\chi^2$ is then given by
$\chi^2_{\mbox{\tiny spec}}-2\log\mathcal{L}_{\text{\tiny DN}}$.
Figure~\ref{figoscdnspec3} shows allowed regions of oscillation
parameters from SK-IV data with the external constraint from reactor
neutrino data on $\theta_{13}$ at the $1$, $2$, $3$, $4$, and $5$ $\sigma$
confidence level. SK-IV determines $\sin^2\theta_{12}$ to be
$0.327^{+0.026}_{-0.031}$, as well as $\Delta m_{21}^2$ to be
$\left(3.2^{+2.8}_{-0.2} \right) \times 10^{-5}$ eV$^2$. A secondary region appears at about
the $3 \sigma$ level at $\Delta m_{21}^2\approx8 \times 10^{-8}$eV$^2$.
Small mixing is only very marginally allowed at about the $5 \sigma$
confidence level.

We combined the SK-IV constraints with those of previous SK
phases, as well as other solar neutrino experiments \cite{snopaper,othersolar}.
For the combined SK fit, the spectrum and rate $\chi^2$ is
\begin{equation}
\chi^2_{\mbox{\tiny spec}}=\underset{\small\nu,\epsilon_p,\rho_p,\beta,\eta}{\mbox{Min}}
\left(\sum_{p=1}^4\chi^2_{p,\alpha_p}+\tau^2+\sum_{p=1}^{4}(\epsilon_p^2+\rho_p^2)+\Phi\right).
\end{equation}
Each SK phase is represented by a separate day/night likelihood ratio,
where the flux and nuisance parameters are taken from the combined fit.
Fig.~\ref{figoscdnspec3} shows the SK combined allowed areas
based on rate, spectrum, and day/night variation. SK selects large
mixing ($0.5>\sin^2\theta_{12}>0.2$) over small mixing by more than
five standard deviations and very strongly (3.6 $\sigma$) favors 
the $\Delta m^2_{21}$ of the large mixing angle (LMA) solution
(below $2\cdot 10^{-4}$eV$^2$ and above $2\cdot 10^{-5}$eV$^2$)
over any other oscillation parameters. SK determines $\sin^2\theta_{12}$ to be
$0.334^{+0.027}_{-0.023}$, as well as
$\Delta m_{21}^2$ to be $\left( 4.8^{+1.5}_{-0.8} \right) \times10^{-5}$ eV$^2$.

\begin{figure*}[t]
\includegraphics[width=8.8cm,clip]{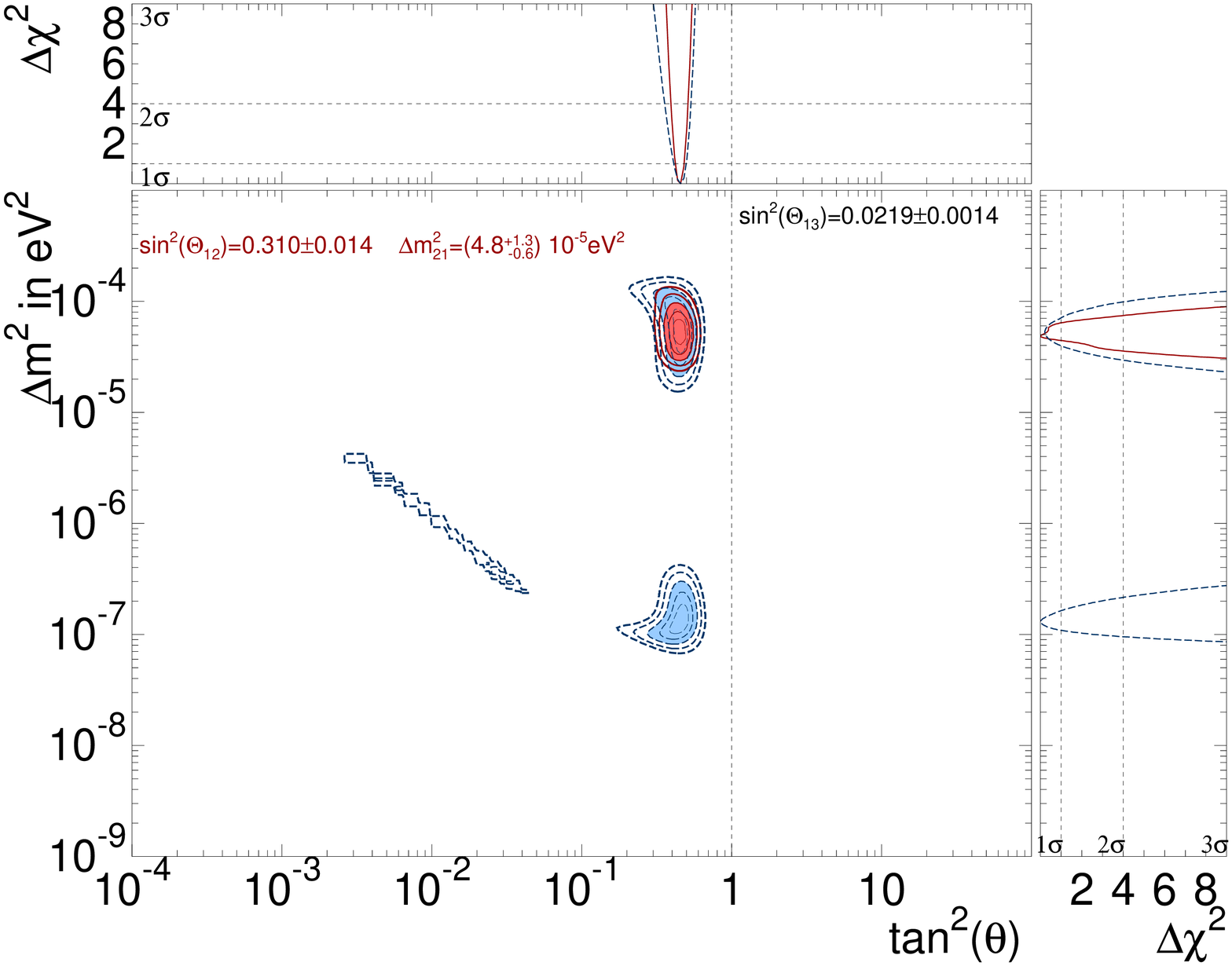}
\includegraphics[width=8.8cm,clip]{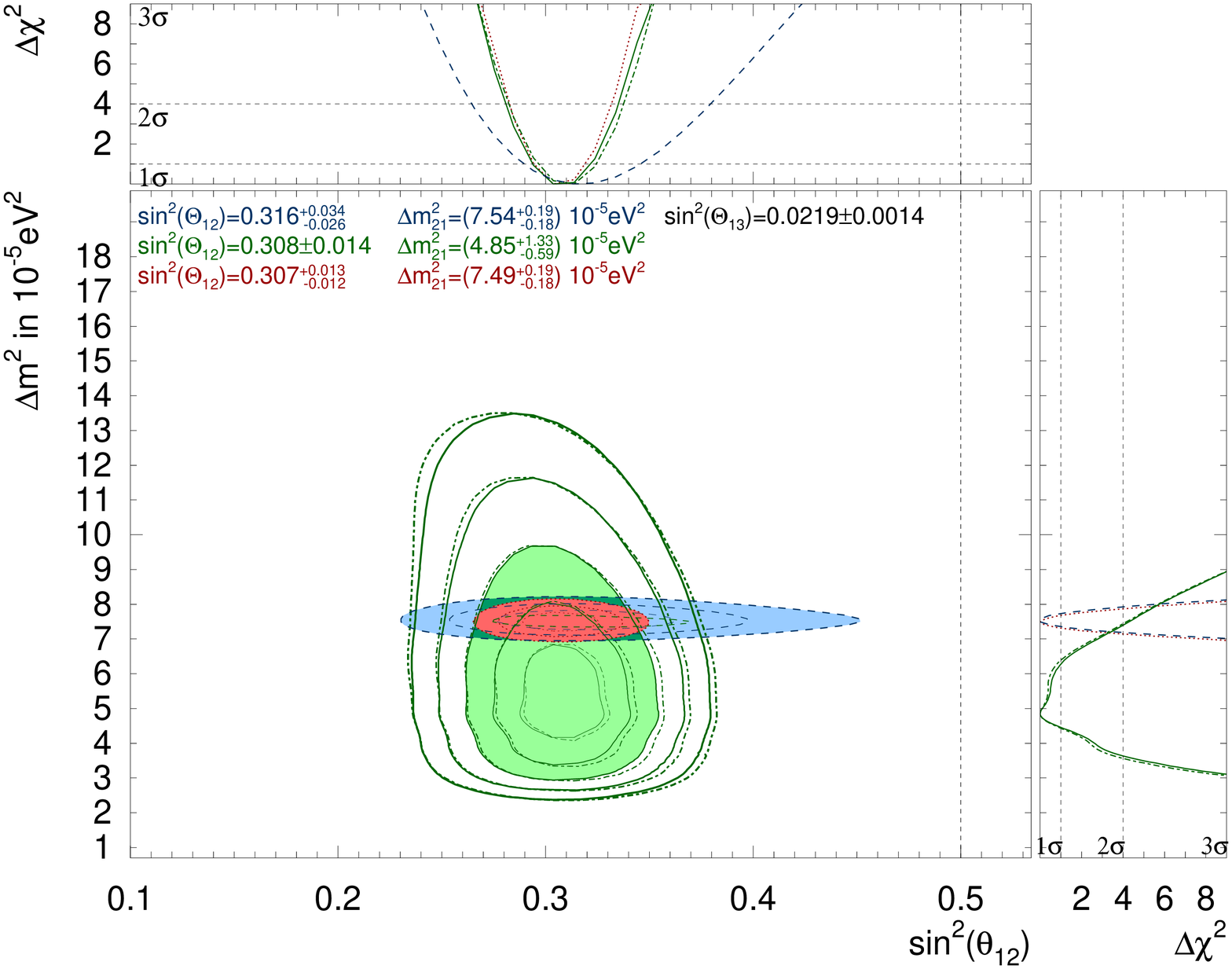}
\caption{Left: comparison of the oscillation parameter determination
of the SK and SNO combined analysis (red) to the oscillation
constraints of SNO by itself (blue).
Right: allowed contours of $\Delta m^2_{21}$ vs. $\sin^2\theta_{12}$
from solar neutrino data (green), KamLAND data (blue), and the
combined result (red). For comparison, the almost
identical result of the SK+SNO combined fit is shown by the dashed
dotted lines.
The filled regions give the 3 $\sigma$ confidence level results,
the other contours shown are at the 1 and  2 $\sigma$ confidence
level (for the solar analyses, 4 and 5 $\sigma$ confidence level
contours are also displayed). 
$\theta_{13}$ is constrained by
$\left(\frac{\sin^2\theta_{13}-0.0219}{0.0014}\right)^2$.}
\label{figoscotherexp}
\end{figure*}

Fig.~\ref{figoscotherexp} compares the SK+SNO combined constraints to
those based on SNO data alone~\cite{snopaper}. While SNO's measurement
of the mixing angle is more precise ($\sin^2\theta_{12}=0.299^{+0.023}_{-0.020}$)
than SK's, its $\Delta m^2_{21}$ constraints are poorer
($\left( 5.6^{+1.9}_{-1.4} \right) \times10^{-5}$ eV$^2$). Also, SNO very slightly favors
the Low solution (near $10^{-7}$ eV$^2$) and allows small
mixing at the 3.6 $\sigma$ level. The combined analysis of SK and
SNO is particularly powerful: as SNO and SK both measure $^8$B
neutrinos in a very similar energy range but in a different way
and with different systematic effects, the
combined analysis profits from correlations and is better than a mere
addition of $\chi^2$'s.
The SK+SNO combined analysis measures $\sin^2\theta_{12}=0.310\pm0.014$
and $\Delta m_{21}^2= \left(4.8^{+1.3}_{-0.6}\right)\times10^{-5}$eV$^2$.
Oscillation parameter values outside the LMA are very strongly
excluded: the solar mixing angle lies within
$0.12\le\sin^2\theta_{12}\le0.45$ at about the 7.5 $\sigma$ C.L.,
$\Delta m_{21}^2<1.33\times10^{-5}$eV$^2$ (which includes the
``small mixing angle'' and ``low'' regions) is ruled out at the 5.5 $\sigma$
C.L., and $\Delta m_{21}^2>1.9\times10^{-4}$eV$^2$ is excluded at
7.5 $\sigma$ C.L. The {\it hep} flux constraint used by SNO
is $(7.9\pm1.2)\times10^3$/(cm$^2$sec) from the solar standard model~\cite{ssm}.
The SK and SNO combined analysis also uses this tighter constraint.

\begin{figure}[t]
\includegraphics[width=8.8cm,clip]{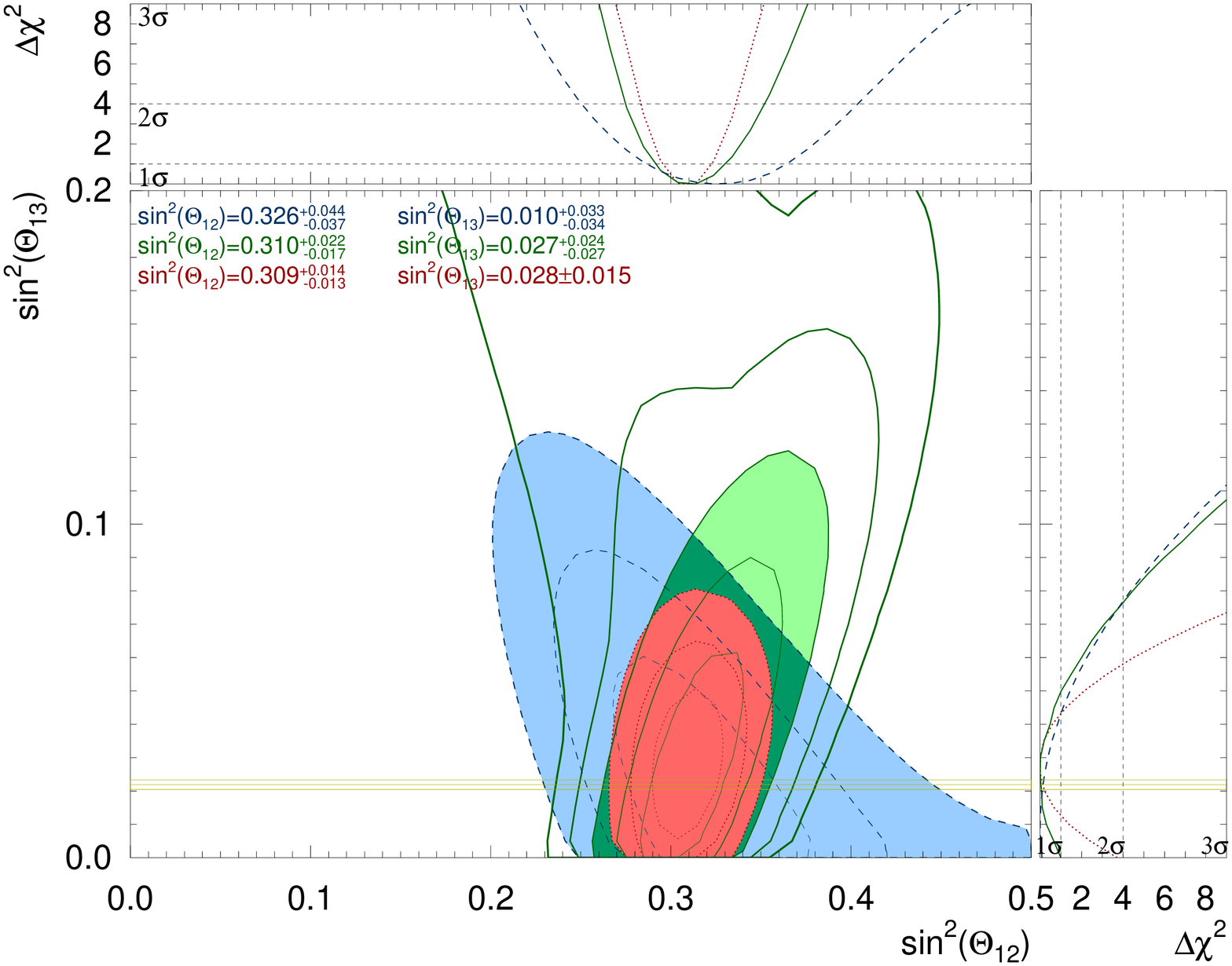}
\caption{Allowed contours of $\sin^2\theta_{13}$ vs.
$\sin^2\theta_{12}$ from solar neutrino data (green)
 at 1, 2, 3, 4 and 5 $\sigma$ and KamLAND measurements (blue) at
the 1, 2 and 3 $\sigma$ confidence levels. Also shown is the
combined result in red. The yellow band is the $\theta_{13}$ measurement
from reactor neutrino data~\cite{reactorexp}.}
\label{figoscglobalangle}
\end{figure}

The combined allowed contours based on SK, SNO~\cite{snopaper} and
other solar neutrino experiments'~\cite{othersolar} data, KamLAND's
constraints and the combination of the two are shown in
Fig.~\ref{figoscotherexp} and
Fig.~\ref{figoscglobalangle}. SK and SNO dominate the combined fit to all
solar neutrino data.  This can be seen from the two almost identical sets
of green contours in Fig.~\ref{figoscotherexp}.
In the right panel of this figure, some tension between the
solar neutrino and reactor anti-neutrino measurements
of the solar $\Delta m^2_{21}$ is evident, stemming from the SK day/night
measurement. Even though the
expected day/night amplitude agrees within $\sim1.1\sigma$ with the fitted
amplitude for any $\Delta m^2_{21}$, in either the KamLAND or the SK range,
the SK data slightly favor the shape of the day/night variation predicted
by values of $\Delta m^2_{21}$ that are smaller than
KamLAND's. Fig.~\ref{figoscglobalangle} shows the results of the $\theta_{13}$
unconstrained fit. Solar neutrinos by themselves weakly favor a non-zero
$\theta_{13}$ by about one standard deviation because for low energy solar
neutrinos the survival probability (e.g $^7Be$) is about
$(1-\frac{1}{2}\sin^2(2\theta_{12}))\cos^4(\theta_{13})$
while the MSW effect causes a high energy ($^8$B) solar neutrino survival
probability of $\sin^2(\theta_{12})\cos^4(\theta_{13})$. This results in
a correlation of $\sin^2(\theta_{12})$ and $\sin^2(\theta_{13})$ for high
energy neutrinos and an anti-correlation for low energy neutrinos.
KamLAND reactor neutrino data has the same anti-correlation as the low
energy solar neutrinos because in both cases matter effects play a minor
role. Therefore the significance of non-zero $\theta_{13}$ increases in the
solar+KamLAND data combined fit to about two $\sigma$, favoring
$\sin^2\theta_{13}=0.028\pm0.015$.

\section{Conclusion}
The fourth phase of SK measured the solar $^8$B neutrino-electron
elastic scattering-rate with the highest precision yet.
SK-IV measured a solar neutrino flux of
$(2.308\pm0.020(\text{stat.})^{+0.039}_{-0.040}(\text{syst.}))\times10^6/(\text{cm}^{2}\text{sec})$
assuming no oscillations.
When combined with the
results from the previous three phases, the SK combined flux is
($2.345\pm0.014$(stat.)$\pm0.036$(syst.))$\times10^6$ /(cm$^{2}$sec).
A quadratic fit of the electron-flavor survival probability as a function of
energy to all SK data, as well as a combined fit with SNO solar neutrino data,
very slightly favor the presence of spectral distortions, but are still
consistent with an energy-independent electron neutrino flavor content.
The SK-IV solar neutrino elastic scattering day/night rate asymmetry is
measured as ($-3.6\pm1.6$(stat.)$\pm0.6$(syst.))$\%$.  Combining this with
other SK phases, the SK solar zenith angle
variation data gives the first significant indication for matter-enhanced
neutrino oscillation. This leads SK to having the world's
most precise measurement of $\Delta m_{21}^2=\left(4.8^{+1.5}_{-0.8}\right)\times10^{-5}$ 
eV$^2$, using neutrinos
rather than anti-neutrinos.  There is a slight tension of 1.5 $\sigma$
between this value and KamLAND's measurement using reactor
anti-neutrinos. The tension increases to 1.6 $\sigma$, if other solar
neutrino data are included.  
The SK-IV solar neutrino data determine the solar mixing angle as
$\sin^2\theta_{12}=0.327^{+0.026}_{-0.031}$, all SK solar data measures
this angle to be $\sin^2\theta_{12}=0.334^{+0.027}_{-0.023}$,
the determined squared splitting is $\Delta m^2_{21}=4.8^{+1.5}_{-0.8}\times10^{-5}$ eV$^2$.
A $\theta_{13}$ constrained fit to all solar
neutrino data and KamLAND yields $\sin^2\theta_{12}=0.307^{+0.013}_{-0.012}$ and
$\Delta m_{21}^2=\left(7.49^{+0.19}_{-0.18}\right)\times10^{-5}$
eV$^2$. 
When this constraint is removed, solar neutrino experiments and KamLAND measure
$\sin^2\theta_{13}=0.028 \pm 0.015$, a value in good agreement
with reactor neutrino measurements.

\section{Acknowledgments}
The authors gratefully acknowledge the cooperation of the Kamioka Mining and
Smelting Company. Super-K has been built and operated from funds provided by
the Japanese Ministry of Education, Culture, Sports, Science and Technology,
the U.S. Department of Energy, and the U.S. National Science Foundation. This
work was partially supported by the  Research Foundation of Korea (BK21 and
KNRC), the Korean Ministry of Science and Technology, the National Science
Foundation of China (Grant NO. 11235006), the European Union 
H2020 RISE-GA641540-SKPLUS, and 
the National Science Centre, Poland (2015/17/N/ST2/04064, 2015/18/E/ST200758).

\clearpage
\appendix
\setcounter{figure}{0} \renewcommand{\thefigure}{A.\arabic{figure}} 
\setcounter{table}{0} \renewcommand{\thetable}{A.\arabic{table}} 

\section{Revised SK-III results}

Since the publication of the previous report~\cite{sk3}, two mistakes were found.
One is in how energy-dependent systematic errors are calculated and the
other is related to the flux calculation in SK-III.
The estimates of the energy-correlated uncertainties in the main text
of that report are
based on the Monte Carlo (MC) simulated $^8$B solar neutrino events.
It is found that this evaluation method was not accurate enough.
The statistical error of the MC simulation distorted the shapes of the 
energy-correlated uncertainties systematically.

\begin{figure}[!hbt]
\includegraphics[width=8.0cm,clip]{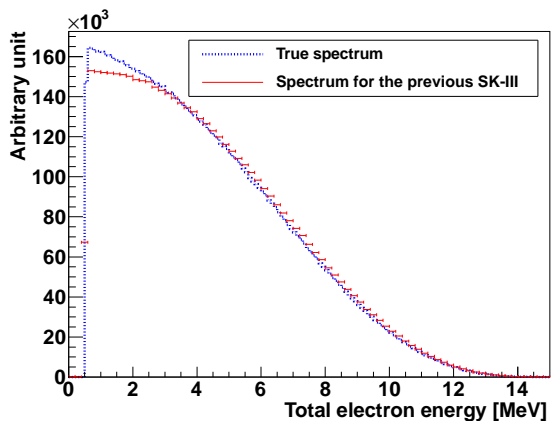}
\caption{Energy spectrum shapes of recoil electrons from $^8$B solar neutrinos
 for SK-III.
   The blue dotted and red solid lines show the true theoretical calculation 
   and incorrect spectrum used in the SK-III analysis in the previous report~\protect{\cite{sk3}}.
   }
   \label{fig:hitmis-sk3}
\end{figure}

The energy dependence of the differential interaction cross-section
between neutrinos and electrons was accidentally eliminated
only for the SK-III flux calculation in the main text.
Figure \ref{fig:hitmis-sk3} shows the energy distributions of recoil
electrons from $^8$B solar neutrinos.
The blue dotted histogram shows the true energy spectrum shape from a
theoretical calculation considering the detector resolutions.
The red solid plot shows the energy spectrum shape used in the SK-III analysis
in the previous report.
The expected total flux was normalized correctly, but the expected 
$^8$B energy spectrum shape was improperly distorted in the analysis. 

These mistakes have been fixed in this paper.
In this appendix, the revised SK-III solar neutrino results are described.
The latest oscillation results, including both revised SK-III data 
and SK-IV data, are reported in the main text of this report.

\subsection{Systematic uncertainties}

The energy-correlated systematic uncertainties are obtained by counting the number of 
events in the solar neutrino MC simulation with artificially shifted energy scale,
energy resolution and $^8$B solar neutrino energy spectrum.
In the SK-III analysis in the previous report, this estimation was done with the generated 
solar neutrino MC events.
However, in the high energy region, not enough MC events were generated
to accurately estimate the small systematic errors.
In the current analysis, this estimation is performed with a theoretical calculation
considering the detector resolutions,
thus eliminating the statistical effects introduced by the small MC
statistics.

\begin{figure}[thb]
\includegraphics[width=7.0cm,clip]{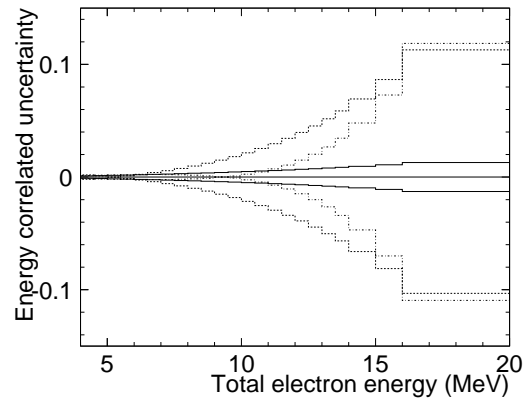}
\caption{Revised energy-correlated systematic uncertainties in SK-III. 
   The solid, dotted, and dashed lines show the uncertainties of the $^8$B
   spectrum, the energy scale, and the energy resolution, respectively.
 This is a revision of Fig.\ 25 in the previous paper~\protect{\cite{sk3}}.}
   \label{fig:ecorr-sk3}
\end{figure}

The revised results of the energy-correlated systematic uncertainties
are shown in Fig.\ \ref{fig:ecorr-sk3}.
In this update, the uncertainty from $^8$B spectrum shape was improved.

\begin{table}[!hbt]
\label{tab:totalsys-sk3}
\caption{ Revised summary of the systematic uncertainty of the total flux 
 in E$_{\rm total}$ = 5.0--20.0 MeV in SK-III.
 This is a revision of Table IV in the previous paper~\protect{\cite{sk3}}.}
\begin{center}
\begin{tabular}{l c} \hline\hline
Source               &  Total Flux       \\ \hline
Energy scale         &$\pm 1.4\%   $ \\ 
Energy resolution    &$\pm 0.2\%   $  \\ 
$^8$B spectrum       &$\pm 0.4\%   $            \\ 
Trigger efficiency   &$\pm 0.5\%  $   \\ 
Angular resolution   &$\pm 0.67\%   $ \\ 
Vertex shift         &$\pm 0.54\%  $ \\ 
Event quality cuts   &  \\
- Reconstruction Goodness &$\pm 0.4\%   $  \\ 
- Hit pattern        &$\pm 0.25\%  $  \\ 
- Second vertex      &$\pm 0.45\%  $  \\ 
Spallation cut       &$\pm 0.2\%   $ \\ 
Gamma-ray cut        &$\pm 0.25\%  $ \\
Cluster hit cut      &$\pm 0.5\%  $  \\ 
Background shape     &$\pm 0.1\%   $  \\ 
Signal extraction    &$\pm 0.7\%  $   \\ 
Livetime             &$\pm 0.1\% $ \\
Cross section        &$\pm 0.5\%  $   \\ \hline
Total                &$\pm 2.2\%  $\\ \hline\hline
\end{tabular}
\end{center}
\end{table}

\begin{table*}[!t]
 \label{tab:rates-sk3}
 \begin{center}
  \caption{Revised observed energy spectra expressed in units of 
event/kton/year in SK-III in each recoil electron total energy region.  
The errors in the observed rates are statistical only.  The expected rates 
neglecting oscillations are for a flux value of 
$5.79\times 10^\text{6}$ cm$^{-2}$sec$^{-1}$.  $\theta_{z}$ is the angle between 
the z-axis of the detector and the vector from the Sun to the detector.  
This is a revision of Table VI in the previous paper~\protect{\cite{sk3}}.}
  \begin{tabular}{c c c c c c}
   \hline
   Energy    & \multicolumn{3}{c}{Observed Rate} & \multicolumn{2}{c}{Expected Rate}\\
   (MeV)     &  ALL & DAY & NIGHT & $^8$B & hep\\
             & $ -1 \leq \cos\theta_{\rm z} \leq 1 $ 
             & $ -1 \leq \cos\theta_{\rm z} \leq 0 $ 
             & $  0 <    \cos\theta_{\rm z} \leq 1 $  & &\\ 
   \hline
  \hline
 $ 5.0- 5.5$ & $ 82.3^{+ 10.3}_{-  9.9}$ & $ 93.4^{+ 15.7}_{- 14.9}$ & $ 72.6^{+ 13.7}_{- 13.0}$ & 189.7 & 0.334 \\
 $ 5.5- 6.0$ & $ 66.4^{+  6.4}_{-  6.1}$ & $ 73.7^{+  9.8}_{-  9.3}$ & $ 59.9^{+  8.4}_{-  7.9}$ & 172.2 & 0.321 \\
 $ 6.0- 6.5$ & $ 62.9^{+  4.9}_{-  4.7}$ & $ 55.3^{+  7.0}_{-  6.5}$ & $ 70.4^{+  7.1}_{-  6.7}$ & 155.2 & 0.310 \\
 $ 6.5- 7.0$ & $ 54.8^{+  2.7}_{-  2.6}$ & $ 50.8^{+  3.8}_{-  3.7}$ & $ 58.7^{+  3.8}_{-  3.7}$ & 134.3 & 0.289 \\
 $ 7.0- 7.5$ & $ 53.8^{+  2.5}_{-  2.4}$ & $ 55.6^{+  3.6}_{-  3.5}$ & $ 52.1^{+  3.5}_{-  3.3}$ & 117.1 & 0.271 \\
 $ 7.5- 8.0$ & $ 40.4^{+  2.2}_{-  2.1}$ & $ 39.6^{+  3.1}_{-  3.0}$ & $ 41.1^{+  3.1}_{-  2.9}$ & 101.2 & 0.257 \\
 $ 8.0- 8.5$ & $ 36.4^{+  1.9}_{-  1.8}$ & $ 37.2^{+  2.7}_{-  2.6}$ & $ 35.7^{+  2.6}_{-  2.5}$ &  85.8 & 0.240 \\
 $ 8.5- 9.0$ & $ 30.5^{+  1.7}_{-  1.6}$ & $ 28.4^{+  2.3}_{-  2.2}$ & $ 32.6^{+  2.4}_{-  2.2}$ &  71.7 & 0.223 \\
 $ 9.0- 9.5$ & $ 22.4^{+  1.4}_{-  1.3}$ & $ 19.8^{+  1.9}_{-  1.8}$ & $ 24.9^{+  2.1}_{-  1.9}$ &  58.5 & 0.205 \\
 $ 9.5-10.0$ & $ 19.1^{+  1.2}_{-  1.2}$ & $ 17.7^{+  1.7}_{-  1.6}$ & $ 20.3^{+  1.8}_{-  1.7}$ &  47.1 & 0.186 \\
 $10.0-10.5$ & $ 14.3^{+  1.0}_{-  1.0}$ & $ 15.0^{+  1.5}_{-  1.4}$ & $ 13.6^{+  1.4}_{-  1.3}$ &  37.0 & 0.169 \\
 $10.5-11.0$ & $ 13.7^{+  1.0}_{-  0.9}$ & $ 14.7^{+  1.4}_{-  1.3}$ & $ 12.9^{+  1.3}_{-  1.2}$ &  28.5 & 0.151 \\
 $11.0-11.5$ & $9.41^{+ 0.79}_{- 0.73}$ & $ 9.36^{+ 1.17}_{- 1.03}$ & $ 9.44^{+ 1.11}_{- 0.98}$ & 21.45 & 0.134 \\
 $11.5-12.0$ & $5.63^{+ 0.64}_{- 0.57}$ & $ 5.24^{+ 0.90}_{- 0.76}$ & $ 6.04^{+ 0.94}_{- 0.81}$ & 15.76 & 0.118 \\
 $12.0-12.5$ & $4.91^{+ 0.57}_{- 0.50}$ & $ 4.08^{+ 0.79}_{- 0.66}$ & $ 5.69^{+ 0.85}_{- 0.73}$ & 11.21 & 0.102 \\
 $12.5-13.0$ & $3.03^{+ 0.44}_{- 0.38}$ & $ 2.67^{+ 0.61}_{- 0.49}$ & $ 3.38^{+ 0.65}_{- 0.53}$ &  7.79 & 0.088 \\
 $13.0-13.5$ & $1.92^{+ 0.35}_{- 0.29}$ & $ 1.59^{+ 0.47}_{- 0.35}$ & $ 2.25^{+ 0.55}_{- 0.43}$ &  5.22 & 0.074 \\
 $13.5-14.0$ & $1.32^{+ 0.29}_{- 0.23}$ & $ 1.13^{+ 0.39}_{- 0.27}$ & $ 1.48^{+ 0.47}_{- 0.35}$ &  3.39 & 0.062 \\
 $14.0-15.0$ & $2.15^{+ 0.36}_{- 0.30}$ & $ 2.00^{+ 0.51}_{- 0.40}$ & $ 2.31^{+ 0.53}_{- 0.42}$ &  3.49 & 0.092 \\
 $15.0-16.0$ & $0.832^{+0.234}_{-0.175}$ & $0.381^{+0.289}_{-0.158}$ & $1.208^{+0.385}_{-0.275}$ & 1.227 & 0.059 \\
 $16.0-20.0$ & $0.112^{+0.130}_{-0.064}$ & $0.244^{+0.238}_{-0.117}$ & $0.000^{+0.123}_{-0.401}$ & 0.513 & 0.068 \\
\hline
 \end{tabular}
 \end{center}
\end{table*}

\begin{figure}[tbh]
 \includegraphics[width=7.0cm,clip]{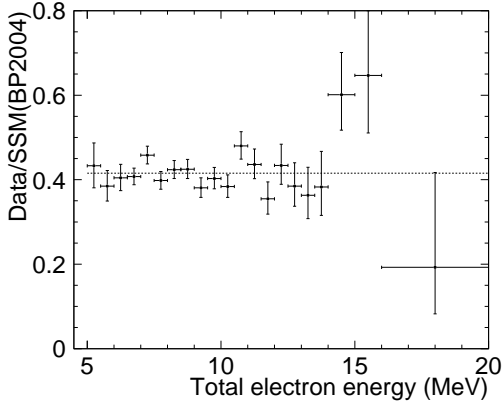}
 \caption{Revised ratio of observed and expected energy spectra in SK-III. 
 The dashed line represents the revised SK-III average.
 This is a revision of Fig.\ 27 in the previous paper~\protect{\cite{sk3}}.}
 \label{fig:energy_spectrum-sk3}
\end{figure}
The systematic uncertainties on total flux in SK-III are also revised.
The revised uncertainties are summarized in Table~\ref{tab:totalsys-sk3}.
The $^8$B spectrum error was underestimated in the analysis in the main
text of~\cite{sk3}.
The revised systematic uncertainty on the total flux in E$_{\rm total}$ = 5.0--20.0 MeV
in SK-III is estimated to be 2.2$\%$.
\vspace{5mm}

\subsection{$^8$B solar neutrino flux results}

The observed number of solar neutrino events is also updated.
In this analysis, the extracted number of $^8$B solar neutrinos
with the ES reaction in E$_{\rm total}$ = 5.0--20.0 MeV
for a live time of 548 days of SK-III data
was 8148$^{+133}_{-131}$(stat) $\pm176$(sys).
The corresponding $^8$B flux is obtained to be: 
\begin{eqnarray*}
 2.404 \pm 0.039 (\textrm{stat.}) \pm 0.053 (\textrm{sys.}) \times 10^ 6~\textrm{cm}^{-2} \textrm{sec}^{-1}.
\end{eqnarray*}
Fixing the cross section problem, a 3.4\% increase was observed.

The observed and expected fluxes are re-estimated in each energy region.
Table~\ref{tab:rates-sk3} shows the revised event rate in each energy region.
Figure~\ref{fig:energy_spectrum-sk3} shows the revised observed energy spectrum 
divided by the $5.79\times10^{6}$ cm$^{-2}$sec$^{-1}$ flux value without 
oscillations.

\clearpage

\setcounter{figure}{0} \renewcommand{\thefigure}{B.\arabic{figure}} 
\setcounter{table}{0} \renewcommand{\thetable}{B.\arabic{table}} 

\section{Parametrized Survival Probability Fit}
\label{appsec:peefit}

\begin{figure*}[thb]
\includegraphics[width=8.5cm,clip]{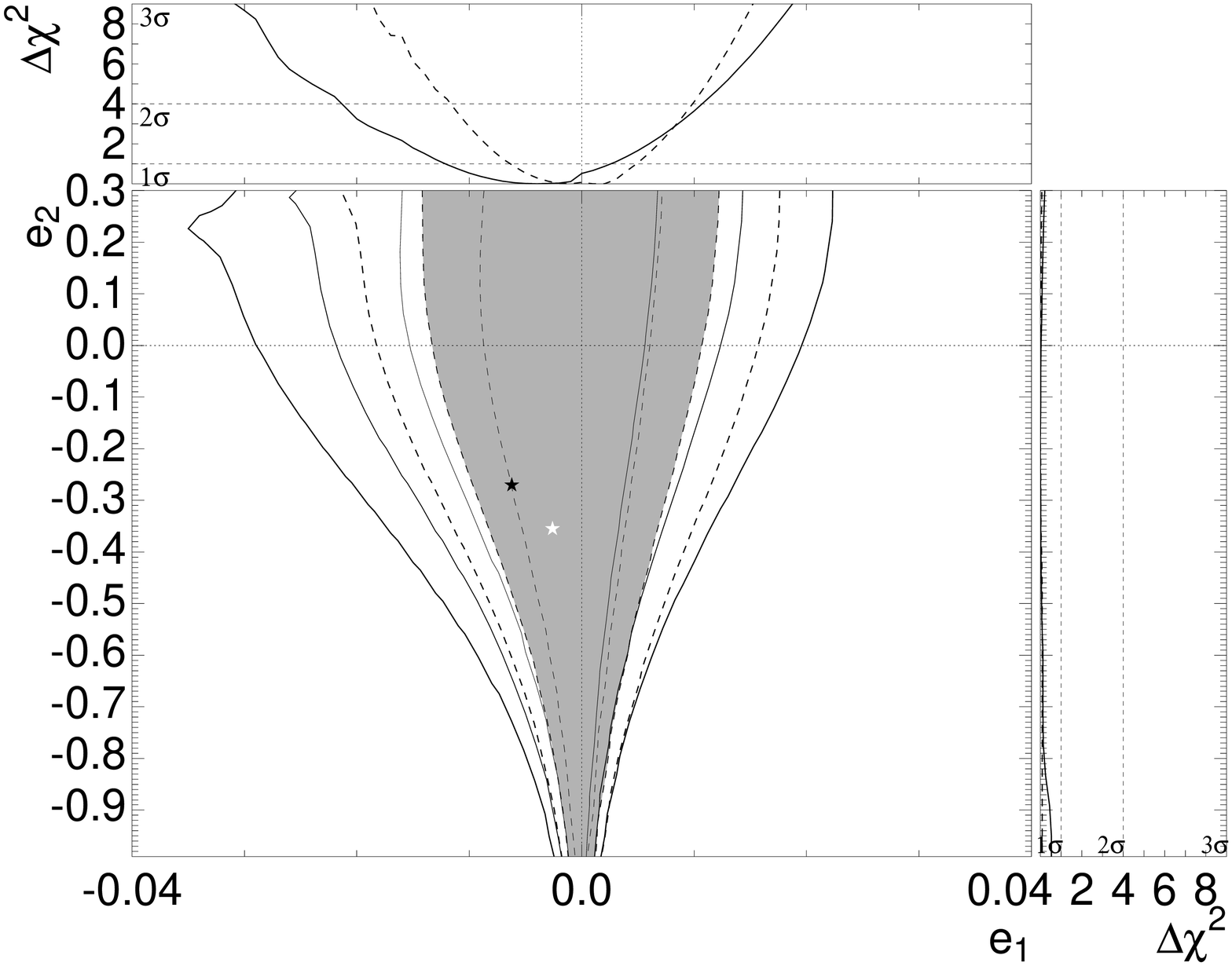}
\includegraphics[width=8.5cm,clip]{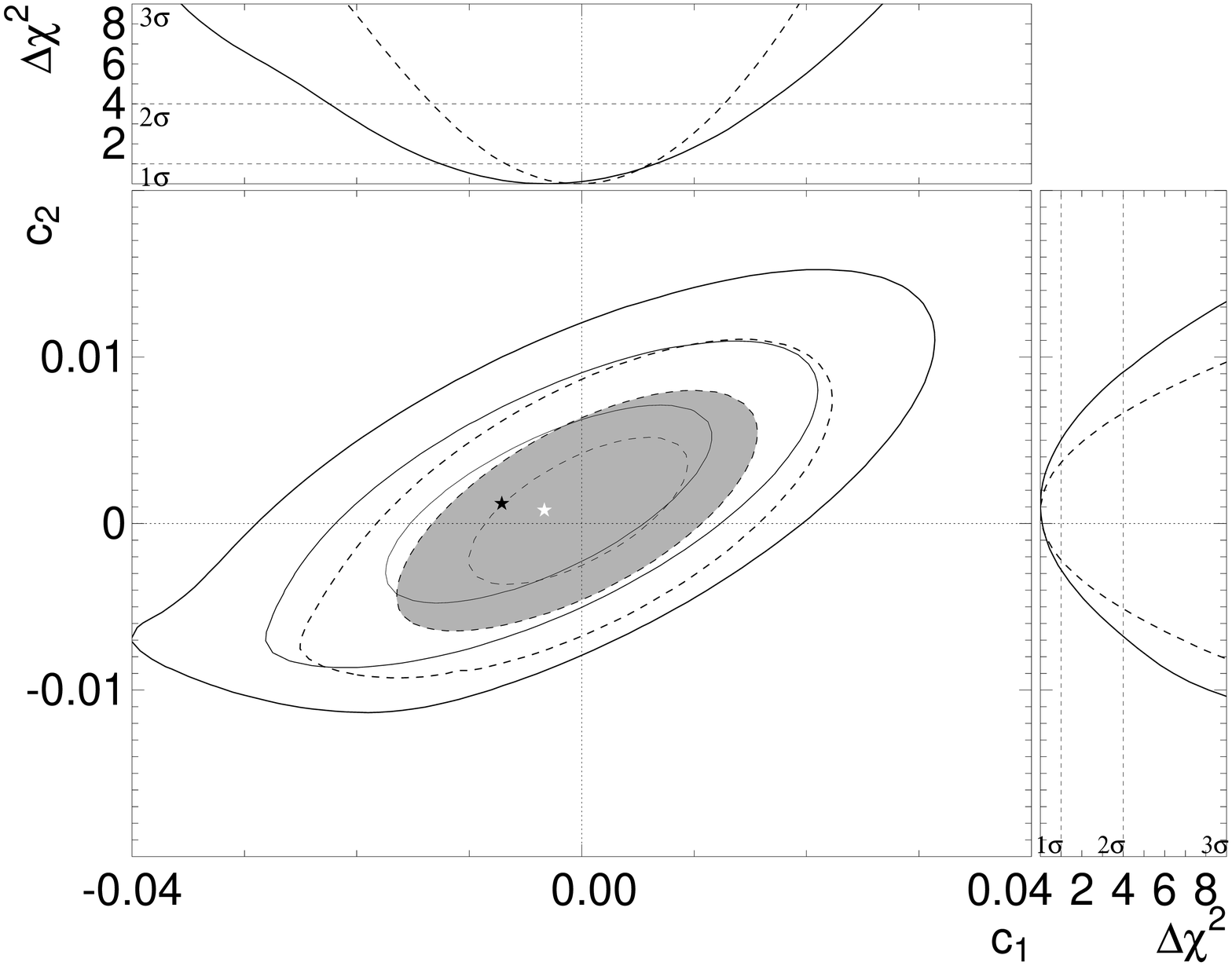}
\caption{Allowed areas of the shape parameters ($e_1$ and $e_2$ on left,
$c_1$ and $c_2$ on the right) of an exponential (left) and quadratic
(right) fit to the survival probability $P_{ee}$ of SK-IV
(solid lines) and all SK data (dashed lines) at the 1, 2
(filled region) and 3 $\sigma$ confidence levels. The
oscillation parameter set corresponding to the SK (or all solar neutrino) data
best-fit is indicated by the white star. The solar+KamLAND
best-fit (black star) is also shown.}
\label{figspecshapecoef}
\end{figure*}

\begin{figure*}[thb]
\includegraphics[width=8.8cm,clip]{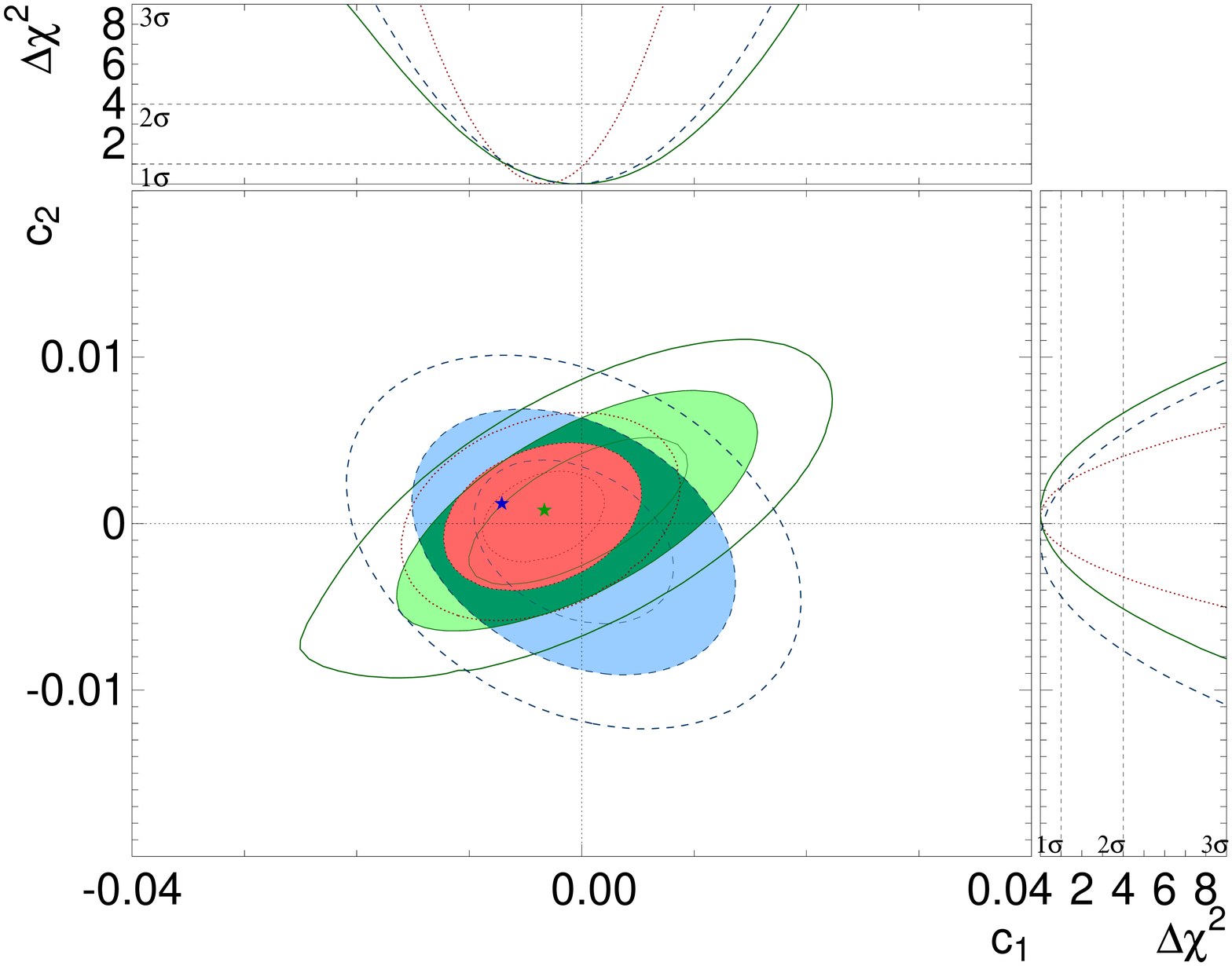}
\includegraphics[width=8.8cm,clip]{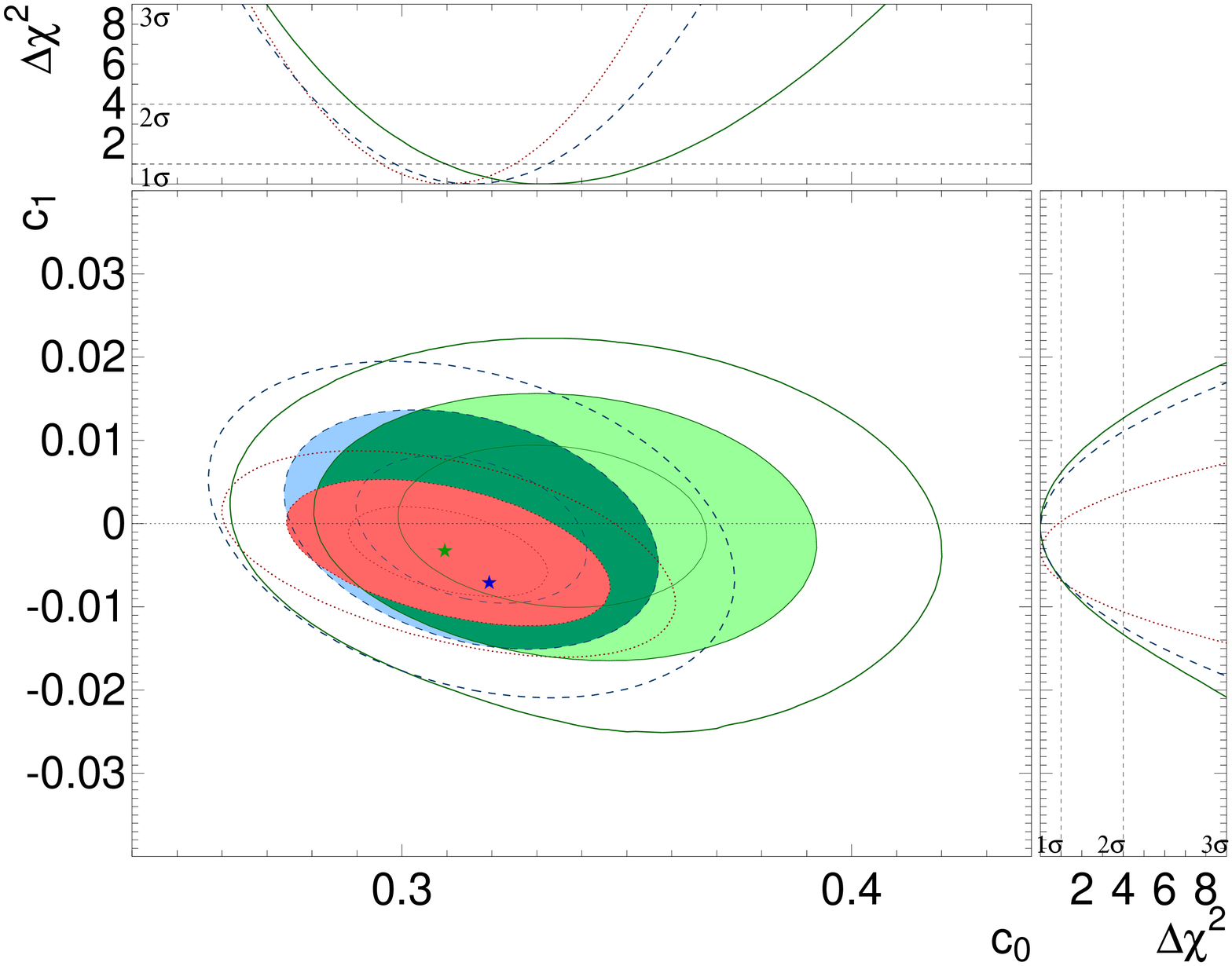}
\caption{Left: Allowed areas of the shape parameters ($c_1$ and $c_2$) of a
quadratic fit to the survival probability $P_{ee}$ of SK (solid green) and SNO
(dashed blue) data at the 1, 2 (filled region) and 3 $\sigma$ confidence levels. 
Right: Allowed areas of the slope ($c_1$) and baseline ($c_0$) of a
quadratic fit to the survival probability $P_{ee}$ of SK (solid green) and SNO
(dashed blue) data at the 1, 2 (filled region) and 3 $\sigma$ confidence
levels. Also shown is a combined fit (dotted red). The oscillation parameter set
corresponding to the SK (all solar neutrino) data best-fit is indicated by the
dark green (light blue) star. The solar+KamLAND best-fit (dark blue) is also
shown.}
\label{figspecc2vsc1andc1vsc0}
\end{figure*}

We fit the SK spectral data to the exponential, quadratic, and cubic survival probability
in the same manner as we fit them to the MSW prediction. Fig.~\ref{figspecshapecoef}
shows the resulting allowed areas of the exponential coefficients $e_1$ and $e_2$. The
``baseline'' (average $P_{ee}$) $e_0$ is profiled; the $e_0$ constraint results from the
comparison of the electron elastic scattering rate in SK and the SNO
neutral-current interaction rate on deuterium. The contours deviate from a multivariate
Gaussian. As there is no significant deviation from an undistorted spectrum, the
data impose no constraint on $e_2$, the ``steepness'' of the exponential.
Table~\ref{tabspecpeecoef} uses the
best quadratic form approximation of the $\chi^2$ of the fit as a function of the
parameters to extract the values, uncertainties and correlations. 
Fig.~\ref{figspecc2vsc1andc1vsc0} shows the allowed shape parameters ($c_1$ and $c_2$)
and the allowed slope ($c_1$) versus the baseline ($c_0$) of the quadratic fit.
The SK-IV contours show some deviations from a multivariate Gaussian at
$3\sigma$, while the SK combined result is consistent with it.
Overlaid in blue are the constraints from the SNO measurements. The
corresponding coefficients of Table~\ref{tabspecpeecoef} differ slightly from those
in~\cite{snopaper} which fits both the survival probability to a quadratic function and the
energy dependent day/night asymmetry to a linerar function. Here, we assume
the energy dependence of the day/night effect calculated from
standard earth matter effects. The resulting reduction in the degree of freedom leads
to somewhat tighter constraints as well as a slight shift in the best fit value.
The precision of the SK constraint is similar to that based on SNO data,
and also statistically consistent.  Since SK's
correlation between $c_1$ and $c_2$ is opposite to that of SNO's, a combined
fit is rather powerful in constraining the shape.
The $c_1-c_2$ correlation is slightly smaller. The addition of SK data
to SNO data not only significantly increases the precision of the
$c_0$ determination, but the uncertainties on the shape are
reduced.

\begin{table*}[]
\caption{SK exponential and polynomial best-fit coefficients and their
correlations. Also given are SNO's quadratic fit coefficients
(slightly different than the published value since the day/night
asymmetry is not fit) as well as SK and SNO combined measured quadratic fit
coefficients and their respective correlations.}
\begin{tabular}{l c c c c c c c}
\hline\hline
Data Set        & $e_0$            & $e_1$               & $e_2$ &
$e_0$-$e_1$ corr. \cr
\hline
SK-IV           & $0.326\pm0.024$  & $-0.0029\pm0.0073$ & no constraint &
$+0.202$  \cr
SK              & $0.336\pm0.023$  & $-0.0014\pm0.0051$ & no constraint &
$+0.077$  \cr
\hline
                & \multicolumn{3}{c}{quadratic function} &
                  \multicolumn{4}{c}{cubic function} \cr
                & $c_0$            & $c_1$               & $c_2$       &
$c_0$           & $c_1$            & $c_2$               & $c_3$ \cr
\hline
SK-IV           & $0.324\pm0.025$  & $-0.0030\pm0.0097$ & $0.0012\pm0.0040$ &
$0.313\pm0.028$ & $-0.018\pm0.021$ & $0.0059\pm0.0074$  & $0.0021\pm0.0028$ \cr
\hline
$c_0$           & $1$              & $-0.125$           & $-0.412$ &
$1$             & $+0.388$         & $-0.602$           & $-0.488$  \cr
$c_1$           & $-0.125$         & $1$                & $+0.6830$ & 
$+0.388$        & $1$              & $-0.580$           & $-0.892$ \cr
$c_2$           & $-0.412$         & $+0.683$           & $1$ &
$-0.602$        & $-0.580$         & $1$                & $+0.839$ \cr
$c_3$ & & & &
$-0.488$        & $-0.892$         & $+0.839$           & $1$ \cr
\hline
SK              & $0.334\pm0.023$  & $-0.0003\pm0.0065$ & $0.0008\pm0.0029$ &
$0.313\pm0.024$ & $-0.031\pm0.016$ & $0.0097\pm0.0051$  & $0.0044\pm0.0020$ \cr
\hline
$c_0$           & $1$              & $-0.131$           & $-0.345$ &
$1$             & $+0.258$         & $-0.449$           & $-0.327$  \cr
$c_1$           & $-0.131$         & $1$                & $+0.649$ & 
$+0.258$        & $1$              & $-0.599$           & $-0.916$ \cr
$c_2$           & $-0.345$         & $+0.649$           & $1$ &
$-0.449$        & $-0.599$         & $1$                & $+0.814$ \cr
$c_3$ & & & &
$-0.327$        & $-0.916$         & $+0.814$           & $1$ \cr
\hline
 & & & & $c_0$-$c_1$ corr. & $c_0$-$c_2$ corr. & $c_1$-$c_2$ corr.\cr
\hline
SNO     & $0.315\pm0.017$ & $-0.0007\pm0.0059$ & $-0.0011\pm0.0033$ &
$-0.301$    & $-0.391$    & $-0.312$ \cr
SK+SNO  & $0.311\pm0.015$ & $-0.0034\pm0.0036$ & $+0.0004\pm0.0018$ &
$-0.453$    & $-0.407$    & $+0.301$ \cr
\hline\hline
\end{tabular}
\label{tabspecpeecoef}
\end{table*}

\setcounter{figure}{0} \renewcommand{\thefigure}{C.\arabic{figure}} 
\setcounter{table}{0} \renewcommand{\thetable}{C.\arabic{table}} 

\section{Tables}

\begin{table*}[]
\begin{center}
\caption{
The observed event rates in each energy bin (events/year/kton), at 1 
AU. The errors are statistical errors only. The reduction efficiencies are
corrected and the expected event rates are for a flux of
5.25$\times10^6$ /(cm$^{2}$sec).}
\begin{tabular}{c c c c c c} \hline\hline
Energy & \multicolumn{3}{c}{Observed Rate} & \multicolumn{2}{c}{Expected Rate}\\
  (MeV)   &  ALL              &     DAY                  & NIGHT              &  $^8$B & $hep$ \\ 
             & $ -1 \leq \cos\theta_{\rm z} \leq 1 $   & $ -1 \leq \cos\theta_{\rm z} \leq 0 $  & $  0 <    \cos\theta_{\rm z} \leq 1 $  & &\\ 
\hline
 $ 3.49- 3.99$ & $ 92.2^{+ 10.8}_{- 10.6}$ & $ 96.0^{+ 16.8}_{- 16.3}$ & $ 81.5^{+ 14.0}_{- 13.6}$ & 196.8 & 0.346 \\
 $ 3.99- 4.49$ & $ 76.7^{+  5.2}_{-  5.1}$ & $ 64.6^{+  7.9}_{-  7.6}$ & $ 85.2^{+  6.9}_{-  6.7}$ & 182.8 & 0.335 \\
 $ 4.49- 4.99$ & $ 82.1^{+  3.4}_{-  3.3}$ & $ 79.4^{+  5.1}_{-  5.0}$ & $ 84.6^{+  4.6}_{-  4.5}$ & 167.8 & 0.323 \\
 $ 4.99- 5.49$ & $ 69.3^{+  2.1}_{-  2.1}$ & $ 65.9^{+  3.1}_{-  3.0}$ & $ 72.5^{+  3.0}_{-  2.9}$ & 153.3 & 0.312 \\
 $ 5.49- 5.99$ & $ 59.6^{+  1.6}_{-  1.6}$ & $ 58.3^{+  2.3}_{-  2.2}$ & $ 60.5^{+  2.2}_{-  2.2}$ & 137.8 & 0.298 \\
 $ 5.99- 6.49$ & $ 54.2^{+  1.4}_{-  1.4}$ & $ 51.0^{+  2.1}_{-  2.0}$ & $ 56.9^{+  2.0}_{-  2.0}$ & 121.9 & 0.282 \\
 $ 6.49- 6.99$ & $ 47.8^{+  1.3}_{-  1.3}$ & $ 45.7^{+  1.9}_{-  1.8}$ & $ 49.9^{+  1.9}_{-  1.8}$ & 106.8 & 0.266 \\
 $ 6.99- 7.49$ & $ 40.6^{+  1.2}_{-  1.1}$ & $ 41.8^{+  1.7}_{-  1.7}$ & $ 39.5^{+  1.6}_{-  1.6}$ &  92.1 & 0.250 \\
 $ 7.49- 7.99$ & $ 35.7^{+  1.0}_{-  1.0}$ & $ 35.0^{+  1.5}_{-  1.5}$ & $ 36.1^{+  1.5}_{-  1.4}$ &  78.0 & 0.232 \\
 $ 7.99- 8.49$ & $ 29.1^{+  0.9}_{-  0.9}$ & $ 28.6^{+  1.3}_{-  1.3}$ & $ 28.9^{+  1.3}_{-  1.2}$ &  65.2 & 0.214 \\
 $ 8.49- 8.99$ & $ 24.0^{+  0.8}_{-  0.8}$ & $ 24.1^{+  1.2}_{-  1.1}$ & $ 23.7^{+  1.1}_{-  1.1}$ &  53.4 & 0.197 \\
 $ 8.99- 9.49$ & $ 18.5^{+  0.7}_{-  0.7}$ & $ 17.9^{+  1.0}_{-  0.9}$ & $ 19.2^{+  1.0}_{-  0.9}$ &  42.9 & 0.179 \\
 $ 9.49-9.99$ & $ 14.5^{+  0.6}_{-  0.6}$ & $ 14.5^{+  0.9}_{-  0.8}$ & $ 14.4^{+  0.8}_{-  0.8}$ &  33.8 & 0.162 \\
 $9.99-10.5$ & $ 10.7^{+  0.5}_{-  0.5}$ & $ 10.2^{+  0.7}_{-  0.7}$ & $ 11.1^{+  0.7}_{-  0.7}$ &  26.0 & 0.144 \\
 $10.5-11.0$ & $8.43^{+ 0.43}_{- 0.41}$ & $ 7.73^{+ 0.61}_{- 0.56}$ & $ 9.23^{+ 0.64}_{- 0.60}$ & 19.55 & 0.128 \\
 $11.0-11.5$ & $6.60^{+ 0.37}_{- 0.35}$ & $ 6.60^{+ 0.54}_{- 0.49}$ & $ 6.72^{+ 0.53}_{- 0.49}$ & 14.34 & 0.112 \\
 $11.5-12.0$ & $4.40^{+ 0.30}_{- 0.28}$ & $ 3.83^{+ 0.41}_{- 0.37}$ & $ 4.89^{+ 0.44}_{- 0.40}$ & 10.24 & 0.097 \\
 $12.0-12.5$ & $3.04^{+ 0.25}_{- 0.23}$ & $ 3.04^{+ 0.35}_{- 0.31}$ & $ 3.06^{+ 0.36}_{- 0.32}$ &  7.10 & 0.083 \\
 $12.5-13.0$ & $2.14^{+ 0.20}_{- 0.18}$ & $ 2.41^{+ 0.31}_{- 0.27}$ & $ 1.93^{+ 0.29}_{- 0.25}$ &  4.80 & 0.070 \\
 $13.0-13.5$ & $1.47^{+ 0.17}_{- 0.15}$ & $ 1.48^{+ 0.25}_{- 0.21}$ & $ 1.47^{+ 0.25}_{- 0.21}$ &  3.11 & 0.059 \\
 $13.5-14.5$ & $1.59^{+ 0.17}_{- 0.15}$ & $ 1.54^{+ 0.25}_{- 0.21}$ & $ 1.63^{+ 0.25}_{- 0.22}$ &  3.18 & 0.088 \\
 $14.5-15.5$ & $0.469^{+0.102}_{-0.082}$ & $0.486^{+0.151}_{-0.112}$ & $0.493^{+0.161}_{-0.121}$ & 1.117 & 0.056 \\
 $15.5-19.5$ & $0.186^{+0.072}_{-0.051}$ & $0.150^{+0.108}_{-0.065}$ & $0.203^{+0.113}_{-0.071}$ & 0.464 & 0.064 \\
\hline \hline
\end{tabular}
\label{tab:spectrumsignal}
\end{center}
\end{table*}

\begin{table*}[]
\caption{Elastic scattering rate ratios and energy-uncorrelated
uncertainties (statistical plus systematic) for each SK phase.}
\centerline{\begin{tabular}{l c c c c}
\hline\hline
Energy (MeV) & SK-I & SK-II & SK-III & SK-IV\cr \hline
3.49-3.99   & $-$                       & $-$                       &
                $-$ & $0.468^{+0.060}_{-0.059}$ \cr
3.99-4.49   & $-$                       & $-$                       &
                $0.448^{+0.100}_{-0.096}$ & $0.419\!\pm\!0.030$ \cr
4.49-4.99   & $0.453^{+0.043}_{-0.042}$ & $-$                       &
                $0.472^{+0.058}_{-0.056}$ & $0.488\!\pm\!0.023$ \cr
4.99-5.49   & $0.430^{+0.023}_{-0.022}$ & $-$                       &
                $0.420^{+0.039}_{-0.037}$ & $0.451\!\pm\!0.014$ \cr
5.49-5.99   & $0.449\!\pm\!0.018$       & $-$                       &
                $0.457^{+0.035}_{-0.034}$ & $0.432\!\pm\!0.012$ \cr
5.99-6.49   & $0.444\!\pm\!0.015$       & $-$                       &
                $0.433^{+0.023}_{-0.022}$ & $0.444\!\pm\!0.015$ \cr
6.49-6.99   & $0.461^{+0.016}_{-0.015}$ & $0.439^{+0.050}_{-0.048}$ &
                $0.504^{+0.025}_{-0.024}$ & $0.447\!\pm\!0.015$ \cr
6.99-7.49   & $0.476\!\pm\!0.016$       & $0.448^{+0.043}_{-0.041}$ &
                $0.424^{+0.024}_{-0.023}$ & $0.440\!\pm\!0.015$ \cr
7.49-7.99   & $0.457^{+0.017}_{-0.016}$ & $0.461^{+0.037}_{-0.036}$ &
                $0.467^{+0.024}_{-0.023}$ & $0.455\!\pm\!0.014$ \cr
7.99-8.49   & $0.431^{+0.017}_{-0.016}$ & $0.473^{+0.036}_{-0.035}$ &
                $0.469^{+0.026}_{-0.025}$ & $0.439^{+0.015}_{-0.014}$ \cr
8.49-8.99   & $0.454^{+0.018}_{-0.017}$ & $0.463^{+0.036}_{-0.034}$ &
                $0.420^{+0.026}_{-0.025}$ & $0.445^{+0.016}_{-0.015}$ \cr
8.99-9.49   & $0.464\!\pm\!0.019$       & $0.499^{+0.038}_{-0.037}$ &
                $0.444^{+0.029}_{-0.027}$ & $0.430\!\pm\!0.016$ \cr
9.49-9.99   & $0.456^{+0.021}_{-0.020}$ & $0.474^{+0.038}_{-0.036}$ &
                $0.423^{+0.031}_{-0.029}$ & $0.426^{+0.018}_{-0.017}$ \cr
9.99-10.5   & $0.409\!\pm\!0.021$       & $0.481^{+0.041}_{-0.039}$ &
                $0.529^{+0.037}_{-0.035}$ & $0.408^{+0.019}_{-0.018}$ \cr
10.5-11.0   & $0.472^{+0.025}_{-0.024}$ & $0.452^{+0.043}_{-0.040}$ &
                $0.481^{+0.041}_{-0.037}$ & $0.432^{+0.023}_{-0.021}$ \cr
11.0-11.5   & $0.439^{+0.028}_{-0.026}$ & $0.469^{+0.046}_{-0.043}$ &
$0.391^{+0.044}_{-0.040}$ & $0.461^{+0.026}_{-0.025}$ \cr
11.5-12.0   & $0.460^{+0.033}_{-0.031}$ & $0.482^{+0.052}_{-0.048}$ &
                $0.479^{+0.055}_{-0.049}$ & $0.423^{+0.029}_{-0.027}$ \cr
12.0-12.5   & $0.465^{+0.039}_{-0.036}$ & $0.419^{+0.054}_{-0.049}$ &
                $0.425^{+0.061}_{-0.053}$ & $0.425^{+0.035}_{-0.032}$ \cr
12.5-13.0   & $0.461^{+0.048}_{-0.043}$ & $0.462^{+0.063}_{-0.057}$ &
                $0.400^{+0.073}_{-0.061}$ & $0.445^{+0.043}_{-0.039}$ \cr
13.0-13.5  & $0.582^{+0.064}_{-0.057}$ & $0.444^{+0.070}_{-0.062}$ &
                $0.422^{+0.093}_{-0.074}$ & $0.465^{+0.055}_{-0.049}$ \cr
13.5-14.5 & $0.475^{+0.059}_{-0.052}$ & $0.430^{+0.066}_{-0.059}$ &
                $0.663^{+0.110}_{-0.093}$ & $0.485^{+0.054}_{-0.048}$ \cr
14.5-15.5 & $0.724^{+0.120}_{-0.102}$ & $0.563^{+0.100}_{-0.087}$ &
                $0.713^{+0.201}_{-0.150}$ & $0.418^{+0.090}_{-0.074}$ \cr
15.5-19.5 & $0.575^{+0.173}_{-0.130}$ & $0.648^{+0.123}_{-0.103}$ &
                $0.212^{+0.248}_{-0.122}$ & $0.338^{+0.140}_{-0.099}$ \cr
\hline\hline
\end{tabular}}
\label{tabspecdata}
\end{table*}

\end{document}